\documentclass[preprint]{elsarticle}
\pdfoutput=1
\usepackage[latin1]{inputenc}
\usepackage{psfrag,epsfig}      
\usepackage{amssymb,amsmath,mathrsfs,amsthm}
\usepackage{array}
\usepackage{mathtools}
\usepackage{longtable,multirow}
\usepackage{color}             
\usepackage{enumerate}          
\usepackage{extarrows}
\usepackage[hang]{caption}
\usepackage[titletoc,title]{appendix}
\usepackage{sansmath}
\usepackage{MnSymbol}
\usepackage{stackengine}
\usepackage{relsize}
\usepackage{float}
\usepackage{xcolor}	
\usepackage{graphicx}
\usepackage[labelfont=bf]{caption}
\usepackage{xfrac}%defines sfrac
\usepackage[obeyFinal]{todonotes} % final -> todo vanish
\usepackage{siunitx}
\usepackage{tikzscale,pgfplots}
\usepackage{wasysym}
\usepackage[final]{changes}
\definechangesauthor[name=First reviewer, color=blue]{1r}
\definechangesauthor[name=Second reviewer, color=orange]{2r}

\newcommand{\stkout}[1]{\ifmmode\text{\sout{\ensuremath{#1}}}\else\sout{#1}\fi}
\setdeletedmarkup{\stkout{#1}}

\usepackage[colorlinks]{hyperref}
\usepackage{cleveref}

\pgfplotsset{compat=1.13}
 
\newlength\figH
\newlength\figW
\pgfkeys{/pgf/number format/.cd,fixed,precision=4}
\graphicspath{{Figures/}{figures/}}

\newtheorem{theorem}    {Theorem}[section]

\theoremstyle{definition}

\newtheorem{examplex}[theorem]{$\triangleright\;$Ejemplo}

\theoremstyle{remark}
\newtheorem{remark}              {Remark}

\newcommand{\defined}{:=}
\newcommand{\mbs}[1]{\boldsymbol{#1}}

\newcommand{\pd}[2]{\frac{\partial #1}{\partial #2}}

\newcommand{\floor}{\mathop{\mathtt{floor}}}
\renewcommand{\texttt}[1]{{\color{blue}{\ttfamily #1}}}

\pgfplotsset{
every axis/.append style={
scale ticks above exponent=1
, scaled x ticks =false, scale ticks below exponent=0},
}
\makeatletter
\renewcommand{\todo}[2][]{\tikzexternaldisable\@todo[#1]{#2}\tikzexternalenable}

\makeatother
\pgfplotsset{soldot/.style={color=blue,only marks,mark=*}} \pgfplotsset{holdot/.style={color=blue,fill=white,only marks,mark=*}}

\newboolean{external}
\setboolean{external}{true} 				
\ifthenelse{\boolean{external}}{\usetikzlibrary{external}\tikzexternalize[prefix=tikz/,force remake=false]}{}
\ifthenelse{\boolean{external}}{}{}

\journal{Journal of Computational Physics}
\begin{document}
\begin{frontmatter}
\title{Energy-momentum conserving integration schemes for molecular dynamics}
\author[1]{Mark Schiebl}
\ead{mark.schiebl@kit.edu}
\address[1]{Institut f\"ur Mechanik, Karlsruher Institut f\"ur Technologie,
  Geb\"aude 10.30, Otto-Ammann-Platz 9, D-76131 Karlsruhe, Germany}
\author[2,3]{Ignacio Romero\corref{cor1}}
\ead{ignacio.romero@upm.es}
\cortext[cor1]{Corresponding author}
\address[2]{Universidad Polit\'ecnica de Madrid; Jos\'e Guti\'errez Abascal,2; 28006 Madrid; Spain} 
\address[3]{IMDEA Materials; Eric Kandel,2; 28906 Madrid; Spain}
%
%\maketitle
\begin{abstract}
  We address the formulation and analysis of energy and momentum conserving
  time integration schemes in the context of particle dynamics, and in particular atomic systems. The
  article identifies three critical aspects of these models that
  demand a careful analysis when discretized: first, the treatment of periodic boundary conditions;
  second, the formulation of
  approximations of systems with three-body interaction forces;
  third, their extension to atomic systems with functional potentials.
  These issues, and in particular their interplay with Energy-Momentum integrators,
  are studied in detail. Novel expressions for these time integration schemes are
  proposed and numerical examples are given to illustrate their performance.
\end{abstract}

  
\end{frontmatter}

\section{Introduction}
Energy and momentum conserving algorithms are a frequently used class of structure preserving
integration schemes  \cite{hairer2006geometric,leimkuhler2004simulating}. Their remarkable
robustness and their good qualitative accuracy have made them popular choices for simulating the
governing equations of particle dynamics \cite{labudde1976energy,labudde1975energy}, nonlinear solid
mechanics \cite{simo1992discrete,Gonzalez:2000wx}, nonlinear shells and rods
\cite{Simo:1994tx,Simo:1995tz,Zhong:1998iu,Romero:2002uw}, multibody dynamics
\cite{GarciaOrden:2000tc,BeUh07}, gradient systems \cite{McLachlan:1999tm}, and  general PDEs
\cite{Celledoni:2012ff}.

Given the good properties of these methods, it is remarkable that they have not received more
attention in the field of molecular dynamics, or in general, molecular thermodynamics. The governing
equations of the latter are essentially Hamiltonian and fit seamlessly in the framework developed
since the 70s for integrating this type of problems, while preserving the energy and the momenta. It
would seem natural that integration schemes designed to preserve the main invariants of the motion
would give accurate predictions of the thermodynamic averages, of interest in many practical and
theoretical situations \cite{Toxvaerd1983,rapaport2004art,Skeel2007,allen2017computer}, but have
rarely been studied \cite{saluena2014molecular}. Instead, molecular dynamics codes seem to favor the
use of the explicit Verlet method or symplectic methods. While the latter have good properties in
terms of computational cost, accuracy, and geometry preservation, they lack energy conservation, a
key invariant most important in the simulation of microcanonical ensembles. \added[id=1r]{A thorough
  investigation of the accuracy of energy and momentum conserving schemes in capturing
  the statistical behavior of atomistic systems for long periods of simulation is lacking. Preliminary
  results \cite{saluena2014molecular} are promising, but much testing and validation is still
  required.}

The development and implementation of energy and momentum conserving algorithms in the context of
molecular dynamics has specific issues that affect the discretization of the equations and their
analysis, issues that do not appear in their application to nonlinear solids, shells, rods, etc., or any of
the other systems for which the use of these methods is widespread. The first critical issue is the
treatment of periodic boundary conditions. These are almost invariably required for the study of
average properties in particle systems \cite{allen2017computer,griebel2007numerical}, and demand a
careful analysis, especially to ascertain whether they spoil the conserving properties of the method
or not. Taking this into account requires to consider the geometry and topology of the
periodic configuration space, and might affect also the accuracy of the integration scheme.

The second issue that needs to be carefully dealt with is the use of conserving schemes in the
context of three-body potentials. These functions are employed in modeling angle interactions in
atomic bonding \cite{brooks1983charmm,stillinger1985computer}, and their impact on the global
behavior of some systems is so critical that needs to be accounted for. In fact, atomic systems with
potentials of this type allow for large relative motions among the particles, and this is precisely
the arena where conserving schemes have shown their superiority with regard to other implicit
integrators.

The third aspect that requires a detailed analysis is the application of conserving schemes to
mechanical systems in which the potential is based on cluster functionals
\cite{daw1984embedded,daw1989model,baskes1992modified,daw1993embedded,tadmor2011modeling}.  These
effective potentials are often required for the correct modeling of complex binding among metallic
atoms and need again a careful study when used in combination with conserving schemes.
While pair potentials of the Lennard-Jones type \cite{lennard1924determination}
have been employed together with Energy-Momentum conserving schemes \cite{labudde1976energy},
their formulation for cluster potentials needs to be specifically addressed.

In this article we formulate energy and momentum conserving schemes for the
simulation of the dynamics of atomic systems. Some of the methods discussed have
already been employed in the literature, and we identify new ones. In all cases,
we explain how the three critical issues identified before (periodic boundary
conditions, three-body potentials, functional potentials) affect their formulation
since none of the three have been studied, to our knowledge, before.

The rest of the article has the following structure. In \Cref{sec-projection} we
review the basic topology of periodic systems so as to clearly define
the distance function. \Cref{sec-particle} introduces particle dynamics, with
special attention to its Hamiltonian structure. Next, in \Cref{sec:pair-potentials},
conserving time integration schemes are presented for particle systems in periodic
domains, restricted to those with pair potentials. These are extended to systems with
angle potentials in \Cref{sec:three-body-potent} and to atomic systems
with functional potentials in \Cref{sec:embedded-atom-method}. In the
three cases of study, numerical simulations are provided to confirm the conservation
properties of the fully discrete method. The article concludes in \Cref{sec:conclusion}
with a summary of the main findings.

%%% Local Variables:
%%% TeX-command-extra-options: "-shell-escape"
%%% mode: latex
%%% TeX-master: "driver"
%%% End:

\section{The topology of periodic domains}
\label{sec-projection}
In this article we study the dynamic motion of systems of particles.
This type of problems is of great importance for the simulation of
matter at the atomic scale and a very large body of references study
details pertaining to their numerical solution and the information
that can be extracted from these simulations.

Many particle systems of interest are formulated in periodic domains. These
allow to study large systems by only discretizing representative volumes, much
smaller in size, while hopefully not losing too much information. In this
section we gather some topological and geometrical facts of periodic domains
that will be necessary to analyze numerical methods.

We start by considering a periodic three-dimensional box
$\mathcal{B}$ of side length~$L$, noting that all the results are
applicable to systems in one and two dimensions, with the corresponding
modifications. This box is isomorphic to the torus $\mathbb{T}^3$ (see
e.g. \cite{allen2017computer,rapaport2004art}), which itself can be identified with the product
manifold $S^1\times S^1\times S^1$.  Hence, each point $\mbs{\xi}\in\mathbb{T}^3$ can be uniquely
characterised by three angles $(\alpha,\beta,\gamma)$ and  the complete manifold is
covered by a single chart.

Numerical methods defined on $\mathbb{T}^3$ pose difficulties that can be alleviated by mapping
this set into a more convenient one. For that, let us first define the following equivalence
relation on $\mathbb{R}^3$: two points $\mbs{x},\mbs{y}\in\mathbb{R}^3$ are defined to
be equivalent, and indicated as
$\mbs{x}\sim \mbs{y}$, if there exists a triplet of integers $\mbs{z}\in\mathbb{Z}^3$ such that
$\mbs{x}= \mbs{y}+ L \mbs{z}$. Using this equivalence relation, we can define
the quotient space $\mathbb{P}\defined\mathbb{R}^3/\mathbb{Z}^3$ that is homeomorphic to
the torus.  In what follows the equivalent class of a point
$\mbs{x}\in\mathbb{R}^3$  will be denoted as $[\mbs{x}]\in\mathbb{P}$.

When dealing with systems of particles in periodic domains one has to choose one of the two
homeomorphic descriptions described above, namely, the torus and $\mathbb{P}$.
From the computational point of view, employing the latter has many advantages.  The first one is
that given a distance on $\mathbb{R}^3$, this quotient space naturally inherits a distance, and thus
a topology.  In terms of the standard Euclidean distance
$d(\cdot,\cdot):\mathbb{R}^3\times\mathbb{R}^3\to \mathbb{R}^+\cup\{0\}$ we can define
$d_T(\cdot,\cdot):\mathbb{P}\times\mathbb{P}\to \mathbb{R}^+\cup\{0\}$ by the relation
\begin{equation}
	d_T([\mbs{x}], [\mbs{y}]) = \inf_{\mbs{x}\in[\mbs{x}], \mbs{y}\in[\mbs{y}]} d( \mbs{x}, \mbs{y}) \ .
	\label{eq-distance-torus}
\end{equation}
Abusing slightly the notation, from this point we will write $d_T(\mbs{x},\mbs{y})$ instead of
$d_T([\mbs{x}],[\mbs{y}])$.

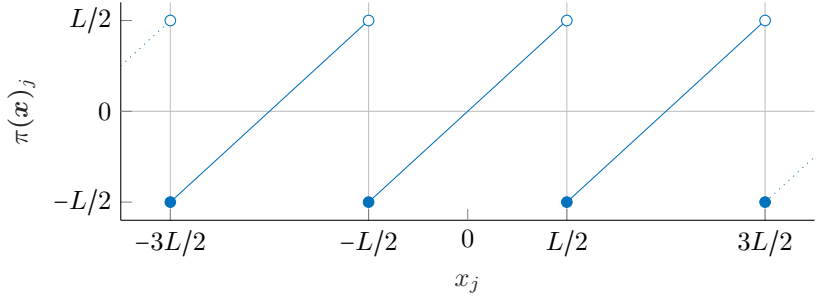
\begin{figure}[H]
  \centering
  \setlength{\figH}{0.15\textheight}
  \setlength{\figW}{0.8\textwidth}
\definecolor{mycolor1}{rgb}{0.00000,0.44700,0.74100}%
\definecolor{mycolor2}{rgb}{0.85000,0.32500,0.09800}%
\definecolor{mycolor3}{rgb}{0.92900,0.69400,0.12500}%
\definecolor{mycolor4}{rgb}{0.49400,0.18400,0.55600}%
\begin{tikzpicture}

\pgfmathsetmacro{\rmin}{0.8}
\pgfmathsetmacro{\rmax}{2}
\pgfmathsetmacro{\rcut}{1.8}
\begin{axis}[%
samples=400,
width=0.951\figW,
height=\figH,
at={(0\figW,0\figH)},
scale only axis,
xmin=-3.5,
xmax=3.5,
ymin=-1.2,
ymax=1.2,
restrict y to domain=-3:3,
xtick={-3,-1,1,3},
    xticklabels={$-3L/2$,$-L/2$,$L/2$,$3L/2$},
ytick={-1,1},
    yticklabels={$-L/2$,$L/2$},
axis background/.style={fill=white},
ylabel style={font=\color{white!15!black}},
ylabel={$\pi(\mbs{x})_j$},
xlabel style={font=\color{white!15!black}},
xlabel={$x_j$},
xmajorgrids = true,
axis x line*=bottom,
axis y line*=left,
extra x ticks={0},
extra x tick style={
        grid=none},
extra x tick labels={0},
extra y ticks={0},
extra y tick style={
        grid=major},
extra y tick labels={0},
legend style={legend pos = outer north east, legend cell align=left, align=left, draw=white!15!black}
]

\addplot[domain=-3:-1,mycolor1] {x+2};
\addplot[domain=-1:1,mycolor1] {x};
\addplot[domain=1:3,mycolor1] {x-2};
\addplot[dotted,domain=-3.5:-3,mycolor1] {x+4};
\addplot[dotted,domain=3:3.5,mycolor1] {x-4};
\addplot[only marks,mark=*,color=mycolor1,fill=white,forget plot] coordinates{(1,1)(3,1)(-1,1)(-3,1)};
\addplot[only marks,mark=*,color=mycolor1,fill=mycolor1,forget plot] coordinates{(-3,-1)(-1,-1)(1,-1)(3,-1)};

\end{axis}
\end{tikzpicture}%
  \caption{Graph of the projection operator $\mbs{\pi}$
    restricted to one of the three coordinates of Euclidean space.}
  \label{projection}
\end{figure}

The second advantage of such a choice is that, for each equivalent class $[\mbs{x}]$, there exists a
unique point $\bar{\mbs{x}}\in [ \mbs{x} ] \cap [-L/2,L/2)^3$ that serves as identifier of the whole
class which, in practical terms, implies that all operations need to be performed as with standard
points in a cubic box. This identifier can be found using a projection operator
\begin{equation}
  \mbs{\pi} : \mathbb{R}^3 \to \mathcal{B}
  \label{eq-projector1}
\end{equation}
defined as
\begin{equation}
  \added[id=1r]{\bar{x}_j = \pi(\mbs{x})_j = x_j - \floor\left(\frac{x_j+L/2}{L}\right) L,}
  \label{eq-projection}
\end{equation}
where $x_j, j=1,2,3$ denote the Cartesian coordinates of the point $\mbs{x}$ and
\added[id=1r]{$\floor:\mathbb{R}\to\mathbb{Z}$ is the function that
  gives the largest integer smaller than or equal to a given real number.}
See \Cref{projection} for an illustration of the projection operator.

The projection map $\mbs{\pi}$ determines some of the properties and limitations
of the numerical methods employed on systems with periodic boundary conditions,
and it is helpful to summarize some of its properties:
\begin{enumerate}
  
\item $\mbs{\pi}$ is a nonlinear,  surjective, projection, that is
$\mbs{\pi}\circ\mbs{\pi}=\mbs{\pi}$.

\item The point $\mbs{\pi}(\mbs{x})$ is the closest one to the origin among all points in
$[\mbs{x}]$, that is,
  \begin{equation}
    \mbs{\pi}(\mbs{x}) =
    \arg\inf_{\mbs{x}\in[\mbs{x}]} d(\mbs{x},\mbs{0})
    \label{eq-distance-from-origin}
  \end{equation}

\item In general, for arbitrary $\mbs{x},\mbs{y}\in\mathbb{R}^3$,
  \begin{equation}
    d_T(\mbs{x},\mbs{y})
    = \left| \mbs{\pi}( \mbs{x} - \mbs{y}) \right| \ne \left| \mbs{\pi}( \mbs{x})  - \mbs{\pi}(\mbs{y}) \right| .
    \label{eq-prop}
  \end{equation}

\item The map $\mbs{\pi}$ is $C^{\infty}$ except on the planes $x_i= L/2 + k_i L$, with $k_i\in
\mathbb{Z}$ and $i=1,2,3$, where it is discontinuous.  Away from these planes, the gradient
$\nabla\mbs{\pi}$ is the identity $\mbs{I}:\mathbb{R}^3\to\mathbb{R}^3$.   A one-dimensional
illustration of the gradient $\nabla\mbs{\pi}$ is given in \Cref{Dprojection}.

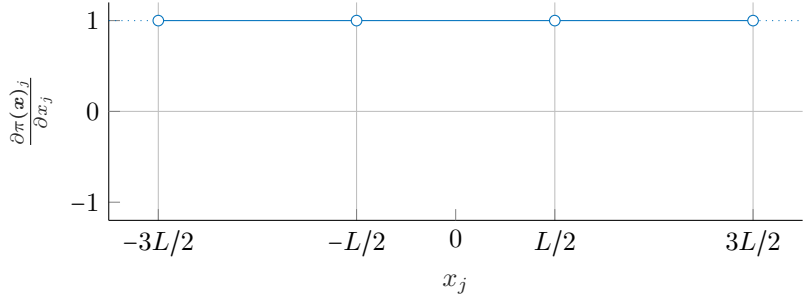
\begin{figure}[H]
  \centering
  \setlength{\figH}{0.15\textheight}
  \setlength{\figW}{0.8\textwidth}
\definecolor{mycolor1}{rgb}{0.00000,0.44700,0.74100}%
\definecolor{mycolor2}{rgb}{0.85000,0.32500,0.09800}%
\definecolor{mycolor3}{rgb}{0.92900,0.69400,0.12500}%
\definecolor{mycolor4}{rgb}{0.49400,0.18400,0.55600}%
\begin{tikzpicture}

\pgfmathsetmacro{\rmin}{0.8}
\pgfmathsetmacro{\rmax}{2}
\pgfmathsetmacro{\rcut}{1.8}
\begin{axis}[%
samples=400,
width=0.951\figW,
height=\figH,
at={(0\figW,0\figH)},
scale only axis,
xmin=-3.5,
xmax=3.5,
ymin=-1.2,
ymax=1.2,
restrict y to domain=-3:3,
xtick={-3,-1,1,3},
    xticklabels={$-3L/2$,$-L/2$,$L/2$,$3L/2$},
ytick={-1,1},
    yticklabels={$-1$,$1$},
axis background/.style={fill=white},
ylabel style={font=\color{white!15!black}},
ylabel={$\pd{\pi(\mbs{x})_j}{x_j}$},
xlabel style={font=\color{white!15!black}},
xlabel={$x_j$},
xmajorgrids = true,
axis x line*=bottom,
axis y line*=left,
extra x ticks={0},
extra x tick style={
        grid=none},
extra x tick labels={0},
extra y ticks={0},
extra y tick style={
        grid=major},
extra y tick labels={0},
legend style={legend pos = outer north east, legend cell align=left, align=left, draw=white!15!black}
]

\addplot[domain=-3:3,mycolor1] {1};
\addplot[dotted,domain=-3.5:-3,mycolor1] {1};
\addplot[dotted,domain=3:3.5,mycolor1] {1};
\addplot[only marks,mark=*,color=mycolor1,fill=white,forget plot] coordinates{(1,1)(3,1)(-1,1)(-3,1)};

\end{axis}
\end{tikzpicture}%
  \caption{Graph of the gradient of the projection operator $\nabla\mbs{\pi}$ restricted to one of the three coordinates of Euclidean space.}
  \label{Dprojection}
\end{figure}

\item For any $\mbs{x}\in\mathbb{R}^3$
  \begin{equation}
    \label{eq:proj-skewprop}
    \mbs{\pi}(\mbs{x})=-\mbs{\pi}(-\mbs{x})\ .
  \end{equation}
\item For any $\mbs{x},\mbs{y}\in\mathbb{R}^3$ and $\mbs{a}\in\mathbb{R}^3$
  \begin{equation}
    \label{eq:noether-transl}
    d_T(\mbs{x}+{\mbs{a}},\mbs{y}+\mbs{a})=d_T(\mbs{x},\mbs{y})\ .
  \end{equation}
  However, in general, if $\mbs{Q}\in SO(3)$,
  \begin{equation}
    d_T(\mbs{x},\mbs{y}) \ne d_T(\mbs{Q}\mbs{x},\mbs{Q}\mbs{y})\ .
    \label{eq:noether-rot}
  \end{equation}
  Hence, in contrast with the Euclidean distance, the function $d_T(\cdot,\cdot)$
  is not invariant under the action of the special orthogonal group.

\end{enumerate}

%%% Local Variables:
%%% TeX-command-extra-options: "-shell-escape"
%%% mode: latex
%%% TeX-master: "driver"
%%% End:

\section{Particle dynamics: basic description}
\label{sec-particle}
In this section we provide the basic ingredients that describe
the dynamics of particulate systems. This dynamical system
is governed by Hamilton's equations of motion, and most of its
complexity comes from the particle interactions, as given
by the potential energy. Here we present a fairly general class
of potentials that will be examined more carefully in \Cref{sec:pair-potentials,sec:three-body-potent,sec:embedded-atom-method}.

\subsection{System description}
Let us consider a system of $N$ particles labeled $a=1,2,\ldots,N$ 
moving inside a periodic box $\mathcal{B}$. Let the
mass of the  $a$-th particle be denoted as $m^a$, its position as $\mbs{x}^a$, and its
velocity as $\mbs{v}^a=\dot{\mbs{x}}^a$, where
the dot indicates the derivative with respect to time.

A system of particles such as the one introduced possesses a kinetic energy defined by
\begin{equation}
  T := \sum_{a=1}^N \frac{1}{2} m^a |\mbs{v}^a|^2\ ,
  \label{eq-kinetic}
\end{equation}
and a potential energy
\begin{equation}
 V=\hat{V}(\{\mbs{x}^a\}_{a=1}^N)\ ,
  \label{eq-potential}
\end{equation}
modeling the energetic interactions
among all the particles. This potential energy is always a model of the true
interatomic interaction and as such there exists
a large number of simple \emph{effective} potentials that have proven their value in different
contexts (gases, fluids, organic molecules, metals, etc.).

\subsection{Equations of motion}
The dynamics of systems of particles
are governed by Hamilton's equations of motion, namely,
\begin{equation}
    \dot{{\mbs{x}}}^{a}  = \frac{1}{m^a} \mbs{p}^{a} \ ,\qquad
    \dot{\mbs{p}}^{a}    = -\pd{}{{\mbs{x}}^{a}}\hat{V}(\{\mbs{x}^b\}_{b=1}^N)\ ,
  \label{eq-motion}
\end{equation}
where $a=1,\ldots,N$, and $\mbs{p}^a = m^a \mbs{v}^a$ is the momentum of particle~$a$. The gradient
of the potential energy is related to the forces acting on the particles, and we define
\begin{equation}
  \mbs{f}^a
  \defined
  -\pd{}{{\mbs{x}}^{a}}\hat{V}(\{\mbs{x}^b\}_{b=1}^N)
  \label{eq-force}
\end{equation}
to be the resultant of all forces applied on particle~$a$.

These standard equations need to be carefully studied
since the topology of $\mathcal{B}$ is not identical to that of Euclidean space
and the notion of derivative has to be re-examined. For the moment being,
let us assume that this object is well-defined, deferring until \Cref{sec:pair-potentials}
a more detailed inspection.

% The gradient of the potential energy can be simplified to
% \begin{equation}
% 	\label{eq:force-gen}
% 	-\pd{}{\mbs{x}^a}V=\sum_{\substack{b=1\\a \neq b}}^N\mbs{f}^{ab}
% \end{equation}
% using
% \begin{equation}
% 	\label{eq:force-gen-2}
% 	\mbs{f}^{ab}=\varphi^{ab}\frac{\mbs{\pi}(\mbs{x}^b-\mbs{x}^a)}{|\mbs{\pi}(\mbs{x}^b-\mbs{x}^a)|},\quad \varphi^{ab}=\pd{V}{d_T[\mbs{x}^a][\mbs{x}^b]}
% \end{equation}
% \todo{general cut-radius}
% which holds for arbitrary atomistic models, for which the interatomic potential only depends on the interatomic distances and consequently satisfy the weak and strong law of action and reaction (see \cite{tadmor2011modeling}).

% The system at hand contains conserved quantities which we will specify in the following.

\subsection{Conserved quantities}
Eqs.~\eqref{eq-motion} describe the motion of systems of particles
that often possess first integrals, that is, conserved quantities along their trajectories. These
quantities are of great relevance to understand the qualitative dynamics of the system, to develop
controls, etc. These \emph{momentum maps} are related to the
symmetries of the equations, according to Noether's theorem (see e.g. \cite{goldstein2002classical}).
We review them very briefly, since they have a direct impact in the formulation of
conserving schemes.

We consider only potential energies with translational invariant, that is, functions $\hat{V}$
such that for every vector $\mbs{c}\in\mathbb{R}^3$ satisfy
\begin{equation}
  \hat{V}(\{ \mbs{x}^a \}_{a=1}^N)
  =
  \hat{V}(\{ \mbs{x}^a + \mbs{c} \}_{a=1}^N)\ .
  \label{eq-trans-invariance}
\end{equation}
Differentiating both sides of this equation with respect to $\mbs{c}$ and
setting later $\mbs{c}=\mbs{0}$ we obtain the relation
\begin{equation}
  \mbs{0}
  =
  \sum_{a=1}^N \pd{\hat{V}}{\mbs{x}^a}(\{ \mbs{x}^b \}_{b=1}^N)
  =
  - \sum_{a=1}^N \mbs{f}^a\ .
  \label{eq-sum-0}
\end{equation}
This invariance condition is related to the conservation of the linear momentum
of the system, defined as
\begin{equation}
  \mbs{L}=\sum_{a=1}^N\mbs{p}^{a}\ .
  \label{eq-cont-total-momentum}
\end{equation}
The time derivative of this quantity follows from the definition of particle momentum
and \cref{eq-sum-0}:
\begin{equation}
  \dot{\mbs{L}}
  =
  \sum_{a=1}^N \dot{\mbs{p}}^{a}
  =
  \sum_{a=1}^N \mbs{f}^a
  =
  \mbs{0}\ .
  \label{eq-cons-cont-momentum}
\end{equation}

Systems of particles moving in the Euclidean space $\mathbb{R}^n$, with $n=2$ or~$3$, often
conserve angular momentum. This is a consequence of the rotational invariance of the
potential energy. However, systems defined on a periodic domain do not preserve it, in general
(see e.g. \cite{kuzkin2015angular,haile1993molecular}). One way to explain this loss of
symmetry is by noting that the projection~\eqref{eq-projection} is not rotationally invariant
(see \cref{eq:noether-rot})
and thus when used in the definition of the potential energy, it spoils the invariance of
the whole system. 

The total energy of the system is given as the sum of kinetic energy $T$ and potential
energy~$V$. The time derivative of the energy can be evaluated using the equations
of motion~\eqref{eq-motion}, giving
\begin{equation}
    \dot{E}
    =
    \sum_{a=1}^N m^{a}\mbs{v}^{a}\cdot\dot{\mbs{v}}^{a}
    -
    \sum_{a=1}^N\mbs{f}^{a}\cdot\dot{\mbs{x}}^{a}\\
    = \sum_{a=1}^N \mbs{f}^a \cdot \mbs{v}^{a}
    -\sum_{a}^N \mbs{f}^{a} \cdot \dot{\mbs{x}}^a
    = 0\ ,
\end{equation}
proving that the total energy must be a first integral of the motion.

Given the relevance of the aforementioned conservation laws, numerical
schemes have been proposed that attempt to preserve them. In particular, Energy-Momentum (EM)
schemes, the ones under study in the present work, have been designed to
integrate the equations of Hamiltonian systems while preserving
both linear and angular momentum, in addition to the total energy.
From the previous discussion, however, it follows that
when dealing with periodic systems, one might focus on the preservation of the
linear momentum and energy, only.

\subsection{The hierarchical definition of the potential energy}
The potential energy of a system of particles is a function with
the general form given in \cref{eq-potential} satisfying
the invariance condition \eqref{eq-trans-invariance}.  A hierarchy
of functions of growing complexity can be defined considering
interactions involving an increasing number of particles. This
is abstractly expressed as
\begin{equation}
  V =
  V_0
  + \sum_{\substack{a,b=1\\a\neq b}}^N V_2( \mbs{x}_a, \mbs{x}_b)
  + \sum_{\substack{a,b,c=1\\a\neq b\neq c}}^N V_3( \mbs{x}_a, \mbs{x}_b, \mbs{x}_c)
  +
  \ldots,
  \label{eq-v-hierarchy}
\end{equation}
where $V_k$ is a function involving $k-$tuples of atoms and
satisfying \cref{eq-trans-invariance}. The formulation
of accurate potentials is an active field of research and
we limit our exposition to the most common types.
The reader may consult standard references for a detailed
motivation and derivation of other types (e.g. \cite{tadmor2011modeling}).

A convenient way of formulating potential functions that
are translationally invariant is to include atomic interactions
only via the \deleted[id=1r]{relative} distance between pairs of particles.
In general, this would mean that the functions $V_k$ employed
in \cref{eq-v-hierarchy} must be of the form
\begin{equation}
  \begin{split}
    V_2( \mbs{x}_a,\mbs{x}_b)
    &= \tilde{V}_2( d(\mbs{x}_a,\mbs{x}_b)) \ ,
    \\
    V_3( \mbs{x}_a,\mbs{x}_b,\mbs{x}_c)
    &= \tilde{V}_3( d(\mbs{x}_a,\mbs{x}_b)
    , d(\mbs{x}_b,\mbs{x}_c), d(\mbs{x}_c,\mbs{x}_a))
    \ ,
  \end{split}
  \label{eq-v-distance}
\end{equation}
and similarly for higher order terms.

\subsection{Dynamics in periodic domains}
\label{sec:dynamics_in_periodic_domains}

Eqs.~\eqref{eq-motion} define the motion of particles in periodic
domains, but special care has to be taken with the definition of
the potential energy and its derivative.

With respect to the potential, we note that when formulating the dynamics of particles in periodic
domains the distance function $d(\cdot,\cdot)$ in \cref{eq-v-distance} should be replaced with the distance
$d_T(\cdot,\cdot)$ defined in \cref{eq-distance-torus}.

An aspect with important practical implications is that hierarchical potential functions of the
form~\eqref{eq-v-hierarchy} are invariably defined employing a \emph{cut\added[id=1r]{-off} radius}
that effectively limits the number of particles that interact with those within that distance, in
the sense of $d_T(\cdot,\cdot)$. Moreover, in order to avoid the singularities in the definition of
the gradient of this distance, the cut\added[id=1r]{-off} radius is always chosen to be strictly smaller than $L/2$
(see \Cref{Dprojection}). Equivalently, the dimension $L$ of the periodic box must be selected
larger than twice the cut\added[id=1r]{-off} radius.  \added[id=1r]{Under this condition, we observe
that a collection of $N$ particles in a periodic box~$\mathcal{B}$ is a mathematical representation
of an infinite domain consisting of boxes of dimension $L\times L\times L$ that repeat themselves in
the three directions of space.}

The formulation of energy and momentum conserving schemes in this kind of
domains must take these two remarks into consideration, and we explore them
in the following sections, starting from the simplest potential function possible.

%%% Local Variables:
%%% TeX-command-extra-options: "-shell-escape"
%%% mode: latex
%%% TeX-master: "driver"
%%% End:

\section{Energy-Momentum methods for periodic systems with pairwise interactions}
\label{sec:pair-potentials}
In this section we study the formulation of energy and momentum conserving algorithms for systems
of particles in periodic domains where the potential energy includes only pairwise interactions. In
terms of practical applications, only the simplest potentials belong to this class (for example,
Lennard-Jones'). They are only accurate for modeling noble gases, but are very often employed
for benchmarking and the study of numerical methods.

\subsection{Equations of motion}
We consider again a system of $N$ particles in a periodic box $\mathcal{B}$ of side $L$ with
equations of motion~\eqref{eq-motion} and an effective potential
\begin{equation}
  V  =
  \frac{1}{2}\sum_{\substack{a,b=1  \\b\neq a}}^N
  \tilde{V}\left(d_T( {\mbs{x}}^{a}, {\mbs{x}}^{b})
  \right) ,
    \label{eq-v-pair}
\end{equation}
where $\tilde{V}:\mathbb{R}^+\cup \{0\} \to\mathbb{R}$.
Using the definitions
\begin{equation}
  \bar{\mbs{r}}^{ab}
  \defined
  \mbs{\pi}(\mbs{x}^b - \mbs{x}^a)
  \qquad
  \mathrm{and}
  \qquad
  \bar{r}^{ab} = d_T(\mbs{x}^a,\mbs{x}^b)
  \defined
  | \bar{\mbs{r}}^{ab}|
  \ ,
  \label{eq-notation-distance}
\end{equation}
the forces~\eqref{eq-force} deriving from a pairwise potential
can be written  as
\begin{equation}
  \mbs{f}^a
  =
  \sum_{\substack{b=1\\a\ne b}}^N \mbs{f}^{ab}
  \ ,
  \qquad
  \mathrm{with}
  \qquad
  \mbs{f}^{ab}
  =
  \tilde{V}'(\bar{r}^{ab}) \frac{ \bar{\mbs{r}}^{ab} }{ \bar{r}^{ab}}\ .
  \label{eq-pair-force}
\end{equation}
For the following sections we further define
\begin{equation}
  {\mbs{r}}^{ab}
  \defined
  \mbs{x}^b - \mbs{x}^a
  \qquad
  \mathrm{and}
  \qquad
  {r}^{ab} = d(\mbs{x}^a,\mbs{x}^b)
  =
  | {\mbs{r}}^{ab}|
  \ .
  \label{eq-notation-distance-nopbc}
\end{equation}
%
  
%

% and $\mbs{f}^{ab}$ is the force on the $a$-th particle due to the presence of the $b$-th particle.
% In the view of eq.~(\ref{eq:force-gen-2}) this gives
% \begin{equation}
% 	\begin{aligned}
% 		\varphi^{ab}=\tilde{V}'\left(|\mbs{\pi}({\mbs{x}}^{a}-{\mbs{x}}^{b})| \right)\ .
% 	\end{aligned}
% \end{equation}
% The restriction just employed, namely, that the projected relative vector $\mbs{\pi}(\mbs{x}^{b}- \mbs{x}^{a} )$ does not fall on the boundary of the periodic box, is not very restrictive.
% \todo{Cut radius needed such that Lagrange formalism is applicable}
% \emph{In practical computations of interacting particles, the inter-particle potential has a cut-off distance which must be significantly smaller than the box sidelength, see Section \ref{sec:cut-dist-cons}.}

\subsection{Time discretization}
\label{sec-discretization-pair}
We consider now the integration in time of the equations of motion \eqref{eq-motion}
of a system in the periodic box $\mathcal{B}$ and effective potential~\eqref{eq-v-pair}.
To
approximate their solution we will employ  implicit time stepping schemes that partition the
integration interval $[0,T]$ into disjoint subintervals $[t_n,t_{n+1}]$ with $t_n=n \Delta t$, and
$\Delta t$ being the time step size, assumed to be constant to simplify the notation.  In the
algorithms defined below, we will use the notation $\mbs{x}_n$ to denote the approximation to
$\mbs{x}(t_n)$, and similarly for the velocity.  Moreover, the symbol $f_{n+\alpha}$ will denote the
convex combination $(1-\alpha) f_n + \alpha f_{n+1}$ for any variable $f$ and $0\le\alpha\le1$.

\subsubsection{Midpoint scheme}
\label{sec-midpoint-pair}
The canonical midpoint rule approximates the equations of motion \eqref{eq-motion} by the implicit formula
\begin{equation}
  \frac{ \mbs{x}^{a}_{n+1} - \mbs{x}^{a}_{n}}{\Delta t}    = \mbs{v}^{a}_{n+1/2} ,
  \qquad
    m^{a} \frac{\mbs{v}^{a}_{n+1} - \mbs{v}^{a}_n}{\Delta t}  = \sum_{\substack{b=1   \\a\neq b}}^N\mbs{f}_{MP}^{ab}\ .
  \label{eq-midpoint-pair}
\end{equation}
Here we introduced $\mbs{f}^{ab}_{MP}$, the midpoint approximation of the force acting
on particle $a$ due to the presence of particle $b$, that is,
\begin{equation}
  \mbs{f}_{MP}^{ab}
  \defined
  \tilde{V}'(\bar{r}^{ab}_{n+1/2})
  \frac{\bar{\mbs{r}}^{ab}_{n+1/2}}{|\bar{\mbs{r}}^{ab}_{n+1/2}|}\ ,
  \label{eq-midpoint-force-pair}
\end{equation}
with
\begin{equation}
  \bar{\mbs{r}}^{ab}_{n+1/2} = \mbs{\pi}(\mbs{x}^b_{n+1/2} - \mbs{x}^a_{n+1/2})\ ,
  \qquad
  \bar{r}^{ab}_{n+1/2} = d_T(\mbs{x}^a_{n+1/2}, \mbs{x}^b_{n+1/2}) .
\end{equation}
As in the continuous case, the condition
\begin{equation}
  \mbs{f}_{MP}^{ab}=-\mbs{f}_{MP}^{ba}
  \label{eq:MP-skew-pair}
\end{equation}
holds due to \cref{eq:proj-skewprop}. The properties of the midpoint rule are well known. For
example, this method preserves the total liner momentum of the system, defined at an instant
$t_n$ as
\begin{equation}
  \mbs{L}_n = \sum_{a=1}^N m^{a} \mbs{v}_n^{a}\ .
  \label{eq-linear-mometum-pair}
\end{equation}
To prove this property it suffices to verify
\begin{equation}
  \frac{\mbs{L}_{n+1}-\mbs{L}_n}{\Delta t}
  =
  \sum_{a=1}^N m^{a}\frac{\mbs{v}_{n+1}^{a}-\mbs{v}_n^{a}}{\Delta t} 
  = \sum_{\substack{a,b=1\\a\neq b}}^N \mbs{f}_{MP}^{ab}=\mbs{0}\ ,
	\label{eq-cons-disc-momentum-pair}
\end{equation}
where we have employed \cref{eq:MP-skew-pair}.

\subsubsection{Energy and momentum conserving discretization}
\label{sec:energy-moment-cons-pair}
It is possible to construct a perturbation of the midpoint rule that, in addition to preserving the
linear momentum of the system, preserves its total energy. The key ingredient of such methods is
the so called \emph{discrete gradient} operator, an approximation to the gradient that guarantees
the strict conservation of energy and momentum along the discrete trajectories generated by the
integrator.

Conserving integrators for problems in molecular dynamics have been studied since the 1970's
\cite{labudde1975energy,labudde1976energy,gonzalez1996design},
although never for systems with periodic boundary conditions, with the exception of
\cite{saluena2014molecular}.
In all of these works, the conserving
schemes are variations of the midpoint rule~\eqref{eq-midpoint-pair} of the form
\begin{equation}
    \frac{ \mbs{x}^{a}_{n+1} - \mbs{x}^{a}_{n}}{\Delta t}
    = \mbs{v}^{a}_{n+1/2},
    \qquad
    % m^{a} \frac{\mbs{v}^{a}_{n+1} - \mbs{v}^{a}_n}{\Delta t} & = -\sum_{\substack{b=1  \\b\neq a}}^N\mathsf{D}_{\mbs{x}^{a}}V(\mbs{r}^{ab}_{n},\mbs{r}^{ab}_{n+1})\\
    m^{a} \frac{\mbs{v}^{a}_{n+1} - \mbs{v}^{a}_n}{\Delta t}
    = -\mathsf{D}_{\mbs{x}^{a}}V ,
    % =\sum_{\substack{b=1 \\a\neq b}}^N\mbs{f}_{algo}^{ab}
  \label{eq-em1}
\end{equation}
where $\mathsf{D}_{\mbs{x}^a}$ is precisely the discrete gradient operator, an algorithmic
approximation to the derivative $\pd{}{\mbs{x}^a}$. In analogy
to expression~\eqref{eq-midpoint-force-pair},
the discrete gradient defines a force contribution, to be
specified later, such that
\begin{equation}
  \mathsf{D}_{\mbs{x}^a} V
  =
  -
  \sum_{\substack{b=1 \\a\neq b}}^N\mbs{f}_{algo}^{ab}
  \ .
  \label{eq-em-atom-force}
\end{equation}

If the following condition holds
\begin{equation}
	\mbs{f}^{ab}_{algo}= \mbs{f}_{MP}^{ab} + \mathcal{O}(\Delta t^2)\ ,
	\label{eq-consistency}
\end{equation}
the second order accuracy of the midpoint rule will be preserved.
If we want the new method to preserve linear momentum we note from the previous section that it
suffices that the pairwise forces $\mbs{f}^{ab}_{algo}$ mimic the symmetry condition~\eqref{eq:MP-skew-pair},
that is,
\begin{equation}
  \mbs{f}^{ab}_{algo}
  =
  -\mbs{f}^{ba}_{algo}\ .
 \label{eq-symmetry}
\end{equation}
      
Any second order perturbation of $\mbs{f}_{MP}^{ab}$ with this property will result in a
second order accurate integrator that preserves linear momentum.  
The ``classical'' EM  method is constructed in such a way, and preserves, in
addition to the total energy, the linear and angular momenta of the system, the latter being
important in domains without periodic boundary conditions \cite{gonzalez1996design}.

For problems in molecular dynamics interacting through pair potentials and posed
on periodic domains, the ``classical'' EM method 
is based on the discrete gradient~\eqref{eq-em-atom-force} with
\begin{equation}
  \mbs{f}^{ab}_{algo}
  =
  \mbs{f}^{ab}_{algo}(\mbs{r}_n^{ab},\mbs{r}_{n+1}^{ab})
  \defined
  \frac{\tilde{V}^{ab}_{n+1} - \tilde{V}^{ab}_n}{\bar{r}^{ab}_{n+1} - \bar{r}^{ab}_{n}}
  \frac{\bar{\mbs{r}}^{ab}_{n+1} + \bar{\mbs{r}}^{ab}_{n}}{\bar{r}^{ab}_{n+1} + \bar{r}^{ab}_{n}}\ ,
  \label{eq-em-simogon}
\end{equation}
where we have introduced the notation
\begin{equation}
  \tilde{V}_{n}^{ab} = \tilde{V}(\bar{r}^{ab}_n)\ ,
  \label{eq-v-short}
\end{equation}
and where the appropriate limit must be taken in \cref{eq-em-simogon} when
$|\bar{r}^{ab}_{n+1}-\bar{r}^{ab}_n|\to0$.  This form of the discrete gradient is frequently cited
in the literature of integration algorithms (see e.g. \cite{saluena2014molecular}) \added[id=1r]{and
is responsible for the conservation properties of the method, also in periodic domains}. However, it
is not the only possible form and, in fact, it can be shown that there are an infinite number of
discrete gradients \cite{romero2012analysis}, some of which can be more easily extended to the
periodic case.

More precisely, a second class of EM schemes follows from a new definition
of the algorithmic approximation of the pairwise forces:
\begin{equation}
  \mbs{f}^{ab}_{algo}(\mbs{r}_n^{ab},\mbs{r}_{n+1}^{ab})
  \defined
  \mbs{f}_{MP}^{ab}
  +
  \frac{\tilde{V}^{ab}_{n+1} - \tilde{V}^{ab}_n
    +
    \mbs{f}_{MP}^{ab}\cdot
    ({\mbs{r}}^{ab}_{n+1} - {\mbs{r}}^{ab}_{n})}{|{\mbs{r}}^{ab}_{n+1} - {\mbs{r}}^{ab}_{n}|}\; \mbs{n},
  \label{eq-em-formula}
\end{equation}
where $\mbs{n}$ is the normalized direction given by
\begin{equation}
	\mbs{n} = \frac{{\mbs{r}}^{ab}_{n+1} - {\mbs{r}}^{ab}_{n}}{|{\mbs{r}}^{ab}_{n+1} - {\mbs{r}}^{ab}_{n}|}\ .
	\label{eq-em-n}
      \end{equation}
This definition corrects the pairwise force between the particles $a$ and $b$ with a ``small'' term
in the direction $\mbs{n}$ depending on \emph{unprojected} relative positions.  This is the result
of the fact that the first equation in \eqref{eq-em1} is not posed in the quotient space but rather
in the full $\mathbb{R}^3$.  It might be argued that such a correction is nonphysical because it is
not defined on the quotient space, where the problem is posed. While this is true, the velocity
equation is not posed on this space from the outset, and the proposed correction results from this
mismatch.

% \todo{why is EM (ii) not working?}
% \bigskip\hrule\bigskip

% The new method is based on the pairwise force (ii):
% \begin{equation}
% 	-\mathsf{D}_{x_a}V( \mbs{r}_n^{ab}, \mbs{r}_{n+1}^{ab}) = \mbs{f}_{MP} + \frac{\tilde{V}_{n+1} - \tilde{V}_n^{ab} - \mbs{f}_{MP}^{ab}\cdot ({\mbs{r}}^{ab}_{n+1} - {\mbs{r}}^{ab}_{n})}{{\lambda}^{ab}_{n+1} - {\lambda}^{ab}_{n}}
% 	\mbs{d}
% 	\label{eq-em-formula2}
% \end{equation}
% and $\mbs{d}$ is the direction defined as
% \begin{equation}
% 	\mbs{d} = \frac{{\mbs{r}}^{ab}_{n+1} + {\mbs{r}}^{ab}_{n}}{\lambda^{ab}_{n+1} + \lambda^{ab}_{n}}\ .
% 	\label{eq-em-d}
% \end{equation}

% \bigskip\hrule\bigskip

The EM force given in \cref{eq-em-formula} is symmetric in $a$ and $b$.
Hence, the method defined by \eqref{eq-em1} and \eqref{eq-em-formula} preserves linear momentum.
To show that the method indeed preserves exactly the total energy exactly,
it suffices to take the dot product of the \eqref{eq-em1}$_1$ with the left hand
side of \eqref{eq-em1}$_2$, and vice versa, and then add the result over all particles, that is,
\begin{equation}
	\begin{aligned}
		\sum_{a=1}^Nm^{a}\frac{ \mbs{x}^{a}_{n+1} - \mbs{x}^{a}_{n}}{\Delta t}\cdot \frac{\mbs{v}^{a}_{n+1} - \mbs{v}^{a}_n}{\Delta t} & =\sum_{a=1}^N m^{a}\mbs{v}^{a}_{n+1/2}\cdot \frac{\mbs{v}^{a}_{n+1} - \mbs{v}^{a}_n}{\Delta t} \\
		% \sum_{a=1}^Nm^{a} \frac{\mbs{v}^{a}_{n+1} - \mbs{v}^{a}_n}{\Delta t}\cdot \frac{\mbs{x}^{a}_{n+1} - \mbs{x}^{a}_n}{\Delta t}   & = -\sum^N_{\substack{a,b=1\\b\neq a}} \mathsf{D}_{\mbs{x}^{a}}V(\mbs{r}^{ab}_{n},\mbs{r}^{ab}_{n+1})\cdot\frac{\mbs{x}^{a}_{n+1} - \mbs{x}^{a}_n}{\Delta t}\ .\\
		\sum_{a=1}^Nm^{a} \frac{\mbs{v}^{a}_{n+1}- \mbs{v}^{a}_n}{\Delta t}\cdot \frac{\mbs{x}^{a}_{n+1} - \mbs{x}^{a}_n}{\Delta t}    & = \sum^N_{\substack{a,b=1\\a\neq b}} \mbs{f}^{ab}_{algo}\cdot\frac{\mbs{x}^{a}_{n+1} - \mbs{x}^{a}_n}{\Delta t}\ .\\
	\end{aligned}
	\label{eq:energy-em-proof-1}
\end{equation}
Subtracting (\ref{eq:energy-em-proof-1})$_2$ from (\ref{eq:energy-em-proof-1})$_1$ gives
\begin{equation}
  \frac{T_{n+1}-T_{n}}{\Delta t}-\sum_{\substack{a,b=1\\a\neq b}}^N
  \mbs{f}^{ab}_{algo}\cdot\frac{\mbs{x}^{a}_{n+1} - \mbs{x}^{a}_n}{\Delta t}=0\ ,
\end{equation}
where the total discrete kinetic energy at time $t_n$ is given by
\begin{equation}
	T_n=\sum_{a=1}^N\frac{1}{2}m^a\mbs{v}^a_n\cdot\mbs{v}^a_n\ .
\end{equation}
In the spirit of \cref{eq-em-formula}, further rewriting results in
\begin{equation}
  \begin{aligned}
    \sum_{\substack{a,b=1\\a\neq b}}^N \mbs{f}_{algo}^{ab}\cdot\frac{\mbs{x}^{a}_{n+1} - \mbs{x}^{a}_n}{\Delta t}
    & =               \frac{1}{\Delta t}\sum_{\substack{a,b=1  \\a< b}}^N\mbs{f}^{ab}_{algo}\cdot(\mbs{x}^{a}_{n+1} - \mbs{x}^{a}_n-\mbs{x}^{b}_{n+1} + \mbs{x}^{b}_n)\\
    & =               -\frac{1}{\Delta t}\sum_{\substack{a,b=1 \\a< b}}^N\mbs{f}_{algo}^{ab}\cdot(\mbs{r}^{ab}_{n+1}-\mbs{r}^{ab}_{n})\ .
  \end{aligned}
\end{equation}
Therefore, a necessary and sufficient condition for
energy conservation is that the following directionality condition
is satisfied
\begin{equation}
  \sum^N_{\substack{a,b=1 \\a<b}}\mbs{f}^{ab}_{algo}\cdot(\mbs{r}^{ab}_{n+1}-\mbs{r}^{ab}_{n})  =
  \frac{1}{2}\sum^N_{\substack{a,b=1\\a\neq b}}\left(\tilde{V}(\bar{r}^{ab}_{n+1})-\tilde{V}(\bar{r}^{ab}_n)\right)\ .
\end{equation}
It is important to emphasize that the proof is based on the inner product between
the algorithmic EM force and the \emph{unprojected} relative position vectors.
Finally, to show that the EM scheme (\ref{eq-em-formula}) is indeed a second-order accurate method it suffices to prove that the correction term in the definition~\eqref{eq-em-formula} is of size $\mathcal{O}(\Delta t^2)$. Making use of the relation
\begin{equation}
	\mbs{f}^{ab}_{MP} = - \pd{\tilde{V}(|\bar{\mbs{r}}^{ab}_{n+1/2}|)}{\mbs{x}^a} = \pd{\tilde{V}(|\bar{\mbs{r}}^{ab}_{n+1/2}|)}{\mbs{r}^{ab}}\ ,
\end{equation}
a Taylor series expansion around the point $\mbs{r}^{ab}_{n+1/2}$ gives
\begin{equation}
	\tilde{V}^{ab}_{n+1} - \tilde{V}^{ab}_n = \mbs{f}^{ab}_{MP} \cdot (\mbs{r}^{ab}_{n+1} - \mbs{r}^{ab}_n) + \mathcal{O}(\Delta t^3)\ .
	\label{eq-taylor}
\end{equation}
Then, since the direction vector of the correction has size
\begin{equation}
	\frac{{\mbs{r}}^{ab}_{n+1} - {\mbs{r}}^{ab}_{n}}{|{\mbs{r}}^{ab}_{n+1} - {\mbs{r}}^{ab}_{n}|} = \mathcal{O}(1)\ ,
\end{equation}
and $|{\mbs{r}}^{ab}_{n+1} - {\mbs{r}}^{ab}_{n}|$ is $\mathcal{O}(\Delta t)$, we conclude that the correction term is indeed $\mathcal{O}(\Delta t^2)$.

\subsubsection{Time reversibility}

\added[id=2r]{Time-reversible (or symmetric) integration schemes are often favored for the approximation of
Hamiltonian systems for two main reasons. First, the Hamiltonian flow itself is symmetric, so
it is desirable that its numerical approximation also possesses this property. Second, symmetric
numerical schemes are known to have several favorable properties \cite{hairer2006geometric},
especially in long-term simulations. The class of EM integration schemes
defined in this section have also this property. This is a direct consequence of the time-reversibility of the
algorithmic approximation of the pairwise forces, namely,}
\begin{equation}
  \added[id=2r]{
  \mbs{f}^{ab}_{algo}(\mbs{r}_n^{ab},\mbs{r}_{n+1}^{ab})
  =
  \mbs{f}^{ab}_{algo}(\mbs{r}_{n+1}^{ab},\mbs{r}_{n}^{ab}),}
  \label{eq-time-symmetry}
\end{equation}
\added[id=2r]{that is trivially satisfied by both~\eqref{eq-em-simogon} and~\eqref{eq-em-formula}}.

\subsection{Interatomic potential}
\label{sec:material-model-LJ}
For the following numerical examples we consider
the well-known Lennard-Jones potential \cite{lennard1924determination} with $r=\bar{r}^{ab}$, that
is,
\begin{equation}
	\label{eq:lennard-jones}
	\tilde{V}(r)=4\epsilon\left[\left(\frac{\sigma}{r}\right)^{12}-\left(\frac{\sigma}{r}\right)^{6}\right]\ ,
\end{equation}
where $\epsilon$ and $\sigma$ are constants.

\subsubsection{Cut-off distance considerations}
\label{sec:cut-dist-cons} In the Lennard-Jones potential, atomic interactions between distant
particles are negligible. For this reason, a cut-off distance $r_c$ is often introduced beyond which
the interaction is completely ignored (see e.g. \cite{allen2017computer,rapaport2004art}).  However,
simply trimming the Lennard-Jones potential beyond the cut-off distance leads to a discontinuity in
this function at $r=r_c$ that might affect the properties of the integration scheme. Since the
derivative of the potential enters the equations of motion \eqref{eq-motion}, this discontinuity
precludes the computation of the interatomic force at $r=r_c$.  The discontinuity can be avoided,
first, by shifting the potential function by the amount $V(r_c)$, leading to the shifted potential
(SP)
\begin{equation}
  \tilde{V}_{\textrm{SP}}\left(r \right)=
  \begin{cases}
    V(r) - V(r_c) & \text{if}\ r<r_c\ ,     \\
    0             & \text{if}\ r\geq r_c\ .
  \end{cases}
\end{equation}
The derivative of this function at $r=r_c$ is still not defined and neither is the force.
To resolve this physical inconsistency one can introduce a shifted and linearly
truncated potential (SF), which is equivalent to a shift in the force
(see e.g. \cite{allen2017computer,haile1993molecular}), and given by
\begin{equation}
  \tilde{V}_{\textrm{SF}}\left(r \right)=
  \begin{cases}
    V(r) - V(r_c) - (r-r_c)V'(r_c) & \text{if}\ r<r_c \ ,    \\
    0                              & \text{if}\ r\geq r_c\ .
  \end{cases}
\end{equation}
One can further introduce a quadratic correction term that yields the shifted and quadratically
truncated potential (STF), which is equivalent to a shift and a linear truncation in the force,
\begin{equation}
  \tilde{V}_{\textrm{STF}}\left(r \right)=
  \begin{cases}
    V(r) - V(r_c) - (r-r_c)V'(r_c)- \frac{1}{2}(r-r_c)^2V''(r_c) & \text{if}\ r<r_c  \ ,   \\
    0                                                            & \text{if}\ r\geq r_c\ .
  \end{cases}
\end{equation}
This potential is twice differentiable. Due to its higher smoothness, it is better suited for
structure-preserving schemes than the standard potential since it eliminates numerical oscillations
in the energy evolution that sometimes appear when employing non-smooth potentials. As mentioned in
\Cref{sec:dynamics_in_periodic_domains}, it is important to stress out that the cut-off distance
must not be greater than $L/2$ for consistency with the minimum image convention.  Due to the
quadratic term in the corrected potential, a cut-off radius of $r_c=5\sigma$ is suggested.

\subsection{Numerical evaluation}
All numerical examples are based on a set of dimensionless units. 

\subsubsection{Accuracy study}
\label{sec:consistency-midpoint}
In the first numerical example,
we consider a two-dimensional box $[-L/2,L/2]^2$ with two particles.
The initial positions and velocities are given, respectively, by
\begin{alignat*}{2}
	\mbs{x}^{1} & =(0,0)^T,\quad &  & \mbs{v}^{1}=(0,0)^T \ ,   \\
	\mbs{x}^{2} & =(1.9,1)^T,\quad &  & \mbs{v}^{2}=(5,0)^T\ ,
\end{alignat*}
where the first particle is constrained to remain on the center of the
box and only the second particle is allowed to move freely.
For this simulation we used the Lennard-Jones potential with a simple spherical cut-off distance of
$r_c=2.5\sigma$.
The numerical values of the remaining parameters of the simulation can be found in \Cref{tab:table1}.
\begin{table}[H]
	\centering
	\caption{Accuracy study: Data used in the simulation}
	\label{tab:table1}
	\begin{tabular}{l c r | l}
          \hline
          \\
          Material parameters      & $\epsilon$            & 20              & Final configuration                                                             \\
                                   & $\sigma$              & 1               & and trajectory                                                                  \\
          Mass                     & $m^{a}$               & 0.06            & \multirow{4}{*}{\includegraphics[scale=0.12]{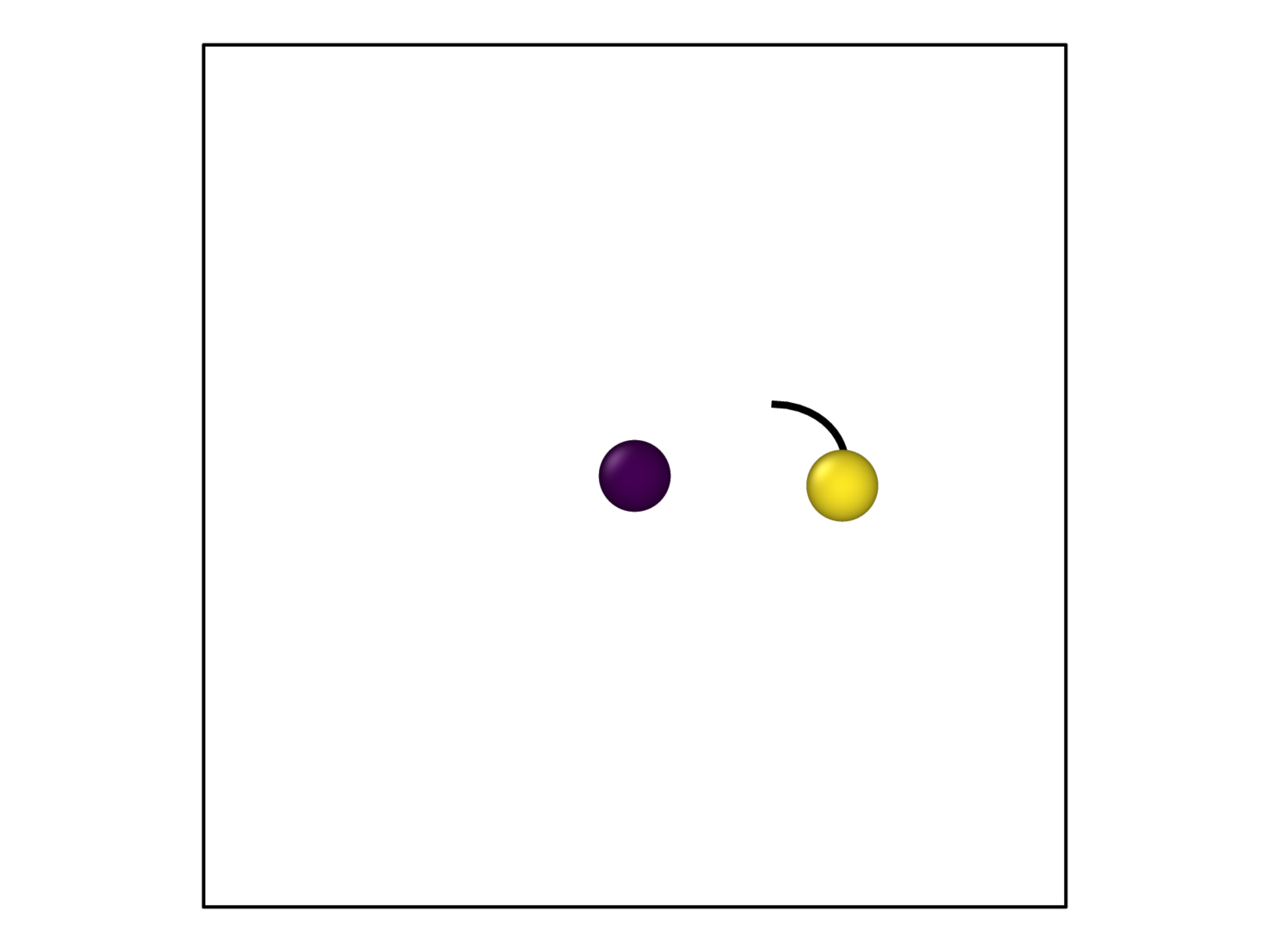}} \\
          Side length              & $L$                   & 12              &                                                                                 \\
          Newton tolerance         & \replaced[id=1r]{-}{$\epsilon$}            & 10$^{-6}$       &                                                                                 \\
          Simulation duration      & $T$                   & 0.8             &                                                                                 \\
          Reference time step size & $\Delta t_\text{ref}$ & 0.0005          &                                                                                 \\
          Time step size           & $\Delta t$            & 0.004, 0.005,   &                                                                                 \\
                                   &                       & 0.00625, 0.008, &                                                                                 \\
                                   &                       & 0.01, 0.125,    &                                                                                 \\
                                   &                       & 0.016, 0.02,    &                                                                                 \\
                                   &                       & 0.025           &     \\
                                                                                       \\
          \hline
	\end{tabular}
\end{table}
We first perform an accuracy analysis of the previously presented integrators and we compare them to
the midpoint rule.  These simulations are performed using ten different time step sizes for each
integrator.  Then we study the relative errors in the position 
and linear momentum, using as reference the midpoint rule solution, and defined respectively as
\begin{equation}
  \textrm{e}^{(MP)}_x=\frac{||\mbs{x}^a-\mbs{x}^a_r||_{2}}{||\mbs{x}^a_r||_{2}},
  \quad
  \textrm{e}^{(MP)}_\textrm{p}=\frac{||\mbs p^a- \mbs p^a_r||_2}{||\mbs{p}^a_r||_{2}},
  \label{error-measures}
\end{equation}
where $\square^a_r$, with $\square\in\{\mbs{x},\mbs{p}\}$, is the solution at time $T$ calculated with
the  midpoint, using the reference time step size $\Delta t_\text{ref} $, and where
$\square^a$ is the solution of the considered scheme at time $T$ for each time step size $\Delta t$.
\Cref{accuracy-study-position,accuracy-study-momentum} confirm that all the
schemes under consideration are second order accurate.

\setlength{\figH}{0.18\textheight}
\setlength{\figW}{0.5\textwidth}
\begin{figure}[H]
	\centering
	\setlength{\figH}{0.24\textheight}
	\setlength{\figW}{0.6\textwidth}
\definecolor{mycolor1}{rgb}{0.00000,0.44700,0.74100}%
\definecolor{mycolor2}{rgb}{0.85000,0.32500,0.09800}%
\definecolor{mycolor3}{rgb}{0.92900,0.69400,0.12500}%
\begin{tikzpicture}

\begin{axis}[%
width=0.951\figW,
height=\figH,
at={(0\figW,0\figH)},
scale only axis,
xmode=log,
xmin=0.001,
xmax=0.1,
xminorticks=true,
xlabel style={font=\color{white!15!black}},
xlabel={${\Delta}\text{t}$},
ymode=log,
ymin=0.000001,
ymax=0.01,
yminorticks=true,
ylabel style={font=\color{white!15!black}},
ylabel={$\text{e}^{(MP)}_x$},
axis background/.style={fill=white},
legend style={legend pos = outer north east, legend cell align=left, align=left, draw=white!15!black}
]
\addplot [color=mycolor1, mark=asterisk, mark options={solid, mycolor1}]
  table[row sep=crcr]{%
0.004	0.000138063803939836\\
0.005	0.000216989310459115\\
0.00625	0.00034023349845796\\
0.008	0.000558326608782416\\
0.01	0.000872808108193634\\
0.0125	0.00136338400866068\\
0.016	0.0022305948336943\\
0.02	0.00347695257644748\\
0.025	0.00541134923487022\\
};
\addlegendentry{Midpoint}

\addplot [color=mycolor2, mark=diamond, mark options={solid, mycolor2}]
  table[row sep=crcr]{%
0.004	3.32905474810458e-05\\
0.005	5.2972367753971e-05\\
0.00625	8.36332165988166e-05\\
0.008	0.000137847784529884\\
0.01	0.000216084808509894\\
0.0125	0.000338215448330399\\
0.016	0.000554630149048422\\
0.02	0.000866484081606667\\
0.025	0.00135254138196923\\
};
\addlegendentry{EM}

\addplot [color=mycolor3, mark=star, mark options={solid, mycolor3}]
  table[row sep=crcr]{%
0.004	8.41244474313736e-06\\
0.005	1.25812847426007e-05\\
0.00625	1.92409528959707e-05\\
0.008	3.11799888959814e-05\\
0.01	4.84374243201709e-05\\
0.0125	7.53957059793905e-05\\
0.016	0.000123260152110027\\
0.02	0.00019230643407395\\
0.025	0.000300135791463045\\
};
\addlegendentry{EM classic}

\addplot [color=black, forget plot]
  table[row sep=crcr]{%
0.01	0.000872808108193634\\
0.0125	0.000872808108193634\\
0.0125	0.00136338400866068\\
};
\node[right, align=left]
at (axis cs:0.012,0.0012) {1.9988};
\node[right, align=left]
at (axis cs:0.010,0.0006) {1};
\end{axis}
\end{tikzpicture}%
	\caption{Accuracy study: Relative error in the position w.r.t midpoint rule}
	\label{accuracy-study-position}
\end{figure}
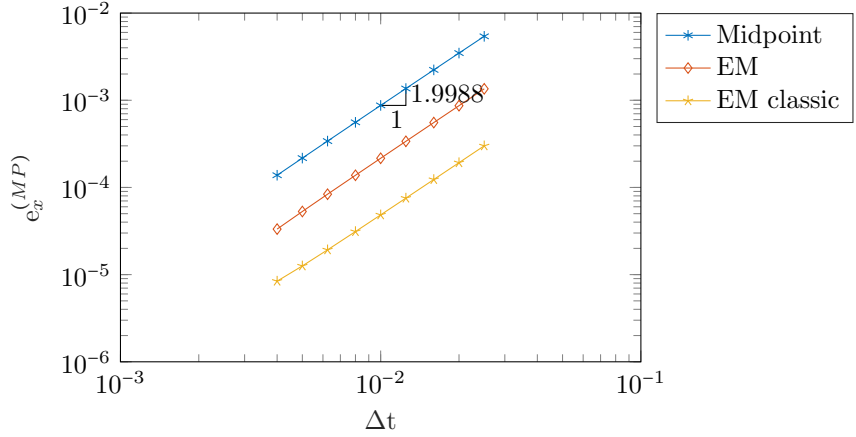
\begin{figure}[H]
	\centering
	\setlength{\figH}{0.24\textheight}
	\setlength{\figW}{0.6\textwidth}
\definecolor{mycolor1}{rgb}{0.00000,0.44700,0.74100}%
\definecolor{mycolor2}{rgb}{0.85000,0.32500,0.09800}%
\definecolor{mycolor3}{rgb}{0.92900,0.69400,0.12500}%
\begin{tikzpicture}

\begin{axis}[%
width=0.951\figW,
height=\figH,
at={(0\figW,0\figH)},
scale only axis,
xmode=log,
xmin=0.001,
xmax=0.1,
xminorticks=true,
xlabel style={font=\color{white!15!black}},
xlabel={${\Delta}\text{t}$},
ymode=log,
ymin=1e-05,
ymax=0.1,
yminorticks=true,
ylabel style={font=\color{white!15!black}},
ylabel={$\text{e}_\text{p}^{(MP)}$},
axis background/.style={fill=white},
legend style={legend pos = outer north east, legend cell align=left, align=left, draw=white!15!black}
]
\addplot [color=mycolor1, mark=asterisk, mark options={solid, mycolor1}]
  table[row sep=crcr]{%
0.004	0.000679577739544522\\
0.005	0.00106797177268311\\
0.00625	0.00167429981571846\\
0.008	0.00274673619385873\\
0.01	0.00429209896384471\\
0.0125	0.00670033558017353\\
0.016	0.0109500837777011\\
0.02	0.0170415028566945\\
0.025	0.0264576755039928\\
};
\addlegendentry{Midpoint}

\addplot [color=mycolor2, mark=diamond, mark options={solid, mycolor2}]
  table[row sep=crcr]{%
0.004	7.84453398733456e-05\\
0.005	0.00012876313975361\\
0.00625	0.000206716946380024\\
0.008	0.000343904654144898\\
0.01	0.000541886373583542\\
0.0125	0.00085074746717889\\
0.016	0.0013982974860261\\
0.02	0.00218693630396039\\
0.025	0.00341566123150894\\
};
\addlegendentry{EM}

\addplot [color=mycolor3, mark=star, mark options={solid, mycolor3}]
  table[row sep=crcr]{%
0.004	4.51992027405763e-05\\
0.005	6.4090334865614e-05\\
0.00625	9.43090524149745e-05\\
0.008	0.000148929405672609\\
0.01	0.000227756572792352\\
0.0125	0.000350681020826396\\
0.016	0.000569765249668587\\
0.02	0.000885942568600428\\
0.025	0.00138098168245359\\
};
\addlegendentry{EM classic}

\addplot [color=black, forget plot]
  table[row sep=crcr]{%
0.01	0.00429209896384471\\
0.0125	0.00429209896384471\\
0.0125	0.00670033558017353\\
};
\node[right, align=left]
at (axis cs:0.012,0.0055) {1.9959};
\node[right, align=left]
at (axis cs:0.01,0.003) {1};
\end{axis}
\end{tikzpicture}%
	\caption{Accuracy study: Relative error in the linear momentum w.r.t midpoint rule}
	\label{accuracy-study-momentum}
\end{figure}
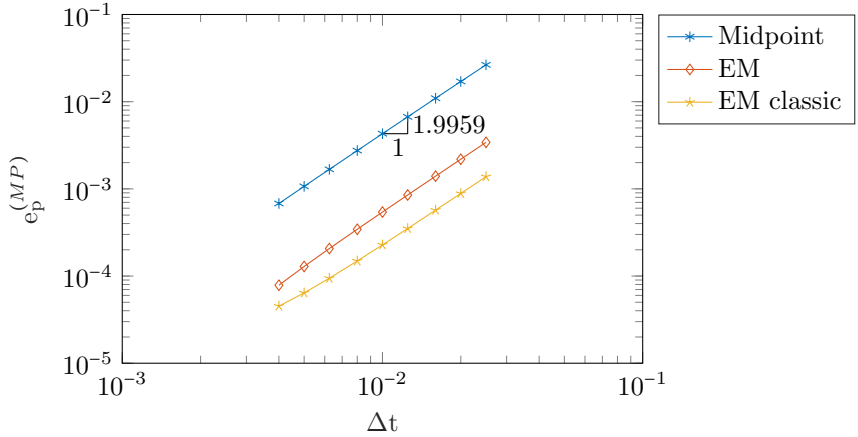

\subsubsection{Energy consistency study}
\begin{figure}[H]
  \centering
  \setlength{\figH}{0.15\textheight}
  \setlength{\figW}{0.5\textwidth}
  \input{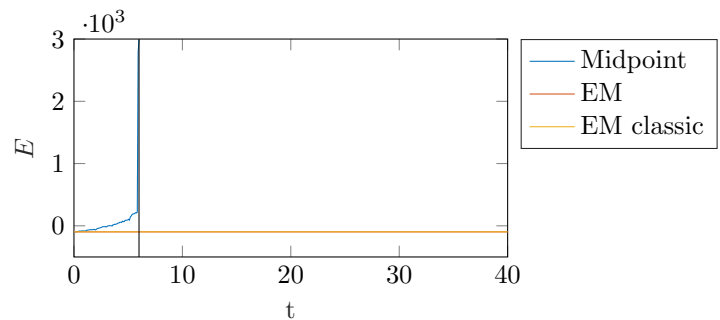}
  \caption{Energy consistency study: Total energy}
  \label{Big-system-energy}
\end{figure}

\begin{figure}[H]
  \centering
  \setlength{\figH}{0.15\textheight}
  \setlength{\figW}{0.5\textwidth}
  \input{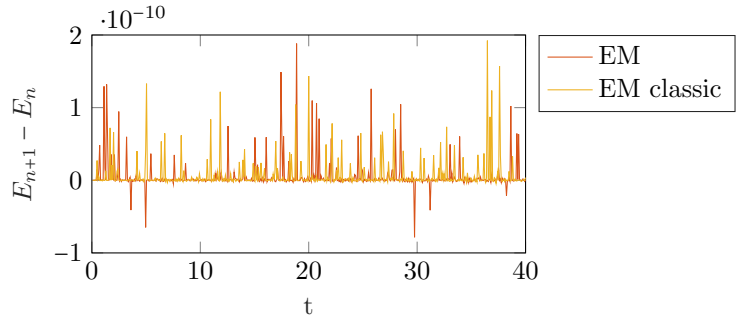}
  \caption{Energy consistency study: Total energy difference}
  \label{Big-system-energy-diff}
\end{figure}

\begin{figure}[H]
	\centering
	\setlength{\figH}{0.15\textheight}
	\setlength{\figW}{0.5\textwidth}
	\input{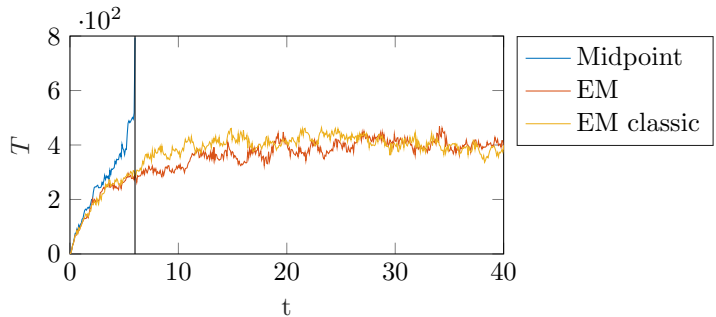}
	\caption{Energy consistency study: Kinetic energy}
	\label{Big-system-kinetic}
\end{figure}

\begin{figure}[H]
	\centering
	\setlength{\figH}{0.15\textheight}
	\setlength{\figW}{0.5\textwidth}
	\input{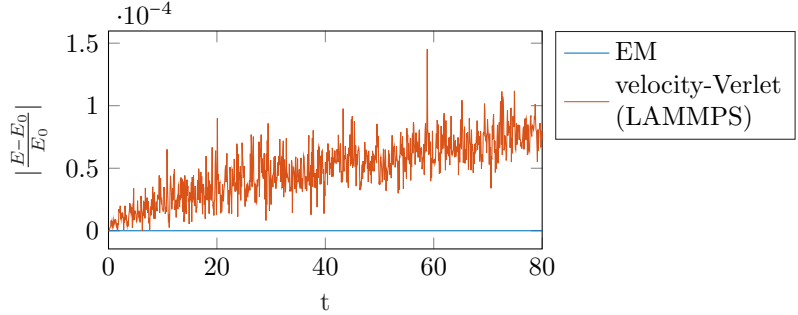}
	\caption{Energy consistency study: Relative energy drift for $\Delta t_{\textrm{EM}}=100\Delta t_{\textrm{vV}}$}
	\label{Big-system-compare}
\end{figure}

The second numerical example investigates the energy
conservation properties of the integrators
described in \Cref{sec:pair-potentials}.  We consider 150 arbitrarily positioned
particles inside a three-dimensional periodic box $[-L/2,L/2]^3$ such that the initial distance between the
particles is greater than $2^{1/6}\sigma$.  Starting from rest, the kinetic energy of the system
will rise until the system is in equilibrium due to the random order of the particles.  For this
simulation, we consider the shifted and quadratically truncated Lennard-Jones potential (STF) with a
spherical cut-off distance of $r_c=5\sigma$. Numerical values for the parameters of the simulation can be found in
\Cref{tab:table2}.
\begin{table}[H]
      	\caption{Energy consistency study: Data used in the simulation}
	\label{tab:table2}
	\begin{tabular}{l c r | l}
		\hline
                                                                                                                                      \\
		Material parameters & $\epsilon$ & 2         & Initial configuration                                                  \\
		                    & $\sigma$   & 1         & \multirow{6}{*}{\includegraphics[scale=0.175]{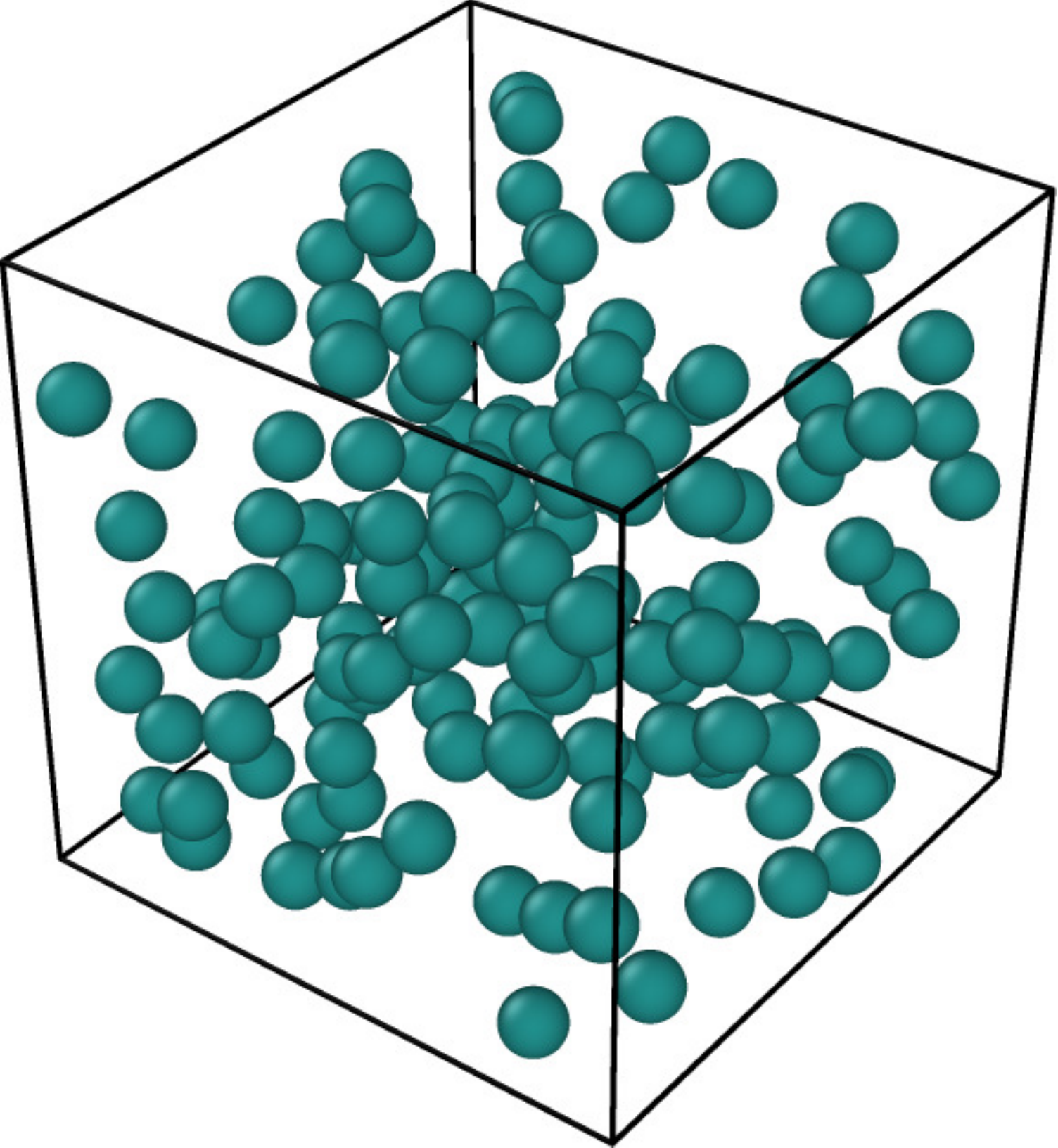}} \\
		Mass                & $m^{a}$    & 1         &                                                                        \\
		Sidelength          & $L$        & 12        &                                                                        \\
		Newton tolerance    & \replaced[id=1r]{-}{$\epsilon$} & 10$^{-9}$ &                                                                        \\
		Simulation duration & $T$        & 40        &                                                                        \\
		Time steps          & $\Delta t$ & 0.08      &                                                                        \\
                                                                                                                                      \\
                                                                                                                                      \\
		\hline
	\end{tabular}
\end{table}
\label{sec:num-conserv}
For this relatively large time step size, the midpoint rule
introduces energy into the system leading to an energy blow-up
and eventually to a termination of the simulation indicated with a vertical line in \Cref{Big-system-energy}.
Both EM methods described, on the other hand, preserve the total energy
up to machine precision for the whole duration of the simulation. See \Cref{Big-system-energy-diff}. 
\Cref{Big-system-kinetic} shows the evolution of the kinetic energy throughout the simulation.
Since this energy is proportional to the temperature in the system, the figure
reveals that only the EM methods
can compute the evolution of the system until it reaches equilibrium,
for the chosen time step size.
\added[id=1r]{
  We compare the newly proposed EM method with an explicit integrator, namely,  the velocity-Verlet
  scheme provided by the molecular dynamics code LAMMPS}\footnote{https://lammps.sandia.gov}\added[id=1r]{\cite{plimpton1993fast}.
For this example we use the shifted and linearly truncated potential (SF) instead of the shifted and quadratically truncated potential (STF), since the former is available in the software package. Moreover, we extended the simulation duration to $T=80$.

Keeping the time step size of the EM method $\Delta t_{\textrm{EM}}=0.08$ constant,
the time step size of the velocity-Verlet integrator $\Delta t_{\textrm{vV}}$ is reduced until the energy
fluctuations become sufficiently small.
As \Cref{Big-system-compare} reveals, the time step size employed for the explicit method is $\Delta t_{\textrm{vV}}=0.0008$,
100 times smaller than the time step size of the EM method. This is the time step size required to keep the relative energy fluctuations in the explicit solution approximately below $0.1\permil$.}
\added[id=1r]{
  Regarding the computational cost per step, the proposed implicit integrator naturally much more expensive that the explicit
  scheme. In order to do a fair comparison, however, the implementation of the EM method should be optimized
  similarly to LAMMPS, but this is not the focus of the present work.}

\added[id=1r]{In addition to the previous investigation, we perform an accuracy comparison between
the EM and the velocity-Verlet methods. We use again the system described in \Cref{tab:table2} and
obtain a reference solution with the explicit method and a small time step  $\Delta
t_{\textrm{ref}}=0.000001$. The displacement and momentum error measures of
eq.~(\ref{error-measures}) will be used to compare the accuracy of the two integrators for the time
step sizes $\Delta t=\{$0.0001, 0.0002, 0.00025, 0.0004, 0.0005, 0.00625, 0.001, 0.0025, 0.004,
0.005, 0.01$\}$ and a final simulation time $T=4$. As shown in \Cref{Big-system-error-position,Big-system-error-momentum},
both schemes are second order accurate in the position and the linear momentum.}

\begin{figure}[H]
	\centering
	\setlength{\figH}{0.24\textheight}
	\setlength{\figW}{0.6\textwidth}
\definecolor{mycolor1}{rgb}{0.00000,0.44700,0.74100}%
\definecolor{mycolor2}{rgb}{0.85000,0.32500,0.09800}%
\begin{tikzpicture}

\begin{axis}[%
width=0.951\figW,
height=\figH,
at={(0\figW,0\figH)},
scale only axis,
xmode=log,
xmin=0.0001,
xmax=0.01,
xminorticks=true,
xlabel style={font=\color{white!15!black}},
xlabel={${\Delta}\text{t}$},
ymode=log,
ymin=0.0001,
ymax=1,
yminorticks=true,
ylabel style={font=\color{white!15!black}},
ylabel={$\text{e}^{(vV)}_x$},
axis background/.style={fill=white},
legend style={legend pos = outer north east, legend cell align=left, align=left, draw=white!15!black}
]
\addplot [color=mycolor1, mark=asterisk, mark options={solid, mycolor1}]
  table[row sep=crcr]{%
0.0001	0.000764454907235304\\
0.0002	0.00294850552806674\\
0.00025	0.00453861938644482\\
0.0004	0.0120355091761453\\
0.0005	0.0178637332285124\\
0.000625	0.0261639423928707\\
0.001	0.0468758561965733\\
0.0025	0.109019556190152\\
0.005	0.179363900381119\\
0.01	0.248358701557789\\
};
\addlegendentry{EM}

\addplot [color=mycolor2, mark=asterisk, mark options={solid, mycolor2}]
  table[row sep=crcr]{%
0.0001	0.000247855230891467\\
0.0002	0.000987947248069413\\
0.00025	0.00153159065654044\\
0.0004	0.00377382557153165\\
0.0005	0.00569099323271938\\
0.000625	0.00849440457067474\\
0.001	0.0197543825597697\\
0.0025	0.0780946962265223\\
0.005	0.148542130436527\\
0.01	0.207722900074053\\
};
\addlegendentry{velocity-Verlet\\ (LAMMPS)}

\addplot [color=black, forget plot]
  table[row sep=crcr]{%
0.0002	0.000987947248069413\\
0.00025	0.000987947248069413\\
0.00025	0.00153159065654044\\
};
\node[right, align=left]
at (axis cs:0.00025,0.0012) {1.9648};
\node[right, align=left]
at (axis cs:0.0002,0.0007) {1};
\end{axis}

\end{tikzpicture}%
	\caption{Energy consistency study: Relative error in the position w.r.t velocity-Verlet scheme (LAMMPS)}
	\label{Big-system-error-position}
\end{figure}
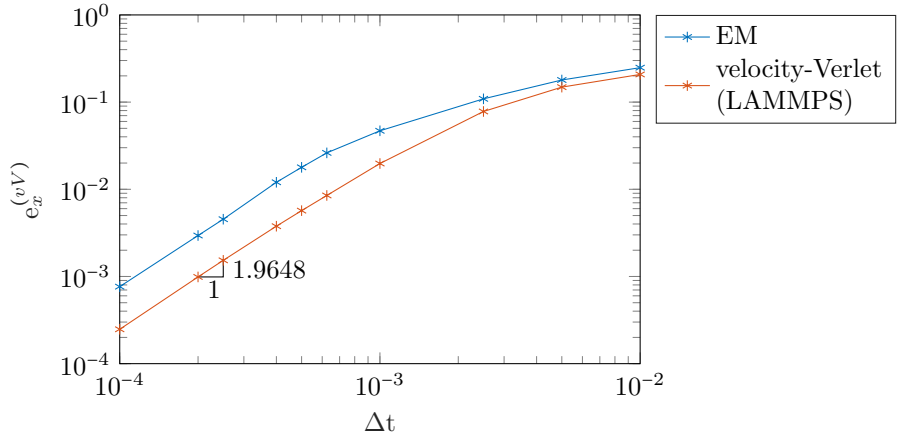
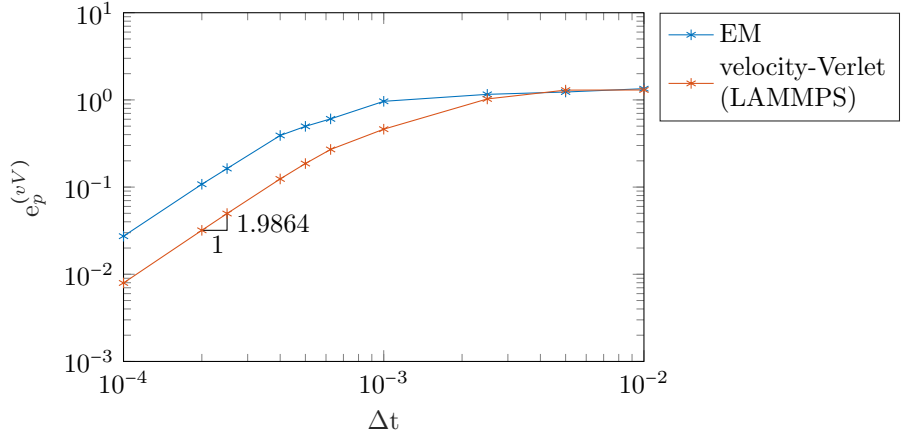
\begin{figure}[H]
	\centering
	\setlength{\figH}{0.24\textheight}
	\setlength{\figW}{0.6\textwidth}
\definecolor{mycolor1}{rgb}{0.00000,0.44700,0.74100}%
\definecolor{mycolor2}{rgb}{0.85000,0.32500,0.09800}%
\begin{tikzpicture}

\begin{axis}[%
width=0.951\figW,
height=\figH,
at={(0\figW,0\figH)},
scale only axis,
xmode=log,
xmin=0.0001,
xmax=0.01,
xminorticks=true,
xlabel style={font=\color{white!15!black}},
xlabel={${\Delta}\text{t}$},
ymode=log,
ymin=0.001,
ymax=10,
yminorticks=true,
ylabel style={font=\color{white!15!black}},
ylabel={$\text{e}^{(vV)}_p$},
axis background/.style={fill=white},
legend style={legend pos = outer north east, legend cell align=left, align=left, draw=white!15!black}
]
\addplot [color=mycolor1, mark=asterisk, mark options={solid, mycolor1}]
  table[row sep=crcr]{%
0.0001	0.0273430865918785\\
0.0002	0.10781024223595\\
0.00025	0.163188348247809\\
0.0004	0.390831098266667\\
0.0005	0.497446840675895\\
0.000625	0.605820306541099\\
0.001	0.963521931745753\\
0.0025	1.15820204208881\\
0.005	1.23193849171196\\
0.01	1.34501370401075\\
};
\addlegendentry{EM}

\addplot [color=mycolor2, mark=asterisk, mark options={solid, mycolor2}]
  table[row sep=crcr]{%
0.0001	0.00795114472193924\\
0.0002	0.0319205107069898\\
0.00025	0.0497243092716872\\
0.0004	0.124000016715146\\
0.0005	0.1861646613189\\
0.000625	0.270241867992502\\
0.001	0.461530697242265\\
0.0025	1.02598848537837\\
0.005	1.29515188888347\\
0.01	1.30022758532974\\
};
\addlegendentry{velocity-Verlet\\ (LAMMPS)}

\addplot [color=black, forget plot]
  table[row sep=crcr]{%
0.0002	0.0319205107069898\\
0.00025	0.0319205107069898\\
0.00025	0.0497243092716872\\
};
\node[right, align=left]
at (axis cs:0.00025,0.038) {1.9864};
\node[right, align=left]
at (axis cs:0.0002,0.022) {1};
\end{axis}

\end{tikzpicture}%
	\caption{Energy consistency study: Relative error in the momentum w.r.t velocity-Verlet scheme (LAMMPS)}
	\label{Big-system-error-momentum}
\end{figure}
%%% Local Variables:
%%% TeX-command-extra-options: "-shell-escape"
%%% mode: latex
%%% TeX-master: "driver"
%%% End:

\section{Energy-Momentum methods for periodic systems with three-body interactions}
\label{sec:three-body-potent}
In \Cref{sec:pair-potentials} we derived the equation of motion of a system of $N$ particles
moving inside a period box $\mathcal{B}$ where the interactions were based on pair
potentials.  For materials with strong covalent-bonding character, however, we further need to
incorporate a bond-angle dependency to the effective potential.  This can be archived by including
three-body terms in the expression of the potential.

\subsection{Equation of motion}
\label{sec:equatio-of-motion-three}
We consider now a system of $N$ particles in a periodic box $\mathcal{B}$ of side $L$ with
equations of motion~\eqref{eq-motion} and an effective potential that depends only on
the interactions among all  triplets of particles. 
Such a potential must be of the form
\begin{equation}
  V =
  \frac{1}{3!}\sum_{\substack{a,b,c=1  \\a\neq b\neq c}}^N
  \tilde{V}\left(
    d_T( {\mbs{x}}^{a}, {\mbs{x}}^{b}),
    d_T( {\mbs{x}}^{a}, {\mbs{x}}^{c}),
    d_T( {\mbs{x}}^{b}, {\mbs{x}}^{c}) \right)\ , 
    \label{eq-v-three}
\end{equation}
where $\tilde{V}:\mathbb{R}^+\cup \{0\}\times\mathbb{R}^+\cup \{0\} \times\mathbb{R}^+\cup \{0\}  \to\mathbb{R}$ is a three-body potential between the $a$-th, $b$-th, and the $c$-th particle.
The total kinetic energy of the system is given by \cref{eq-kinetic}.
The forces acting on the particles defined by \cref{eq-force} can be written using the definitions in \cref{eq-notation-distance} as
\begin{equation}
  \mbs{f}^a
  =
  \sum_{\substack{b=1\\a\ne b}}^N \mbs{f}^{ab}
  \ ,
  \qquad
  \mathrm{with}
  \qquad
  \mbs{f}^{ab}
  \defined
  \varphi^{ab}\frac{ \bar{\mbs{r}}^{ab} }{ \bar{r}^{ab}}\ .
  \label{eq-three-force}
\end{equation}
The strength of the force is now obtained as
\begin{equation}
	\begin{aligned}
		\varphi^{ab} & =\sum_{\substack{c=1 \\a\neq b\neq c}}^N\pd{}{\bar{r}^{ab}}\tilde{V}(\bar{r}^{ab},\bar{r}^{ac},\bar{r}^{bc})\ .
	\end{aligned}
\end{equation}

\subsection{Time discretization}
Next, we consider the integration in time of the equations of motion \eqref{eq-motion} of a system in
the periodic box $\mathcal{B}$ and effective potential~\eqref{eq-v-three} and employ the time
integration strategy outlined in \Cref{sec-discretization-pair}.

\subsubsection{Midpoint scheme}
The canonical midpoint rule approximates the equation of motion by the implicit formula~(\ref{eq-midpoint-pair}), where the midpoint approximation of the force acting on the $a$-th particle in the direction of the $b$-th particle is given by
\begin{equation}
  \begin{aligned}
    \mbs{f}_{MP}^{ab} & = \varphi^{ab}_{MP}\frac{\bar{\mbs{r}}^{ab}_{n+1/2}}{|\bar{\mbs{r}}^{ab}_{n+1/2}|} \\
    \varphi^{ab}_{MP} & =\sum_{\substack{c=1\\a\neq b\neq c}}^N\pd{}{{\bar{r}^{ab}_{n+1/2}}}\tilde{V}\left({\bar{{r}}^{ab}_{n+1/2}},{\bar{{r}}^{ac}_{n+1/2}},{\bar{{r}}^{bc}_{n+1/2}}\right)\ .
		\label{eq-midpoint-force-3}
	\end{aligned}
\end{equation}
As in the continuous case, the weak law of action and reaction is satisfied and therefore the
approximation preserves the total linear momentum of the system. See
\Cref{sec-midpoint-pair}.

\subsubsection{Energy and momentum conserving discretization}
\label{subsub-em-three}
As in \Cref{sec:energy-moment-cons-pair}, it is possible to construct a perturbation of the
midpoint rule which, in addition to preserving the total linear momentum, preserves the total energy.
Instead of the discrete gradient operator the  \emph{partitioned discrete gradient} operator
\cite{gonzalez1996design} will be now employed, which is an approximation similar to the discrete
gradient but is applicable to functions with more than one independent variable.  To this end,
let us rewrite the potential energy function~\eqref{eq-v-three}
in terms of $N(N-1)/2$ independent variables, e.g.
\begin{equation}
  \label{eq:sorted-V}
  \tilde{V} =
  \tilde{V}(\bar{r}^{12},\bar{r}^{13},\dots,\bar{r}^{1N},\bar{r}^{23},\dots,\bar{r}^{N-1,N})
  =
  \tilde{V}(\{\bar{r}^{ab}\})\ ,
\end{equation}
where the double indexed set is given by
\begin{equation}
  \{\bar{r}^{ab}\}=\{\bar{r}^{ab}\ |\ a,b\in(1,\dots,N),\ a<b\}\ .
\end{equation}
To simplify the definition of the partitioned discrete gradient,
it proves useful to re-label the interatomic distances $\bar{r}^{ab}$
using only their position in the array $\{\bar{r}^{ab}\}$.
Therefore, a single indexed set is defined by
\begin{equation}
  \{\bar{r}^\alpha\}=\{\bar{r}^{\alpha}\ |\ \alpha\in(1,\dots,N(N-1)/2)\}\ .
\end{equation}
\added[id=1r]{%
  Note that here an ordering of the $N\choose 2$ pairs $(a,b)$ has been established.
  For example, the map $(a,b)\mapsto\alpha$ could be chosen to
  be lexicographic (e.g., \cite[pg. 43]{Kreher:1999va}).} 
Then, the potential energy can be expressed, abusing slightly the notation, as
\begin{equation}
  \label{eq:sorted-V-2}
  V =
  \tilde{V}(\{\bar{r}^{\alpha}\})\ .
\end{equation}

For a potential function like this one, the discrete gradient
operator is defined as
\begin{equation}
    \label{eq:em-force-3}
    \mathsf{D}_{\mbs{x}^a}V
    =
    -
    \sum_{\substack{b=1 \\a\neq b}}^N\mbs{f}^{ab}_{algo}
    =
    -\sum_{\substack{b=1\\a\neq b}}^N\frac{1}{2}\left(\mbs{f}^{ab}_{n,n+1}+\mbs{f}^{ab}_{n+1,n}\right)\ ,
\end{equation}
with the contributions
\begin{equation}
  \begin{aligned}
    \mbs{f}^{ab}_{n,n+1} & = \mbs{f}_{MP}^{ab}
    + \frac{\tilde{V}_{n,n+1}^{\added[id=1r]{\alpha}}(\bar{r}^{\alpha}_{n+1})
      -
      \tilde{V}_{n,n+1}^{\added[id=1r]{\alpha}}
      (\bar{r}^{\alpha}_{n})-\mbs{f}_{MP}^{ab}\cdot ({\mbs{r}}^{ab}_{n+1} - {\mbs{r}}^{ab}_{n})}{|{\mbs{r}}^{ab}_{n+1} - {\mbs{r}}^{ab}_{n}|}\; \mbs{n}, \\
    \mbs{f}^{ab}_{n+1,n} & = \mbs{f}_{MP}^{ab}
    + \frac{\tilde{V}_{n+1,n}^{\added[id=1r]{\alpha}}
      (\bar{r}^{\alpha}_{n+1})-\tilde{V}_{n+1,n}^{\added[id=1r]{\alpha}}(\bar{r}^{\alpha}_{n})-\mbs{f}_{MP}^{ab}\cdot ({\mbs{r}}^{ab}_{n+1} - {\mbs{r}}^{ab}_{n})}{|{\mbs{r}}^{ab}_{n+1} - {\mbs{r}}^{ab}_{n}|}\; \mbs{n}\ , \\
	\end{aligned}
\end{equation}
for which we introduced the compact notation
\begin{equation}
  \begin{aligned}
    \tilde{V}_{n,n+1}^{\added[id=1r]{\alpha}}(\bar{r}^\alpha) & =\tilde{V}(\bar{r}^1_{n}  ,\dots,\bar{r}^{\alpha-1}_n,\bar{r}^{\alpha},\bar{r}^{\alpha+1}_{n+1},\dots,\bar{r}^{N(N-1)/2}_{n+1}), \\
    \tilde{V}_{n+1,n}^{\added[id=1r]{\alpha}}(\bar{r}^\alpha) & =\tilde{V}(\bar{r}^1_{n+1},\dots,\bar{r}^{\alpha-1}_{n+1},\bar{r}^{\alpha},\bar{r}^{\alpha+1}_{n},\dots,\bar{r}^{N(N-1)/2}_{n})\ . \\
	\end{aligned}
\end{equation}

To show that the proposed integrator exactly preserves the total linear momentum, it suffices to
follow the proof outlined in \Cref{sec:energy-moment-cons-pair} and details are omitted.  Similarly,
to prove the energy conservation property, it is enough to show that 
\begin{equation}
  \sum^N_{\substack{a,b=1 \\a<b}}\mbs{f}^{ab}_{algo}\cdot(\mbs{r}^{ab}_{n+1}-\mbs{r}^{ab}_{n})
  =
  \frac{1}{3!}\sum^N_{\substack{a,b,c=1\\a\neq b\neq c}}
  \left(
  \tilde{V}(\bar{r}^{ab}_{n+1},\bar{r}^{ac}_{n+1},\bar{r}^{bc}_{n+1})
  -
  \tilde{V}(\bar{r}^{ab}_n,\bar{r}^{ac}_n,\bar{r}^{bc}_n)
  \right)
  .
\end{equation}
A straightforward manipulation shows that the proposed method
with algorithmic forces given by \cref{eq:em-force-3} satisfies this condition.

\begin{remark}
  Many three-body potentials are expressed in terms of
  the bond angles $\bar{\theta}^{bac}$ at particle~$a$, between the bonds $ab$ and $ac$.
  Since the angle itself can be written in terms of the distances, that is,
\begin{equation}
  \label{eq:lawcosines}
  \bar{\theta}^{bac}  =g(\bar{r}^{ab},\bar{r}^{ac},\bar{r}^{bc})=
  \arccos\left(\frac{(\bar{r}^{ab})^2+(\bar{r}^{ac})^2-(\bar{r}^{bc})^2}{2(\bar{r}^{ab})(\bar{r}^{ac})}\right)
\end{equation}
the composition $\tilde{V}\circ g$ has again the structure of the potential \eqref{eq:sorted-V}
and thus the partitioned discrete gradient operator defined before can be employed
without modifications.
\end{remark}

\begin{remark}
  In problems without periodic boundary conditions, using
\cref{eq-notation-distance-nopbc}, the EM method reads 
\begin{equation}
  \label{eq:em-force-3np}
  \mathsf{D}_{\mbs{x}^a}V
  =
  -
  \sum_{\substack{b=1 \\a\neq b}}^N\mbs{f}^{ab}_{algo}
  =-
  \sum_{\substack{b=1\\a\neq b}}^N\frac{1}{2}\left(\mbs{f}^{ab}_{n,n+1}+\mbs{f}^{ab}_{n+1,n}\right)\ ,
\end{equation}
with the contributions
\begin{equation}
	\begin{aligned}
		\mbs{f}^{ab}_{n,n+1} & =
                \frac{{\tilde{V}}_{n,n+1}^{\added[id=1r]{\alpha}}({r}^{\alpha}_{n+1})-\tilde{V}_{n,n+1}^{\added[id=1r]{\alpha}}({r}^{\alpha}_{n})}{r^\alpha_{n+1}-r^\alpha_{n}}\;
                \frac{{\mbs{r}}^{ab}_{n+1} + {\mbs{r}}^{ab}_{n}}{r^{ab}_{n+1}+r^{ab}_{n}}, \\
		\mbs{f}^{ab}_{n+1,n} & =
                \frac{{\tilde{V}}_{n+1,n}^{\added[id=1r]{\alpha}}({r}^{\alpha}_{n+1})-\tilde{V}_{n+1,n}^{\added[id=1r]{\alpha}}({r}^{\alpha}_{n})}{r^\alpha_{n+1}-r^\alpha_{n}}\;
                \frac{{\mbs{r}}^{ab}_{n+1} + {\mbs{r}}^{ab}_{n}}{r^{ab}_{n+1}+r^{ab}_{n}}, \\
	\end{aligned}
\end{equation}
where we used the compact notation
\begin{equation}
  \begin{aligned}
    {\tilde{V}}_{n,n+1}^{\added[id=1r]{\alpha}}({r^\alpha}) & ={\tilde{V}}({r}^1_{n}
    ,\dots,{r}^{\alpha-1}_n,{r}^{\alpha},{r}^{\alpha+1}_{n+1},\dots,{r}^{N(N-1)/2}_{n+1}),  \\
    {\tilde{V}}_{n+1,n}^{\added[id=1r]{\alpha}}({r^\alpha}) & =
    {\tilde{V}}({r}^1_{n+1},\dots,{r}^{\alpha-1}_{n+1},{r}^{\alpha},{r}^{\alpha+1}_{n},\dots,{r}^{N(N-1)/2}_{n})\ . \\
  \end{aligned}
\end{equation}
This EM method preserves the total angular momentum in addition to the total energy and
the total linear momentum. It can be used, e.g., for the simulation of bonded three-body interactions between macromolecules.
\end{remark}

\subsection{Interatomic potential}
For our numerical simulation we consider the Stillinger-Weber potential
\cite{stillinger1985computer}, which includes two- and three-body contributions
\begin{equation}
  \label{eq:Stilinger-weber-mat}
  \tilde{V}  =\frac{1}{2!}\sum_{\substack{a,b=1   \\a \neq
      b}}^N\epsilon\tilde{f}_2(\bar{r}^{ab}/\sigma)
  +
  \frac{1}{3!}\sum_{\substack{a,b,c=1 \\a\neq b\neq c}}^N\epsilon\tilde{f}_3(\bar{r}^{ab}/\sigma, \bar{r}^{ac}/\sigma,\bar{r}^{bc}/\sigma)\ .
\end{equation}
The pair contribution is given by
\begin{equation}
  \label{eq:Stilinger-weber-pair}
  \tilde{f}_2(\hat{r})=
  \begin{cases}
    A\left(B\hat{r}^{-q}-\hat{r}^{-p}\right)g_2(\hat{r})& \text{if}\ \hat{r}<a\ , \\
    0& \text{if}\ r\geq a \ ,
  \end{cases}
\end{equation}
where the hats indicate the normalization of the distances by $\sigma$.
After composing with the law of cosines \eqref{eq:lawcosines},
the three-body contribution takes the form
\begin{equation}
  \label{eq:Stilinger-weber-3}
  \tilde{f}_3(\hat{r}^{ab},\hat{r}^{ac},\hat{r}^{bc})
  =
  h(\hat{r}^{ab},\hat{r}^{ac},\hat{r}^{bc})+h(\hat{r}^{ab},\hat{r}^{bc},\hat{r}^{ac})+h(\hat{r}^{ac},\hat{r}^{bc},\hat{r}^{ab})\ ,
\end{equation}
with
\begin{equation}
  \label{eq:Stilinger-weber-h}
  h(\hat{r}^1,\hat{r}^2,\hat{r}^3)=
  \begin{cases}
    \lambda\left(\frac{\left(\hat{r}^1\right)^2+\left(\hat{r}^2\right)^2-\left(\hat{r}^3\right)^2}{2\hat{r}^1\hat{r}^2}
      +\frac{1}{3}\right)^2g_3(\hat{r}^1,\hat{r}^2)&
    \text{if}\
    \hat{r}^1<a\quad\text{and}\quad\hat{r}^2<a\ , \\
                0& \text{otherwise}\ .
        \end{cases}
\end{equation}
Additionally, we introduce the functions
\begin{equation}
  \begin{aligned}
    g_2(\hat{r})&=\exp\left([\hat{r}-a]^{-1}\right) , \\
    g_3(\hat{r}^1,\hat{r}^2)&=\text{exp}\left(\gamma[\hat{r}^1-a]^{-1}+\gamma[\hat{r}^2-a]^{-1}\right)\ .
\end{aligned}
\end{equation}
Here $\theta^{bac}=g(\bar{r}^{ab},\bar{r}^{ac},\bar{r}^{bc})=\text{arccos}(-1/3)\approx
109.47^\circ$ minimizes the function~$h$ given in \cref{eq:Stilinger-weber-h}, that corresponds to
the underlying diamond structure of silicon. From this reasoning it follows that the function $h$
penalizes bond-angles which differ from the ones in this crystal structure.  The parameters in the
potential are $A$, $B$, $a$, $\epsilon$, $\sigma$, $q$, $p$, $\lambda$, and~$\gamma$.

\subsection{Numerical evaluation}
We consider now the numerical solution of systems of particles with the Stillinger-Weber
effective potential. The pairwise contributions to the potential are discretized according to \Cref{sec:pair-potentials}; the remaining three-body
interactions are defined as in \Cref{subsub-em-three}.

\subsubsection{Accuracy study}
\label{sec:consistency-midpoint_three}
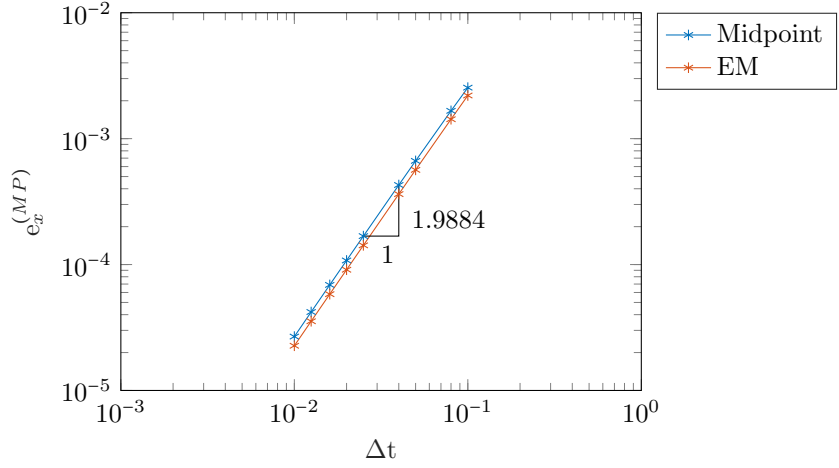
\begin{figure}[H]
	\centering
	\setlength{\figH}{0.26\textheight}
	\setlength{\figW}{0.6\textwidth}
\definecolor{mycolor1}{rgb}{0.00000,0.44700,0.74100}%
\definecolor{mycolor2}{rgb}{0.85000,0.32500,0.09800}%
\begin{tikzpicture}

\begin{axis}[%
width=0.951\figW,
height=\figH,
at={(0\figW,0\figH)},
scale only axis,
xmode=log,
xmin=0.001,
xmax=1,
xminorticks=true,
xlabel style={font=\color{white!15!black}},
xlabel={${\Delta}\text{t}$},
ymode=log,
ymin=1e-05,
ymax=1e-02,
yminorticks=true,
ylabel style={font=\color{white!15!black}},
ylabel={$\text{e}^{(MP)}_x$},
axis background/.style={fill=white},
legend style={legend pos = outer north east, legend cell align=left, align=left, draw=white!15!black}
]
\addplot [color=mycolor1, mark=asterisk, mark options={solid, mycolor1}]
  table[row sep=crcr]{%
0.01	2.67887513008979e-05\\
0.0125	4.19938291881047e-05\\
0.016	6.89294016912103e-05\\
0.02	0.000107750595001252\\
0.025	0.000168259852638578\\
0.04	0.000428394883869509\\
0.05	0.000665587661337871\\
0.08	0.00166254310098797\\
0.1	0.00254135974695313\\
};
\addlegendentry{Midpoint}

\addplot [color=mycolor2, mark=asterisk, mark options={solid, mycolor2}]
  table[row sep=crcr]{%
0.01	2.26423331291582e-05\\
0.0125	3.54611546516687e-05\\
0.016	5.81804357616639e-05\\
0.02	9.0948452877079e-05\\
0.025	0.000142078235848623\\
0.04	0.000362659103317429\\
0.05	0.000564878463920378\\
0.08	0.00142632478674823\\
0.1	0.00220129047749966\\
};
\addlegendentry{EM}

\addplot [color=black, forget plot]
  table[row sep=crcr]{%
0.025	0.000168259852638578\\
0.04	0.000168259852638578\\
0.04	0.000428394883869509\\
};
\node[right, align=left]
at (axis cs:0.042,0.00023) {1.9884};
\node[right, align=left]
at (axis cs:0.028,0.00012) {1};
\end{axis}
\end{tikzpicture}%
	\caption{Accuracy study: Relative error in the position w.r.t midpoint rule}
	\label{accuracy-study-position-three}
\end{figure}

\begin{figure}[H]
	\centering
	\setlength{\figH}{0.26\textheight}
	\setlength{\figW}{0.6\textwidth}
\definecolor{mycolor1}{rgb}{0.00000,0.44700,0.74100}%
\definecolor{mycolor2}{rgb}{0.85000,0.32500,0.09800}%
\begin{tikzpicture}

\begin{axis}[%
width=0.951\figW,
height=\figH,
at={(0\figW,0\figH)},
scale only axis,
xmode=log,
xmin=0.001,
xmax=1,
xminorticks=true,
xlabel style={font=\color{white!15!black}},
xlabel={${\Delta}\text{t}$},
ymode=log,
ymin=0.0001,
ymax=0.1,
yminorticks=true,
ylabel style={font=\color{white!15!black}},
ylabel={$\text{e}_\text{p}^{(MP)}$},
axis background/.style={fill=white},
legend style={legend pos = outer north east, legend cell align=left, align=left, draw=white!15!black}
]
\addplot [color=mycolor1, mark=asterisk, mark options={solid, mycolor1}]
  table[row sep=crcr]{%
0.01	0.00015177304719619\\
0.0125	0.000237909311282551\\
0.016	0.000390482574271878\\
0.02	0.000610343808908923\\
0.025	0.000952949053715426\\
0.04	0.00242464441855083\\
0.05	0.00376481905400872\\
0.08	0.00937901335473552\\
0.1	0.0143008941823212\\
};
\addlegendentry{Midpoint}

\addplot [color=mycolor2, mark=asterisk, mark options={solid, mycolor2}]
  table[row sep=crcr]{%
0.01	0.000310828915829247\\
0.0125	0.000485588449687452\\
0.016	0.000795352911066057\\
0.02	0.00124221311714508\\
0.025	0.00193967994765443\\
0.04	0.00495153380846514\\
0.05	0.00771678710845455\\
0.08	0.0195405106559473\\
0.1	0.0302375807661958\\
};
\addlegendentry{EM}

\addplot [color=black, forget plot]
  table[row sep=crcr]{%
0.025	0.000952949053715426\\
0.04	0.000952949053715426\\
0.04	0.00242464441855083\\
};
\node[right, align=left]
at (axis cs:0.04,0.0014) {1.987};
\node[right, align=left]
at (axis cs:0.027,0.0006) {1};
\end{axis}
\end{tikzpicture}%
	\caption{Accuracy study: Relative error in the linear momentum w.r.t midpoint rule}
	\label{accuracy-study-momentum-three}
\end{figure}
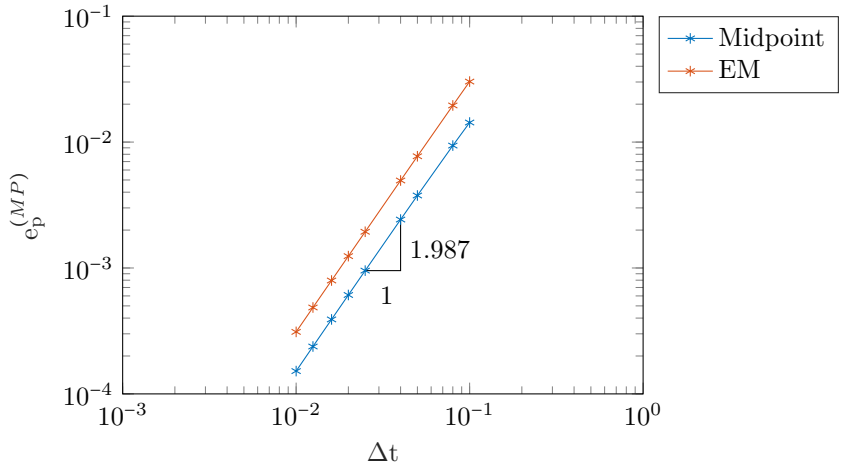

We consider in this example a three-dimensional box $[-L/2,L/2]^3$ filled with~5 particles.
The initial configuration of the system has a particle at the center of the box
and the rest of the particles form bonds with the first one  with angle $\arccos(-1/3)$.
In addition, these four particles are at distance~1 from the center.
Starting from rest, the motion of the system results from the non-vanishing
interacting forces among particles away from the center.
Further parameters of the simulation can be found in \Cref{tab:table3}.
\begin{table}[H]
	\centering
	\caption{Accuracy study: Data used in the simulation}
	\label{tab:table3}
	\begin{tabular}{l c r | l}
		\hline
                                                                                                                                                                      \\
		Material parameters       & $\epsilon$            & 1             & Initial configuration                                                             \\
                                          & $A$                   & 0             &                                                                                   \\
                                          & $B$                   & 0             & \multirow{4}{*}{\includegraphics[scale=0.125]{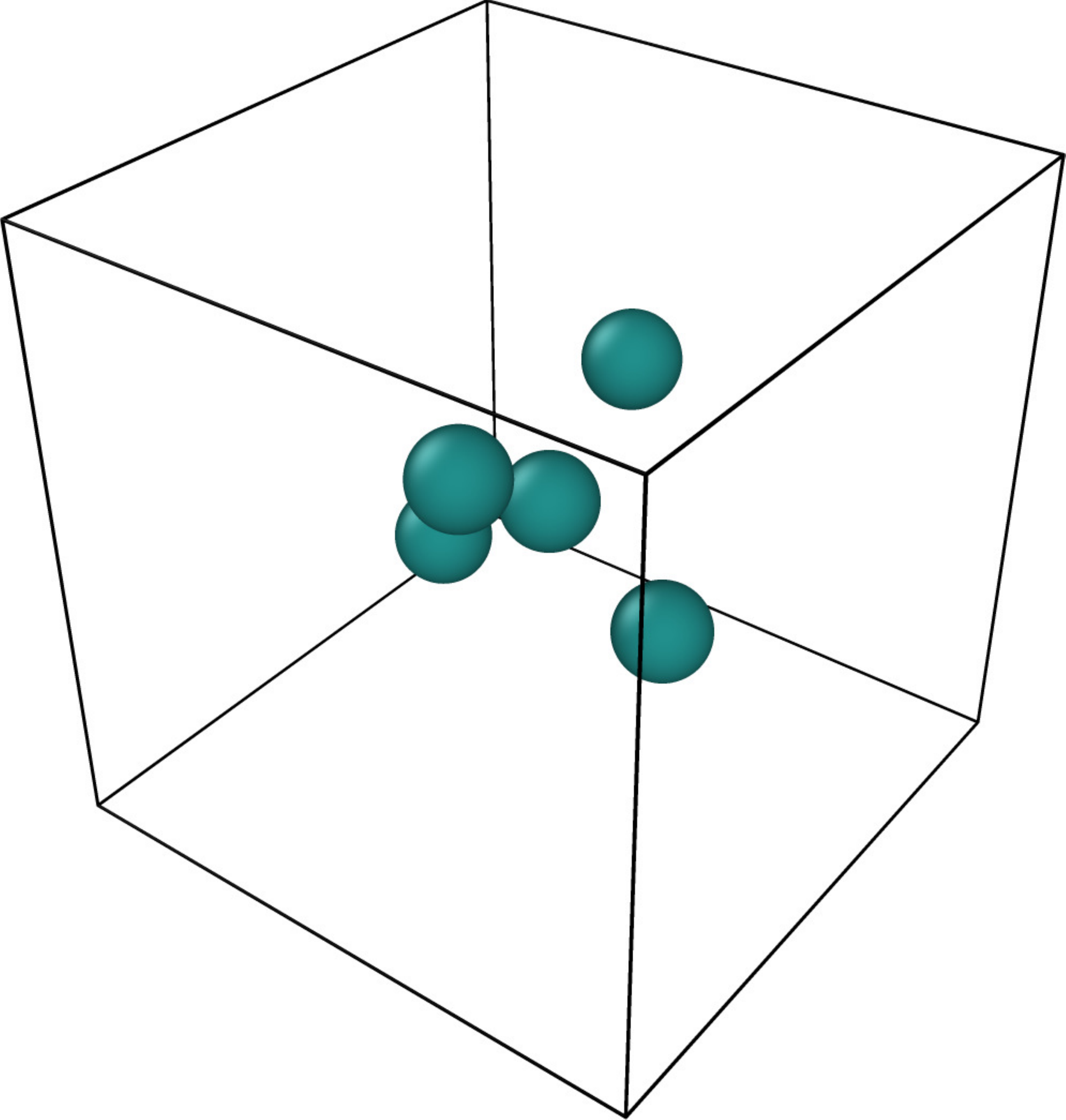}} \\
                                          & $\sigma$              & 1             &                                                                                   \\
                                          & $\lambda$             & 21            &                                                                                   \\
                                          & $p$                   & 0             &                                                                                   \\
                                          & $q$                   & 0             &                                                                                   \\
                                          & $a$                   & 1.8           &                                                                                   \\
                                          & $\gamma$              & 1.2           &                                                                                   \\
		Mass                      & $m^{a}$               & 1             &                                                                                   \\
		Side length               & $L$                   & 4             &                                                                                   \\
		Newton tolerance          & \replaced[id=1r]{-}{$\epsilon$} & 10$^{-8}$     &                                                                                   \\
		Simulation duration       & $T$                   & 0.8           &                                                                                   \\
		 Reference time step size & $\Delta t_\text{ref}$ & 0.001         &                                                                                   \\
		 Time step size           & $\Delta t$            & 0.01, 0.0125, &                                                                                   \\
                                          &                       & 0.016, 0.02,  &                                                                                   \\
                                          &                       & 0.025, 0.04,  &                                                                                   \\
                                          &                       & 0.05, 0.08,   &                                                                                   \\
                                          &                       & 0.1           &                                                                                   \\
                                                                                                                                                                      \\
		\hline
	\end{tabular}
      \end{table}
As in \Cref{sec:consistency-midpoint}, we perform first an accuracy analysis of the EM integrator
using the midpoint rule as a reference.  This study is carried out using ten different time step
sizes for both integrators and employing the error measures
\cref{error-measures}. \Cref{accuracy-study-position-three,accuracy-study-momentum-three} reveal
that both schemes are second order accurate.

\subsection{Energy consistency study}
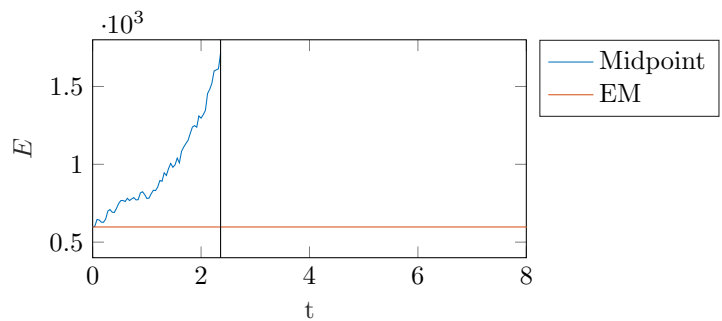
\begin{figure}[H]
  \centering
  \setlength{\figH}{0.15\textheight}
  \setlength{\figW}{0.5\textwidth}
\definecolor{mycolor1}{rgb}{0.00000,0.44700,0.74100}%
\definecolor{mycolor2}{rgb}{0.85000,0.32500,0.09800}%
\begin{tikzpicture}

\begin{axis}[%
width=0.951\figW,
height=\figH,
at={(0\figW,0\figH)},
scale only axis,
xmin=0,
xmax=8,
xlabel style={font=\color{white!15!black}},
xlabel={t},
ymin=400,
ymax=1800,
ylabel style={font=\color{white!15!black}},
ylabel={$E$},
axis background/.style={fill=white},
legend style={legend pos = outer north east, legend cell align=left, align=left, draw=white!15!black}
]
\addplot [color=mycolor1]
  table[row sep=crcr]{%
0	597.370041672624\\
0.04	601.909192029801\\
0.08	645.366366029603\\
0.12	642.50303419172\\
0.16	628.294009749098\\
0.2	627.561852261711\\
0.24	647.868465349177\\
0.28	699.574179364967\\
0.32	709.306025435534\\
0.36	692.59316565008\\
0.4	690.585811736686\\
0.44	717.618196239689\\
0.48	747.812151509372\\
0.52	767.016900890743\\
0.56	767.631537945311\\
0.6	761.083896032278\\
0.64	781.59671027723\\
0.68	767.235496729556\\
0.72	777.848624912706\\
0.76	785.970318505318\\
0.8	771.175686567322\\
0.84	772.980466793055\\
0.88	816.22392275302\\
0.92	823.76626141465\\
0.96	805.531230851908\\
1	780.622938836153\\
1.04	782.726140225919\\
1.08	811.042462662806\\
1.12	833.389408710855\\
1.16	831.24815413595\\
1.2	854.001285006491\\
1.24	895.604975277866\\
1.28	891.070749018008\\
1.32	945.533385061942\\
1.36	928.645768724105\\
1.4	972.776584877765\\
1.44	1005.34595691211\\
1.48	980.90342126933\\
1.52	996.308413740472\\
1.56	1039.44059947638\\
1.6	1009.82433792788\\
1.64	1083.93711177426\\
1.68	1110.74145651701\\
1.72	1133.64447563167\\
1.76	1153.82377934716\\
1.8	1198.21900164467\\
1.84	1239.09353097313\\
1.88	1248.50817672348\\
1.92	1238.62252432384\\
1.96	1310.49604188252\\
2	1296.76818799263\\
2.04	1318.55059959582\\
2.08	1346.34414321503\\
2.12	1454.73952260519\\
2.16	1481.91310803593\\
2.2	1521.42596890245\\
2.24	1599.47183011505\\
2.28	1606.68767336258\\
2.32	1613.1368279963\\
2.36	1713.71540180288\\
};
\addlegendentry{Midpoint}

\addplot [color=mycolor2]
  table[row sep=crcr]{%
0	597.370041672624\\
0.04	597.370041672624\\
0.08	597.370041672625\\
0.12	597.370041672624\\
0.16	597.370041672625\\
0.2	597.370041672623\\
0.24	597.370041672622\\
0.28	597.370041672623\\
0.32	597.370041672623\\
0.36	597.370041672623\\
0.4	597.370041672622\\
0.44	597.370041672623\\
0.48	597.370041672621\\
0.52	597.370041672621\\
0.56	597.370041672622\\
0.6	597.370041672621\\
0.64	597.370041672621\\
0.68	597.370041672621\\
0.72	597.370041672621\\
0.76	597.370041672621\\
0.8	597.370041672621\\
0.84	597.370041672621\\
0.88	597.370041672621\\
0.92	597.370041672621\\
0.96	597.370041672621\\
1	597.370041672621\\
1.04	597.370041672621\\
1.08	597.370041672621\\
1.12	597.370041672621\\
1.16	597.370041672622\\
1.2	597.370041672621\\
1.24	597.370041672621\\
1.28	597.370041672621\\
1.32	597.370041672622\\
1.36	597.370041672639\\
1.4	597.370041672639\\
1.44	597.370041672607\\
1.48	597.370041672606\\
1.52	597.370041672606\\
1.56	597.370041672605\\
1.6	597.370041672606\\
1.64	597.370041672605\\
1.68	597.370041672605\\
1.72	597.370041672606\\
1.76	597.370041672605\\
1.8	597.370041672606\\
1.84	597.370041672605\\
1.88	597.370041672605\\
1.92	597.370041672604\\
1.96	597.370041672605\\
2	597.370041672605\\
2.04	597.370041672604\\
2.08	597.370041672605\\
2.12	597.370041672605\\
2.16	597.370041672605\\
2.2	597.370041672605\\
2.24	597.370041672606\\
2.28	597.370041672605\\
2.32	597.370041672605\\
2.36	597.370041672606\\
2.4	597.370041672606\\
2.44	597.370041672606\\
2.48	597.370041672606\\
2.52	597.370041672606\\
2.56	597.370041672606\\
2.6	597.370041672606\\
2.64	597.370041672605\\
2.68	597.370041672605\\
2.72	597.370041672604\\
2.76	597.370041672604\\
2.8	597.37004167261\\
2.84	597.37004167261\\
2.88	597.370041672611\\
2.92	597.370041672611\\
2.96	597.370041672611\\
3	597.370041672612\\
3.04	597.370041672612\\
3.08	597.370041672612\\
3.12	597.370041672611\\
3.16	597.370041672611\\
3.2	597.370041672612\\
3.24	597.370041672611\\
3.28	597.370041672632\\
3.32	597.370041672632\\
3.36	597.370041672632\\
3.4	597.370041672647\\
3.44	597.370041672647\\
3.48	597.370041672647\\
3.52	597.370041672649\\
3.56	597.370041672648\\
3.6	597.370041672649\\
3.64	597.370041672649\\
3.68	597.370041672649\\
3.72	597.370041672649\\
3.76	597.370041672649\\
3.8	597.370041672649\\
3.84	597.370041672649\\
3.88	597.370041672642\\
3.92	597.370041672641\\
3.96	597.370041672642\\
4	597.370041672642\\
4.04	597.370041672642\\
4.08	597.370041672642\\
4.12	597.370041672642\\
4.16	597.370041672645\\
4.2	597.370041672645\\
4.24	597.370041672645\\
4.28	597.370041672645\\
4.32	597.370041672646\\
4.36	597.370041672646\\
4.4	597.370041672646\\
4.44	597.370041672646\\
4.48	597.370041672646\\
4.52	597.370041672646\\
4.56	597.37004167265\\
4.6	597.37004167265\\
4.64	597.370041672649\\
4.68	597.370041672649\\
4.72	597.370041672651\\
4.76	597.370041672651\\
4.8	597.370041672651\\
4.84	597.37004167262\\
4.88	597.37004167262\\
4.92	597.37004167262\\
4.96	597.37004167262\\
5	597.37004167262\\
5.04	597.37004167262\\
5.08	597.370041672619\\
5.12	597.37004167262\\
5.16	597.370041672619\\
5.2	597.37004167262\\
5.24	597.370041672636\\
5.28	597.370041672635\\
5.32	597.370041672636\\
5.36	597.370041672636\\
5.4	597.370041672635\\
5.44	597.370041672635\\
5.48	597.370041672635\\
5.52	597.370041672635\\
5.56	597.370041672635\\
5.6	597.370041672635\\
5.64	597.370041672643\\
5.68	597.370041672643\\
5.72	597.370041672643\\
5.76	597.370041672644\\
5.8	597.370041672644\\
5.84	597.370041672644\\
5.88	597.370041672644\\
5.92	597.370041672646\\
5.96	597.370041672647\\
6	597.370041672646\\
6.04	597.370041672646\\
6.08	597.370041672646\\
6.12	597.370041672646\\
6.16	597.370041672646\\
6.2	597.370041672646\\
6.24	597.370041672646\\
6.28	597.370041672646\\
6.32	597.370041672646\\
6.36	597.370041672646\\
6.4	597.370041672646\\
6.44	597.370041672646\\
6.48	597.370041672647\\
6.52	597.370041672647\\
6.56	597.370041672647\\
6.6	597.370041672647\\
6.64000000000001	597.370041672648\\
6.68000000000001	597.370041672648\\
6.72000000000001	597.370041672648\\
6.76000000000001	597.370041672648\\
6.80000000000001	597.370041672648\\
6.84000000000001	597.370041672648\\
6.88000000000001	597.370041672647\\
6.92000000000001	597.370041672647\\
6.96000000000001	597.370041672648\\
7.00000000000001	597.370041672647\\
7.04000000000001	597.370041672648\\
7.08000000000001	597.370041672647\\
7.12000000000001	597.370041672648\\
7.16000000000001	597.370041672648\\
7.20000000000001	597.370041672647\\
7.24000000000001	597.370041672647\\
7.28000000000001	597.370041672647\\
7.32000000000001	597.370041672647\\
7.36000000000001	597.370041672647\\
7.40000000000001	597.370041672646\\
7.44000000000001	597.370041672646\\
7.48000000000001	597.370041672647\\
7.52000000000001	597.370041672647\\
7.56000000000001	597.370041672647\\
7.60000000000001	597.370041672647\\
7.64000000000001	597.370041672647\\
7.68000000000001	597.370041672648\\
7.72000000000001	597.370041672648\\
7.76000000000001	597.370041672649\\
7.80000000000001	597.370041672649\\
7.84000000000001	597.370041672649\\
7.88000000000001	597.370041672649\\
7.92000000000001	597.370041672649\\
7.96000000000001	597.370041672648\\
8.00000000000001	597.370041672647\\
8.04	597.370041672648\\
8.08	597.370041672647\\
8.12	597.370041672647\\
8.16	597.370041672647\\
8.2	597.370041672647\\
8.24	597.370041672647\\
8.28	597.370041672647\\
8.32	597.370041672647\\
8.36	597.370041672647\\
8.4	597.370041672646\\
8.44	597.370041672646\\
8.48	597.370041672646\\
8.51999999999999	597.370041672646\\
8.55999999999999	597.370041672643\\
8.59999999999999	597.370041672643\\
8.63999999999999	597.370041672643\\
8.67999999999999	597.370041672642\\
8.71999999999999	597.370041672596\\
8.75999999999999	597.370041672596\\
8.79999999999999	597.370041672596\\
8.83999999999999	597.370041672596\\
8.87999999999999	597.370041672597\\
8.91999999999999	597.370041672597\\
8.95999999999998	597.370041672597\\
8.99999999999998	597.370041672597\\
9.03999999999998	597.370041672598\\
9.07999999999998	597.370041672598\\
9.11999999999998	597.370041672598\\
9.15999999999998	597.370041672599\\
9.19999999999998	597.370041672597\\
9.23999999999998	597.370041672597\\
9.27999999999998	597.370041672596\\
9.31999999999998	597.370041672596\\
9.35999999999998	597.370041672596\\
9.39999999999998	597.370041672596\\
};
\addlegendentry{EM}

\addplot [color=black, forget plot]
  table[row sep=crcr]{%
2.36	400\\
2.36	1800\\
};
\addplot [color=black, forget plot]
  table[row sep=crcr]{%
9.39999999999998	400\\
9.39999999999998	1800\\
};
\end{axis}
\end{tikzpicture}%
  \caption{Energy consistency study: Total energy}
  \label{Big-system-energy-three}
\end{figure}

We consider now 64 atoms inside a three-dimensional box $[-L/2,L/2]^3$, initially arranged
in a perfect diamond cubic lattice structure. This lattice will be disrupted during the simulation
as we consider an initial velocity associated to each atom such that the total
initial kinetic energy is approximately~$768.22$.
\begin{table}[H]
	\centering
	\caption{Energy consistency study: Data used in the simulation}
	\label{tab:table4}
	\begin{tabular}{l c r | l}
		\hline
		\\
		Material parameters & $\epsilon$   & 1                 &  Initial configuration\\
		                    & $A$   &    7.049556277              &  \\
		                    & $B$   &       0.6022245584           &  \multirow{4}{*}{\includegraphics[scale=0.125]{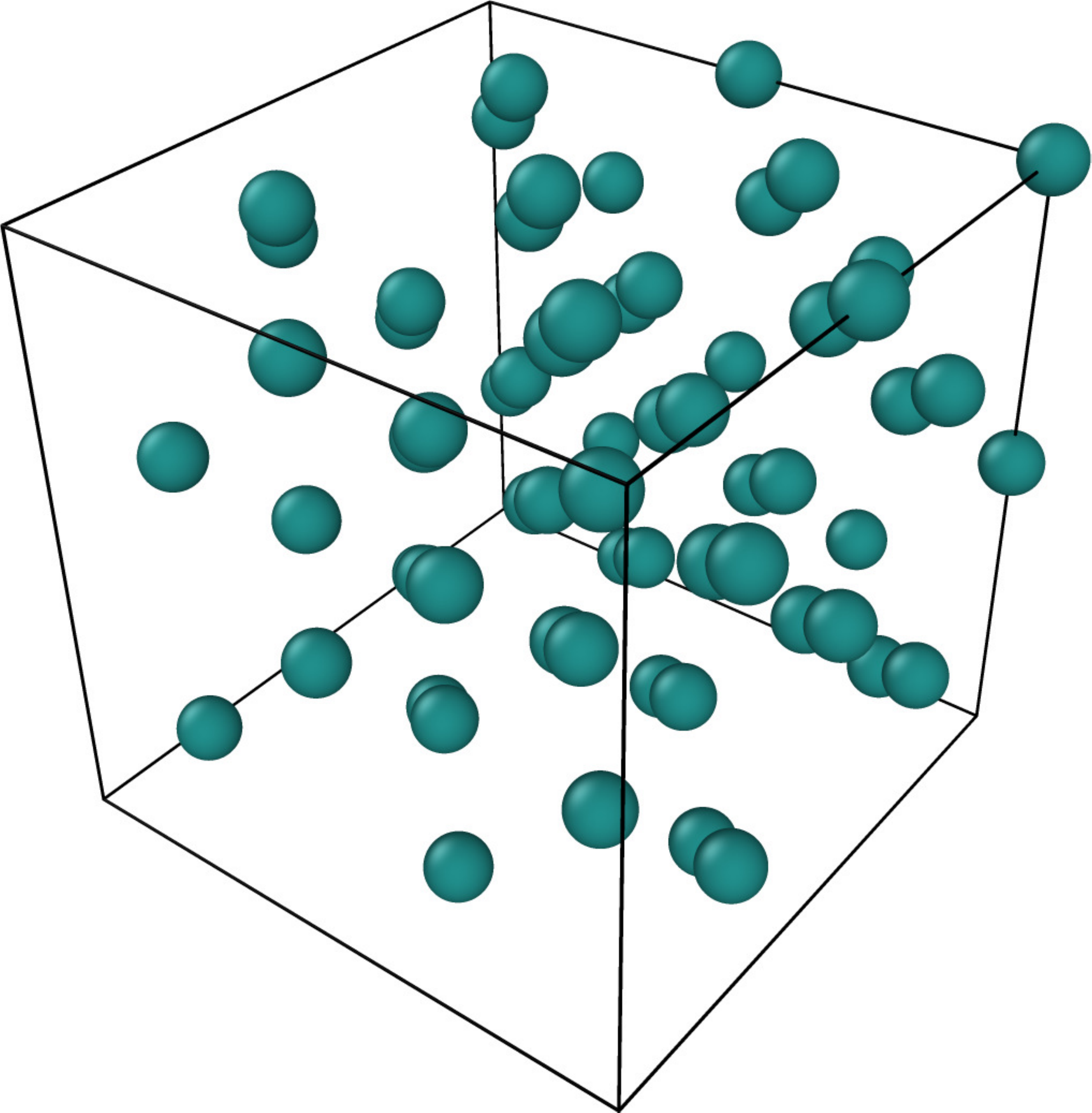}}\\
		                    & $\sigma$   & 1                 &  \\
		                    & $\lambda$   & 210                 &  \\
		                    & $p$   & 4                 &  \\
		                    & $q$   & 0                 &  \\
		                    & $a$   & 2                 &  \\
		                    & $\gamma$   & 1.2                 &  \\
		Mass                & $m^{a}$    & 1                 &                       \\
		Sidelength          & $L$        & 4                &                       \\
		Newton tolerance    & \replaced[id=1r]{-}{$\epsilon$} & 10$^{-9}$         &                       \\
		Simulation duration & $T$        & 16                 &                       \\
		 Time step size                    &  $\Delta t$          & 0.04 &                       \\
		\\
		\hline
	\end{tabular}
\end{table}
For the chosen time step size, the midpoint rule clearly violates
the conservation of the total energy, see \Cref{Big-system-energy-three},
leading to an energy blow-up and finally to a termination of the simulation, indicated by the black
line on the same figure.

\begin{figure}[H]
	\centering
	\setlength{\figH}{0.15\textheight}
	\setlength{\figW}{0.5\textwidth}
\definecolor{mycolor1}{rgb}{0.00000,0.44700,0.74100}%
\definecolor{mycolor2}{rgb}{0.85000,0.32500,0.09800}%
\begin{tikzpicture}

\begin{axis}[%
width=0.951\figW,
height=\figH,
at={(0\figW,0\figH)},
scale only axis,
xmin=0,
xmax=8,
xlabel style={font=\color{white!15!black}},
xlabel={t},
ymin=-200,
ymax=1000,
ylabel style={font=\color{white!15!black}},
ylabel={$V$},
axis background/.style={fill=white},
legend style={legend pos = outer north east, legend cell align=left, align=left, draw=white!15!black}
]
\addplot [color=mycolor1]
  table[row sep=crcr]{%
0	-170.849896374171\\
0.04	-156.293844340908\\
0.08	-139.549099974984\\
0.12	-140.489181228567\\
0.16	-146.473999697682\\
0.2	-146.954146352725\\
0.24	-147.011963347652\\
0.28	-144.872394990133\\
0.32	-146.437039508281\\
0.36	-147.496205444507\\
0.4	-146.802140108216\\
0.44	-144.275392435914\\
0.48	-141.940401168382\\
0.52	-144.440879254842\\
0.56	-143.507862886622\\
0.6	-142.606824704279\\
0.64	-145.047981761199\\
0.68	-145.575494343398\\
0.72	-141.355947434236\\
0.76	-141.442059450684\\
0.8	-145.154189294437\\
0.84	-142.324314726124\\
0.88	-139.062380887979\\
0.92	-136.57576326304\\
0.96	-140.844531426256\\
1	-150.167480823331\\
1.04	-148.246323273541\\
1.08	-143.039386253436\\
1.12	-143.151844807506\\
1.16	-144.350152447802\\
1.2	-140.563246077007\\
1.24	-140.464950456266\\
1.28	-144.395714581227\\
1.32	-142.792290988815\\
1.36	-138.734243919258\\
1.4	-135.585149348399\\
1.44	-122.969403291712\\
1.48	-137.565100823963\\
1.52	-144.935777000925\\
1.56	-139.717618622783\\
1.6	-136.68118776203\\
1.64	-136.395809162883\\
1.68	-133.972485181667\\
1.72	-135.35818715717\\
1.76	-134.03774686608\\
1.8	-127.587833278993\\
1.84	-120.812922004752\\
1.88	-127.395396472775\\
1.92	-136.268427355406\\
1.96	-134.268290651278\\
2	-128.064636959581\\
2.04	-132.799427655212\\
2.08	-137.061800073298\\
2.12	-137.242191182127\\
2.16	-134.867162673376\\
2.2	-137.164442585206\\
2.24	-127.496177273927\\
2.28	-127.605727725225\\
2.32	-132.72519884316\\
2.36	-121.128648800458\\
};
\addlegendentry{$\text{Midpoint V}_\text{2}$}

\addplot [color=mycolor2]
  table[row sep=crcr]{%
0	1.29558111812673e-27\\
0.04	353.175749452166\\
0.08	433.333882653255\\
0.12	422.931356594738\\
0.16	278.066926723476\\
0.2	377.917936582745\\
0.24	318.470521589525\\
0.28	405.597062897914\\
0.32	390.699596223824\\
0.36	321.985237131473\\
0.4	368.647219840209\\
0.44	431.690117111926\\
0.48	453.912132069473\\
0.52	403.331610874044\\
0.56	449.655568024349\\
0.6	407.904642881619\\
0.64	471.703757668591\\
0.68	385.871304240701\\
0.72	386.138343039613\\
0.76	476.155409300682\\
0.8	401.444153185385\\
0.84	389.600128851591\\
0.88	515.145012783373\\
0.92	447.304203584509\\
0.96	493.611613771294\\
1	391.525630041269\\
1.04	443.164986535825\\
1.08	482.285132459473\\
1.12	500.328133586064\\
1.16	437.975598367101\\
1.2	436.378495458462\\
1.24	583.621828799053\\
1.28	419.956827766855\\
1.32	560.891457478661\\
1.36	470.890007447866\\
1.4	567.695183598833\\
1.44	548.76733436114\\
1.48	544.753278086471\\
1.52	545.424067627872\\
1.56	582.020654538576\\
1.6	485.103644929344\\
1.64	655.458587516539\\
1.68	585.336627934451\\
1.72	673.012084689957\\
1.76	673.474672113538\\
1.8	646.189050194932\\
1.84	700.398965291612\\
1.88	723.740745997063\\
1.92	626.078145234155\\
1.96	702.99640066007\\
2	708.638462566312\\
2.04	785.962093517731\\
2.08	648.034973826125\\
2.12	756.752654514206\\
2.16	621.845765974271\\
2.2	773.118280763661\\
2.24	811.302372011532\\
2.28	910.550894537941\\
2.32	768.598552549109\\
2.36	823.138937856712\\
};
\addlegendentry{$\text{Midpoint V}_\text{3}$}

\addplot [color=black, forget plot]
  table[row sep=crcr]{%
2.36	-200\\
2.36	1000\\
};
\end{axis}
\end{tikzpicture}%
	\caption{Energy consistency study: Potential energies using the Midpoint rule}
	\label{Big-system-energy-Pot-mid-three}
\end{figure}
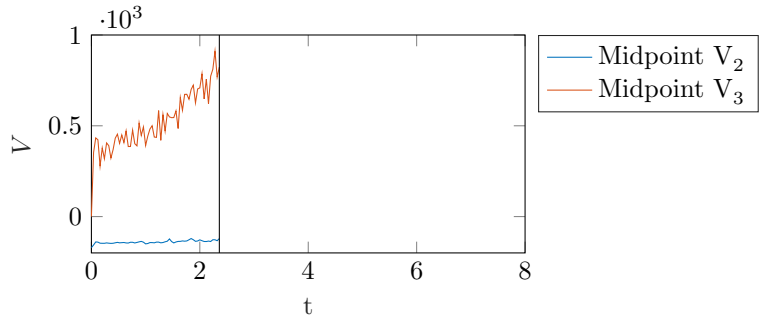

The largest contribution to the algorithmic energy error is due to  the midpoint approximation of the
forces generated from the three-body contribution of the Stillinger-Weber potential.  It can be
observed in \Cref{Big-system-energy-Pot-mid-three} that the potential energy of the two-body
terms remains bounded, while the potential energy of the three-body terms increases unphysically
causing the energy blow-up and, ultimately,  the termination of the simulation.
In contrast, using the EM method both contributions to the potential energy of the system
remain bounded. See \Cref{Big-system-energy-Pot-EM-three}.
\begin{figure}[H]
	\centering
	\setlength{\figH}{0.15\textheight}
	\setlength{\figW}{0.5\textwidth}
	\input{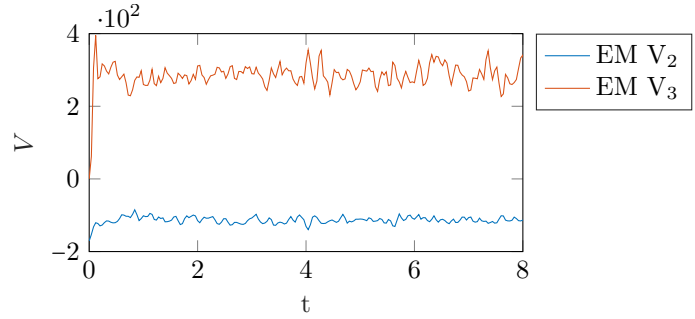}
	\caption{Energy consistency study: Potential energies using the EM method}
	\label{Big-system-energy-Pot-EM-three}
\end{figure}
As expected, the EM method preserves the total energy up to round-off errors. See
\Cref{Big-system-energy-diff-three}.
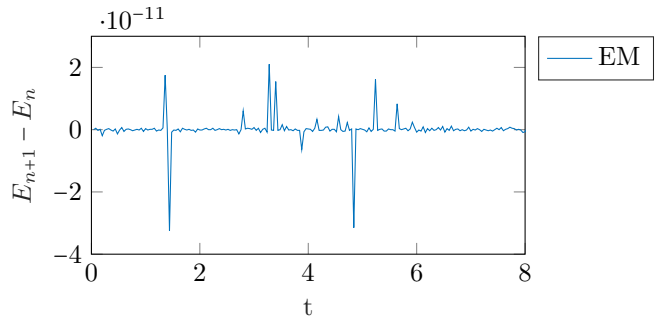
\begin{figure}[H]
  \centering
  \setlength{\figH}{0.15\textheight}
  \setlength{\figW}{0.5\textwidth}
\definecolor{mycolor1}{rgb}{0.00000,0.44700,0.74100}%
\begin{tikzpicture}

\begin{axis}[%
width=0.951\figW,
height=\figH,
at={(0\figW,0\figH)},
scale only axis,
xmin=0,
xmax=8,
xlabel style={font=\color{white!15!black}},
xlabel={t},
ymin=-4e-11,
ymax=3e-11,
ylabel style={font=\color{white!15!black}},
ylabel={$E_{n+1}-E_n$},
axis background/.style={fill=white},
legend style={legend pos = outer north east, legend cell align=left, align=left, draw=white!15!black}
]
\addplot [color=mycolor1]
  table[row sep=crcr]{%
0.04	0\\
0.08	4.54747350886464e-13\\
0.12	-2.27373675443232e-13\\
0.16	1.13686837721616e-13\\
0.2	-1.93267624126747e-12\\
0.24	-2.27373675443232e-13\\
0.28	1.13686837721616e-13\\
0.32	3.41060513164848e-13\\
0.36	-1.13686837721616e-13\\
0.4	-4.54747350886464e-13\\
0.44	2.27373675443232e-13\\
0.48	-1.36424205265939e-12\\
0.52	-1.13686837721616e-13\\
0.56	6.82121026329696e-13\\
0.6	-5.6843418860808e-13\\
0.64	1.13686837721616e-13\\
0.68	2.27373675443232e-13\\
0.72	0\\
0.76	-3.41060513164848e-13\\
0.8	-1.13686837721616e-13\\
0.84	1.13686837721616e-13\\
0.88	-1.13686837721616e-13\\
0.92	4.54747350886464e-13\\
0.96	-5.6843418860808e-13\\
1	3.41060513164848e-13\\
1.04	-1.13686837721616e-13\\
1.08	1.13686837721616e-13\\
1.12	1.13686837721616e-13\\
1.16	4.54747350886464e-13\\
1.2	-4.54747350886464e-13\\
1.24	-1.13686837721616e-13\\
1.28	-2.27373675443232e-13\\
1.32	5.6843418860808e-13\\
1.36	1.75077730091289e-11\\
1.4	-1.13686837721616e-13\\
1.44	-3.24007487506606e-11\\
1.48	-7.95807864051312e-13\\
1.52	-1.13686837721616e-13\\
1.56	-2.27373675443232e-13\\
1.6	2.27373675443232e-13\\
1.64	-6.82121026329696e-13\\
1.68	4.54747350886464e-13\\
1.72	1.13686837721616e-13\\
1.76	-1.13686837721616e-13\\
1.8	1.13686837721616e-13\\
1.84	-2.27373675443232e-13\\
1.88	-1.13686837721616e-13\\
1.92	-7.95807864051312e-13\\
1.96	2.27373675443232e-13\\
2	-1.13686837721616e-13\\
2.04	-1.13686837721616e-13\\
2.08	2.27373675443232e-13\\
2.12	4.54747350886464e-13\\
2.16	0\\
2.2	0\\
2.24	4.54747350886464e-13\\
2.28	-2.27373675443232e-13\\
2.32	0\\
2.36	3.41060513164848e-13\\
2.4	0\\
2.44	1.13686837721616e-13\\
2.48	-1.13686837721616e-13\\
2.52	1.13686837721616e-13\\
2.56	-2.27373675443232e-13\\
2.6	0\\
2.64	-1.13686837721616e-13\\
2.68	0\\
2.72	-1.36424205265939e-12\\
2.76	0\\
2.8	6.02540239924565e-12\\
2.84	2.27373675443232e-13\\
2.88	4.54747350886464e-13\\
2.92	3.41060513164848e-13\\
2.96	1.13686837721616e-13\\
3	6.82121026329696e-13\\
3.04	-2.27373675443232e-13\\
3.08	4.54747350886464e-13\\
3.12	-9.09494701772928e-13\\
3.16	1.13686837721616e-13\\
3.2	4.54747350886464e-13\\
3.24	-7.95807864051312e-13\\
3.28	2.1032064978499e-11\\
3.32	-1.13686837721616e-13\\
3.36	-2.27373675443232e-13\\
3.4	1.54614099301398e-11\\
3.44	-2.27373675443232e-13\\
3.48	0\\
3.52	1.59161572810262e-12\\
3.56	-3.41060513164848e-13\\
3.6	1.02318153949454e-12\\
3.64	-1.13686837721616e-13\\
3.68	0\\
3.72	-3.41060513164848e-13\\
3.76	2.27373675443232e-13\\
3.8	-2.27373675443232e-13\\
3.84	-2.27373675443232e-13\\
3.88	-6.3664629124105e-12\\
3.92	-6.82121026329696e-13\\
3.96	3.41060513164848e-13\\
4	2.27373675443232e-13\\
4.04	1.13686837721616e-13\\
4.08	-5.6843418860808e-13\\
4.12	4.54747350886464e-13\\
4.16	3.29691829392686e-12\\
4.2	-2.27373675443232e-13\\
4.24	-2.27373675443232e-13\\
4.28	-1.13686837721616e-13\\
4.32	7.95807864051312e-13\\
4.36	9.09494701772928e-13\\
4.4	-4.54747350886464e-13\\
4.44	1.13686837721616e-13\\
4.48	1.13686837721616e-13\\
4.52	-6.82121026329696e-13\\
4.56	4.20641299569979e-12\\
4.6	0\\
4.64	-5.6843418860808e-13\\
4.68	-4.54747350886464e-13\\
4.72	2.27373675443232e-12\\
4.76	-2.27373675443232e-13\\
4.8	2.27373675443232e-13\\
4.84	-3.14912540488876e-11\\
4.88	2.27373675443232e-13\\
4.92	0\\
4.96	3.41060513164848e-13\\
5	1.13686837721616e-13\\
5.04	-1.13686837721616e-13\\
5.08	-6.82121026329696e-13\\
5.12	5.6843418860808e-13\\
5.16	-5.6843418860808e-13\\
5.2	1.13686837721616e-13\\
5.24	1.61435309564695e-11\\
5.28	-2.27373675443232e-13\\
5.32	1.13686837721616e-13\\
5.36	2.27373675443232e-13\\
5.4	-4.54747350886464e-13\\
5.44	-4.54747350886464e-13\\
5.48	2.27373675443232e-13\\
5.52	0\\
5.56	0\\
5.6	-4.54747350886464e-13\\
5.64	8.29913915367797e-12\\
5.68	3.41060513164848e-13\\
5.72	0\\
5.76	2.27373675443232e-13\\
5.8	4.54747350886464e-13\\
5.84	-3.41060513164848e-13\\
5.88	2.27373675443232e-13\\
5.92	2.38742359215394e-12\\
5.96	6.82121026329696e-13\\
6	-7.95807864051312e-13\\
6.04	-1.13686837721616e-13\\
6.08	-4.54747350886464e-13\\
6.12	9.09494701772928e-13\\
6.16	-9.09494701772928e-13\\
6.2	3.41060513164848e-13\\
6.24	-3.41060513164848e-13\\
6.28	4.54747350886464e-13\\
6.32	-2.27373675443232e-13\\
6.36	0\\
6.4	6.82121026329696e-13\\
6.44	-1.13686837721616e-13\\
6.48	1.02318153949454e-12\\
6.52	-7.95807864051312e-13\\
6.56	3.41060513164848e-13\\
6.6	-2.27373675443232e-13\\
6.64000000000001	1.13686837721616e-12\\
6.68000000000001	-3.41060513164848e-13\\
6.72000000000001	1.13686837721616e-13\\
6.76000000000001	2.27373675443232e-13\\
6.80000000000001	-3.41060513164848e-13\\
6.84000000000001	0\\
6.88000000000001	-3.41060513164848e-13\\
6.92000000000001	-1.13686837721616e-13\\
6.96000000000001	7.95807864051312e-13\\
7.00000000000001	-4.54747350886464e-13\\
7.04000000000001	1.13686837721616e-13\\
7.08000000000001	-1.13686837721616e-13\\
7.12000000000001	3.41060513164848e-13\\
7.16000000000001	0\\
7.20000000000001	-5.6843418860808e-13\\
7.24000000000001	1.13686837721616e-13\\
7.28000000000001	-1.13686837721616e-13\\
7.32000000000001	-3.41060513164848e-13\\
7.36000000000001	-1.13686837721616e-13\\
7.40000000000001	-4.54747350886464e-13\\
7.44000000000001	0\\
7.48000000000001	4.54747350886464e-13\\
7.52000000000001	0\\
7.56000000000001	6.82121026329696e-13\\
7.60000000000001	-3.41060513164848e-13\\
7.64000000000001	1.13686837721616e-13\\
7.68000000000001	3.41060513164848e-13\\
7.72000000000001	7.95807864051312e-13\\
7.76000000000001	4.54747350886464e-13\\
7.80000000000001	3.41060513164848e-13\\
7.84000000000001	-1.13686837721616e-13\\
7.88000000000001	1.13686837721616e-13\\
7.92000000000001	-1.13686837721616e-13\\
7.96000000000001	-9.09494701772928e-13\\
8.00000000000001	-6.82121026329696e-13\\
8.04	6.82121026329696e-13\\
8.08	-9.09494701772928e-13\\
8.12	-2.27373675443232e-13\\
8.16	-2.27373675443232e-13\\
8.2	0\\
8.24	4.54747350886464e-13\\
8.28	0\\
8.32	-2.27373675443232e-13\\
8.36	-3.41060513164848e-13\\
8.4	-1.13686837721616e-13\\
8.44	-2.27373675443232e-13\\
8.48	-2.27373675443232e-13\\
8.51999999999999	2.27373675443232e-13\\
8.55999999999999	-3.06954461848363e-12\\
8.59999999999999	2.27373675443232e-13\\
8.63999999999999	-9.09494701772928e-13\\
8.67999999999999	-2.27373675443232e-13\\
8.71999999999999	-4.59294824395329e-11\\
8.75999999999999	-3.41060513164848e-13\\
8.79999999999999	1.13686837721616e-13\\
8.83999999999999	0\\
8.87999999999999	5.6843418860808e-13\\
8.91999999999999	2.27373675443232e-13\\
8.95999999999998	-3.41060513164848e-13\\
8.99999999999998	1.13686837721616e-13\\
9.03999999999998	1.13686837721616e-12\\
9.07999999999998	5.6843418860808e-13\\
9.11999999999998	0\\
9.15999999999998	1.13686837721616e-13\\
9.19999999999998	-1.25055521493778e-12\\
9.23999999999998	-3.41060513164848e-13\\
9.27999999999998	-6.82121026329696e-13\\
9.31999999999998	-3.41060513164848e-13\\
9.35999999999998	-3.41060513164848e-13\\
9.39999999999998	6.82121026329696e-13\\
};
\addlegendentry{EM}

\addplot [color=black, forget plot]
  table[row sep=crcr]{%
9.39999999999998	-5e-11\\
9.39999999999998	3e-11\\
};
\end{axis}
\end{tikzpicture}%
  \caption{Energy consistency study: Total energy difference}
  \label{Big-system-energy-diff-three}
\end{figure}
Eventually, as illustrated in \Cref{Big-system-kinetic-three},
the evolution of the kinetic energy reveals that the EM
solution reaches thermodynamic equilibrium for the chosen time step size,
in contrast with the midpoint rule.
\begin{figure}[H]
	\centering
	\setlength{\figH}{0.15\textheight}
	\setlength{\figW}{0.5\textwidth}
\definecolor{mycolor1}{rgb}{0.00000,0.44700,0.74100}%
\definecolor{mycolor2}{rgb}{0.85000,0.32500,0.09800}%
\begin{tikzpicture}

\begin{axis}[%
width=0.951\figW,
height=\figH,
at={(0\figW,0\figH)},
scale only axis,
xmin=0,
xmax=9.39999999999998,
xlabel style={font=\color{white!15!black}},
xlabel={t},
ymin=300,
ymax=1100,
ylabel style={font=\color{white!15!black}},
ylabel={$T$},
axis background/.style={fill=white},
legend style={legend pos = outer north east, legend cell align=left, align=left, draw=white!15!black}
]
\addplot [color=mycolor1]
  table[row sep=crcr]{%
0	768.219938046795\\
0.04	405.027286918544\\
0.08	351.581583351333\\
0.12	360.060858825549\\
0.16	496.701082723303\\
0.2	396.598062031691\\
0.24	476.409907107304\\
0.28	438.849511457186\\
0.32	465.043468719992\\
0.36	518.104133963114\\
0.4	468.740732004694\\
0.44	430.203471563678\\
0.48	435.840420608281\\
0.52	508.126169271542\\
0.56	461.483832807584\\
0.6	495.786077854937\\
0.64	454.940934369837\\
0.68	526.939686832253\\
0.72	533.06622930733\\
0.76	451.25696865532\\
0.8	514.885722676375\\
0.84	525.704652667587\\
0.88	440.141290857627\\
0.92	513.037821093181\\
0.96	452.764148506869\\
1	539.264789618216\\
1.04	487.807476963635\\
1.08	471.796716456768\\
1.12	476.213119932296\\
1.16	537.622708216651\\
1.2	558.186035625035\\
1.24	452.448096935079\\
1.28	615.50963583238\\
1.32	527.434218572095\\
1.36	596.490005195496\\
1.4	540.666550627331\\
1.44	579.548025842684\\
1.48	573.715244006824\\
1.52	595.820123113524\\
1.56	597.137563560589\\
1.6	661.40188076057\\
1.64	564.874333420603\\
1.68	659.377313764228\\
1.72	595.990578098884\\
1.76	614.386854099704\\
1.8	679.617784728731\\
1.84	659.507487686273\\
1.88	652.162827199195\\
1.92	748.812806445088\\
1.96	741.76793187373\\
2	716.194362385896\\
2.04	665.387933733299\\
2.08	835.370969462203\\
2.12	835.229059273116\\
2.16	994.934504735034\\
2.2	885.472130723999\\
2.24	915.665635377444\\
2.28	823.742506549863\\
2.32	977.263474290354\\
2.36	1011.70511274662\\
};
\addlegendentry{Midpoint}

\addplot [color=mycolor2]
  table[row sep=crcr]{%
0	768.219938046795\\
0.04	688.511746367763\\
0.08	423.245525552871\\
0.12	322.841897099634\\
0.16	444.016927701608\\
0.2	445.474841111663\\
0.24	407.692306964133\\
0.28	406.65276918478\\
0.32	416.12679882816\\
0.36	424.384249437996\\
0.4	409.181509886928\\
0.44	397.09324336625\\
0.48	394.213404008885\\
0.52	428.222593174968\\
0.56	432.924234182635\\
0.6	415.029992318493\\
0.64	408.438981843716\\
0.68	435.289565663026\\
0.72	469.557728726746\\
0.76	474.79744130441\\
0.8	449.817331210528\\
0.84	411.033244304345\\
0.88	414.870739484138\\
0.92	432.36925370802\\
0.96	408.896849652521\\
1	438.881134483143\\
1.04	443.719590682958\\
1.08	443.129880183476\\
1.12	420.557467803783\\
1.16	393.692803678696\\
1.2	446.69619233064\\
1.24	453.885191245367\\
1.28	419.555539041755\\
1.32	441.13282154552\\
1.36	428.024237944642\\
1.4	409.396791727225\\
1.44	423.884044876082\\
1.48	429.334286682749\\
1.52	433.999457096094\\
1.56	439.344027211643\\
1.6	435.938551315178\\
1.64	421.321004473787\\
1.68	437.932700051541\\
1.72	443.901817678678\\
1.76	420.443964628011\\
1.8	414.656021928009\\
1.84	442.572900750295\\
1.88	429.670298723578\\
1.92	408.996692479002\\
1.96	425.019840157111\\
2	401.22450111626\\
2.04	405.323972644594\\
2.08	423.051523701364\\
2.12	423.684864102472\\
2.16	444.863984500645\\
2.2	432.051645719506\\
2.24	419.2333882594\\
2.28	415.48944251429\\
2.32	421.611759903776\\
2.36	413.027792929649\\
2.4	416.577459759018\\
2.44	422.013122549904\\
2.48	410.605045890875\\
2.52	429.86173462524\\
2.56	427.638931442456\\
2.6	415.730763530989\\
2.64	411.606902638865\\
2.68	440.233905140812\\
2.72	443.51661807077\\
2.76	423.683258075621\\
2.8	426.183989952074\\
2.84	431.782010678717\\
2.88	410.974504522984\\
2.92	408.390693118976\\
2.96	398.849723760852\\
3	415.218588403821\\
3.04	411.200440787318\\
3.08	387.499582002346\\
3.12	416.483779398994\\
3.16	431.871958860336\\
3.2	445.267279308037\\
3.24	465.20957456325\\
3.28	470.240222487247\\
3.32	433.372965989417\\
3.36	403.860371240865\\
3.4	435.209349396694\\
3.44	445.873919510383\\
3.48	416.972855749034\\
3.52	393.341999050529\\
3.56	431.293567068837\\
3.6	423.116891890504\\
3.64	414.066995112139\\
3.68	411.344006609067\\
3.72	403.237543600155\\
3.76	435.665837192224\\
3.8	452.25620585864\\
3.84	444.796374474913\\
3.88	416.016797047455\\
3.92	421.087887046373\\
3.96	446.017215255805\\
4	408.855703583521\\
4.04	381.89832733921\\
4.08	404.648213599258\\
4.12	424.674725374683\\
4.16	451.126584350148\\
4.2	433.86803234446\\
4.24	374.135418197175\\
4.28	366.070813268131\\
4.32	436.431887966359\\
4.36	443.154230557663\\
4.4	449.062393079446\\
4.44	480.90674293303\\
4.48	446.088341193177\\
4.52	412.93879528059\\
4.56	424.189662245275\\
4.6	420.359809294084\\
4.64	425.78477594189\\
4.68	422.251468703092\\
4.72	429.377434246334\\
4.76	417.664606733846\\
4.8	416.826876164591\\
4.84	432.785409279263\\
4.88	450.838807641621\\
4.92	462.734007601835\\
4.96	454.427860364141\\
5	440.305564991176\\
5.04	408.498209749383\\
5.08	411.645462223633\\
5.12	415.540748271581\\
5.16	399.288642091752\\
5.2	411.500389121917\\
5.24	418.071310939064\\
5.28	459.825908228322\\
5.32	471.158163180586\\
5.36	431.719086353606\\
5.4	414.317755454548\\
5.44	432.309020788115\\
5.48	424.218388959298\\
5.52	452.215503222382\\
5.56	454.017029643335\\
5.6	405.228222699037\\
5.64	416.206145772121\\
5.68	433.474419250531\\
5.72	426.738115464591\\
5.76	454.017482848395\\
5.8	433.018814282557\\
5.84	412.960252832382\\
5.88	419.263310505864\\
5.92	431.930729264964\\
5.96	437.939258109916\\
6	443.415693092103\\
6.04	405.212693010456\\
6.08	369.309640518133\\
6.12	404.050766162366\\
6.16	446.667530007862\\
6.2	442.308150323774\\
6.24	434.913337508909\\
6.28	395.391635872063\\
6.32	366.364807388693\\
6.36	377.658755899854\\
6.4	372.253521148396\\
6.44	380.521193853757\\
6.48	390.130082745097\\
6.52	396.846214463205\\
6.56	384.198392357225\\
6.6	398.934679349464\\
6.64000000000001	436.166600428398\\
6.68000000000001	454.457328399994\\
6.72000000000001	422.167681867318\\
6.76000000000001	403.856008695054\\
6.80000000000001	404.335654739254\\
6.84000000000001	416.628254038221\\
6.88000000000001	416.466230076805\\
6.92000000000001	421.521953906893\\
6.96000000000001	442.628079371529\\
7.00000000000001	461.823471465354\\
7.04000000000001	457.520751485266\\
7.08000000000001	435.645735881846\\
7.12000000000001	439.175950327877\\
7.16000000000001	439.251601387118\\
7.20000000000001	404.026757821746\\
7.24000000000001	423.292067484879\\
7.28000000000001	440.113287803167\\
7.32000000000001	385.022724448345\\
7.36000000000001	367.036262162529\\
7.40000000000001	411.486431118149\\
7.44000000000001	440.958455431993\\
7.48000000000001	421.416752190702\\
7.52000000000001	410.247923464033\\
7.56000000000001	444.419928009177\\
7.60000000000001	471.765301496458\\
7.64000000000001	465.264736764173\\
7.68000000000001	422.990128368932\\
7.72000000000001	434.520056774802\\
7.76000000000001	451.260795975807\\
7.80000000000001	451.221541161177\\
7.84000000000001	437.359487934589\\
7.88000000000001	432.941801890524\\
7.92000000000001	408.414387533515\\
7.96000000000001	380.054840842906\\
8.00000000000001	369.727395143895\\
8.04	398.256357159514\\
8.08	416.03814885689\\
8.12	387.377409181275\\
8.16	431.515979033251\\
8.2	422.885677367386\\
8.24	394.445596246505\\
8.28	405.055362743075\\
8.32	414.032516811813\\
8.36	382.964130207371\\
8.4	384.184812019256\\
8.44	415.966317517188\\
8.48	432.166196667124\\
8.51999999999999	406.127998138371\\
8.55999999999999	381.596892093231\\
8.59999999999999	390.39018101546\\
8.63999999999999	444.174487030132\\
8.67999999999999	434.649705258632\\
8.71999999999999	396.63447676535\\
8.75999999999999	426.606940115042\\
8.79999999999999	463.998986139617\\
8.83999999999999	433.075102247311\\
8.87999999999999	383.762476139645\\
8.91999999999999	389.228069131229\\
8.95999999999998	400.796716014186\\
8.99999999999998	422.348073451601\\
9.03999999999998	459.795908244729\\
9.07999999999998	443.930368367184\\
9.11999999999998	366.528280805025\\
9.15999999999998	348.968438779582\\
9.19999999999998	373.116763896747\\
9.23999999999998	403.389670614169\\
9.27999999999998	441.916993598122\\
9.31999999999998	423.29075493093\\
9.35999999999998	427.894434274164\\
9.39999999999998	455.641338974996\\
};
\addlegendentry{EM}

\addplot [color=black, forget plot]
  table[row sep=crcr]{%
2.36	300\\
2.36	1100\\
};
\end{axis}
\end{tikzpicture}%
	\caption{Energy consistency study: Kinetic energy}
	\label{Big-system-kinetic-three}
\end{figure}
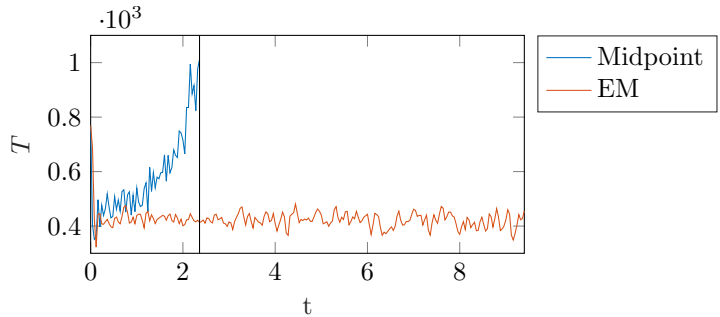

%%% Local Variables:
%%% TeX-command-extra-options: "-shell-escape"
%%% mode: latex
%%% TeX-master: "driver"
%%% End:

\section{Energy-Momentum methods for periodic systems described by the embedded-atom method}
\label{sec:embedded-atom-method}
In addition to three-body potentials of the type described in \Cref{sec:three-body-potent},
more elaborate and accurate potentials include multi-body effects through an environment-dependent variable, resulting
in effective potential that are extremely common, for example, in the simulation of metals.
One of the most popular interatomic potentials of this class is the one employed
in the  embedded-atom method (EAM) \cite{daw1984embedded,baskes1992modified,daw1993embedded},
whose use in the context of conserving schemes is analyzed in this section. 
The interatomic potential in this case is of the form
\begin{equation}
  \label{eq-v-eam}
  V = \frac{1}{2}\sum_{\substack{a,b=1  \\b\neq a}}^N
  \tilde{V}\left(d_T( {\mbs{x}}^{a}, {\mbs{x}}^{b}) \right)
  +
  \sum_{a=1}^N\tilde{F}\left(\bar{\rho}^a\right)\ .
\end{equation}
The EAM potential consists of a pair potential contribution and electronic energies
$\tilde{F}(\bar{\rho}^a)$. The latter is due to the embedding of $a$-th atom in a homogeneous
electron gas of density $\bar{\rho}^a$ \cite{tadmor2011modeling}.  The background electron density
function $\bar{\rho}^a$ is a linear superposition of contributions from each neighbor atom such that
the electronic energy $\tilde{F}(\bar{\rho}^a)$ of the $a$-th atom depends on the interatomic
distance $d_T( {\mbs{x}}^{a}, {\mbs{x}}^{b})$ to each neighbor.

In the case of metals, the environment of each atom is a nearly uniform electron gas and therefore
the embedded-atom approximation is reasonable.  Since we already investigated pair potentials in
\Cref{sec:pair-potentials}, we now investigate the discretization of the
forces due to the electronic energy, noting that the resultant forces in an EAM
potential must include the forces due to the pair potential, as well.

\subsection{Equation of motion}
We consider a system of $N$ particles in a periodic box $\mathcal{B}$ of side $L$ with
equations of motion~(\ref{eq-motion}) and an effective potential that depends only on the electronic energy of each particle, that is
\begin{equation}
  V  = \sum_{a=1}^N\tilde{F}\left(\sum_{\substack{b=1  \\b\neq a}}^N
    g_b\left(d_T( {\mbs{x}}^{a}, {\mbs{x}}^{b}) \right)\right),
\end{equation}
where $\tilde{V}:\mathbb{R}^+\cup \{0\}\to\mathbb{R}$.
Here,  $g_a:\mathbb{R}^+\cup \{0\}\to\mathbb{R}$ is a function of the relative interatomic distance
which represents a spherical electron density field around the isolated $a$-th particle \cite{tadmor2011modeling}.
Using the definitions in \cref{eq-notation-distance}, the background energy density is then given by
\begin{equation}
  \bar{\rho}^a=\sum_{\substack{b=1\\b\neq a}}^Ng_b\left(\bar{r}^{ab}\right)\ .
  \label{eq:background-energy-density}
\end{equation}
The forces acting on the particles are defined by \cref{eq-force} and have the standard form
\begin{equation}
  \mbs{f}^a
  =
  \sum_{\substack{b=1\\a\ne b}}^N \mbs{f}^{ab}
  \ ,
  \qquad
  \mathrm{with}
  \qquad
  \mbs{f}^{ab}
  \defined
  \varphi^{ab}\frac{ \bar{\mbs{r}}^{ab} }{ \bar{r}^{ab}}\ ,
  \label{eq-eam-force}
\end{equation}
where the strength of the force has now the structure
\begin{equation}
  \varphi^{ab}=
  \tilde{F}'\left(\bar{\rho}^a\right)g_b'\left(\bar{r}^{ab}\right)+\tilde{F}'(\bar{\rho}^b)g_a'\left(\bar{r}^{ab}\right)\ .
\end{equation}
We observe that, for this potential contribution, the direction of the interatomic force depends
only on the \replaced[id=1r]{difference}{distance} between the $a$-th and the $b$-th particle, while its strength  is further
determined by the background electron density at the $a$-th and $b$-th particle.

\subsection{Time discretization}
We consider now the integration in time of the equations of motion \eqref{eq-motion} of a system in the periodic box $\mathcal{B}$ and effective potential~\eqref{eq-v-eam} and employ the same time integration strategy as outlined in \Cref{sec-discretization-pair}.

\subsubsection{Midpoint scheme}
The canonical midpoint rule approximates the equation of motion by the implicit
formula~(\ref{eq-midpoint-pair}), where the midpoint approximation of the force acting on the $a$-th
particle in the direction of the $b$-th particle is given by
\begin{equation}
  \begin{aligned}
    \mbs{f}_{MP}^{ab} & = \varphi^{ab}_{MP}
\frac{\bar{\mbs{r}}^{ab}_{n+1/2}}{{|\bar{\mbs{r}}^{ab}_{n+1/2}|}},  \\
\varphi^{ab}_{MP} &
=\tilde{F}'\left(\bar{\rho}^a_{MP}\right)g_b'\left({\bar{{r}}^{ab}_{n+1/2}}\right)+\tilde{F}'(\bar{\rho}^b_{MP})
g_a'\left({\bar{{r}}^{ab}_{n+1/2}}\right),
  \end{aligned}
\end{equation}
and the midpoint approximation of the background energy density  reads
\begin{equation}
	\begin{aligned}
		\bar{\rho}_{MP}^{a}=\sum_{\substack{b=1 \\ b\neq a}}^Ng_b\left(\bar{{r}}^{ab}_{n+1/2} \right)\ .
	\end{aligned}
\end{equation}
As in the continuous case, the weak law of action and reaction is satisfied and
responsible for the conservation of total linear momentum of the system, see \Cref{sec-midpoint-pair}.

\subsubsection{Energy and momentum conserving discretization}
As illustrated in previous sections, it is possible to construct
a perturbation of the midpoint rule which, in addition to preserving the total linear momentum,
preserves the total energy. For that, we follow once more
the steps presented in
\Cref{sec:energy-moment-cons-pair}.  The partitioned discrete gradient will be again with a
slightly different structure due to the nature of the embedded function.  The partitioned discrete
gradient assumes the form
\begin{equation}
    \label{eq:em-force-eam}
    \mathsf{D}_{\mbs{x}^a}V
    =
    -\sum_{\substack{a,b=1 \\a\neq b}}^N\mbs{f}^{ab}_{algo}
    =-
    \sum_{\substack{a,b=1\\a\neq b}}^N\frac{1}{2}\left(\mbs{f}^{ab}_{n,n+1}+\mbs{f}^{ab}_{n+1,n}\right),
\end{equation}
with the contributions
\begin{equation}
  \begin{aligned}
    \mbs{f}^{ab}_{n,n+1} &= \mbs{f}_{MP}^{ab} +
    \frac{\Delta\tilde{F}^{ab}_{n,n+1}-\mbs{f}_{MP}^{ab}\cdot
      (\mbs{r}^{ab}_{n+1}-\mbs{r}^{ab}_{n})}{|\mbs{r}^{ab}_{n+1}-\mbs{r}^{ab}_{n}|}\;
    \mbs{n}\ ,\\
    \mbs{f}^{ab}_{n+1,n} &= \mbs{f}_{MP}^{ab} +
    \frac{\Delta\tilde{F}^{ab}_{n+1,n}-\mbs{f}_{MP}^{ab}\cdot
      (\mbs{r}^{ab}_{n+1}-\mbs{r}^{ab}_{n})}{|\mbs{r}^{ab}_{n+1}-\mbs{r}^{ab}_{n}|}\;
    \mbs{n}\ , \\
	\end{aligned}
\end{equation}
for which we introduced the compact notation
\begin{equation}
  \begin{aligned}
    \Delta\tilde{F}^{ab}_{n,n+1} & =\tilde{F}(\bar{\rho}^{a}_{n,n+1}(\bar{r}^{ab}_{n+1})) -\tilde{F}(\bar{\rho}^a_{n,n+1}(\bar{r}^{ab}_{n})) \\
    & +\tilde{F}(\bar{\rho}^{b}_{n,n+1}(\bar{r}^{ab}_{n+1}))
    -\tilde{F}(\bar{\rho}^b_{n,n+1}(\bar{r}^{ab}_{n})) \ ,\\
    \Delta\tilde{F}^{ab}_{n+1,n} & =\tilde{F}(\bar{\rho}^a_{n+1,n}(\bar{r}^{ab}_{n+1})) -\tilde{F}(\bar{\rho}^a_{n+1,n}(\bar{r}^{ab}_{n}))   \\
    & +\tilde{F}(\bar{\rho}^b_{n+1,n}(\bar{r}^{ab}_{n+1}))
    -\tilde{F}(\bar{\rho}^b_{n+1,n}(\bar{r}^{ab}_{n}))\ ,
	\end{aligned}
\end{equation}
and 
\begin{equation}
  \begin{aligned}
    \bar{\rho}^a_{n,n+1}(\bar{r}^{ab}) &
    =\sum_{d=1}^{b-1}g_a(\bar{r}^{ad}_n)+g_{a}(\bar{r}^{ab})+\sum_{e=b+1}^Ng_a(\bar{r}^{ae}_{n+1}) \
    ,\\
    \bar{\rho}^a_{n+1,n}(\bar{r}^{ab}) &
    =\sum_{d=1}^{b-1}g_a(\bar{r}^{ad}_{n+1})+g_{a}(\bar{r}^{ab})+\sum_{e=b+1}^Ng_a(\bar{r}^{ae}_{n})\
    . \\
  \end{aligned}
\end{equation}
The densities $\bar{\rho}^b_{n,n+1}(\bar{r}^{ab})$ and $\bar{\rho}^b_{n+1,n}(\bar{r}^{ab})$ are
defined similarly.

To show that the integrator exactly preserves the total momentum,
it suffices to follow the proof in \Cref{sec:energy-moment-cons-pair}.
The critical condition that a conserving scheme must satisfy reads now:
\begin{equation}
  \sum^N_{\substack{b=1 \\a < b}} \mbs{f}_{algo}^{ab}\cdot(\mbs{r}^{ab}_{n+1}-\mbs{r}^{ab}_{n})
  =
  \sum_{a=1}^N
  \left(
    \tilde{F}(\bar{\rho}^a_{n+1})-\tilde{F}(\bar{\rho}^a_{n})
  \right)
    \ ,
\end{equation}
which is indeed satisfied by the proposed method.

\subsection{Interatomic potential}
We consider for our subsequent analysis the Lennard-Jones-Baskes (LJB) EAM model
\cite{baskes1992modified,baskes1999many}, which is the extension of the Lennard-Jones material model
into the many-body regime of the EAM formalism \cite{srinivasan2004lennard}.  The total potential
energy of the LJB model is given by
\begin{equation}
  \label{eq:LJB-EAM-mat}
  \tilde{V}  = \frac{1}{2}\sum_{\substack{a,b=1\\a \neq b}}^NV(\bar{r}^{ab})
  +\sum_{\substack{a=1}}^N\tilde{F}(\bar{\rho}^a)\ .
\end{equation}
The two body part has been introduced in \Cref{sec:material-model-LJ}.
For the EAM contribution we further define
\begin{equation}
  \begin{aligned}
    \label{eq:material-eam}
    \tilde{F}(\bar{\rho}^a) & =
    \frac{1}{2}\epsilon AZ_1\bar{\rho}^a\left(\textrm{ln}\left(\bar{\rho}^a\right)-1\right) , \\
    \bar{\rho}^a &=
    \frac{1}{Z_1}\sum_{\substack{b=1 \\b\neq a}}^Ng_b\left({|\bar{\mbs{r}}^{ab}|}\right) , \\
    g_b\left({|\bar{\mbs{r}}^{ab}|}\right)
    &=
        \begin{cases}
                \exp\left(-\beta\left(\sigma^{-1}|\bar{\mbs{r}}^{ab}|-1\right)\right)& \text{if}\ |\bar{\mbs{r}}^{ab}|<r_c \\
                0& \text{otherwise} \ ,
        \end{cases}
  \end{aligned}
\end{equation}
where $\epsilon$, $\sigma$, $\beta$, $A$, $r_c$ and $Z_1$ are material parameters.

\subsection{Numerical evaluation}
Since we already investigated pair potentials in
\Cref{sec:pair-potentials}, we  focus on the EAM part of the LJB material model for the
following numerical evaluations.

\subsubsection{Accuracy study}
A three-dimensional box $[-L/2,L/2]^3$ is now considered with 5 particles inside it. The particles form a unit
body-centered cubic (BCC) cell of side $2$, centered within the box.  Starting from rest, the system
builds up kinetic energy due to the fact that the particles in the exterior of the BCC crystal
are not in equilibrium.  Further parameters of the simulation can be found
in \Cref{tab:table5}.
\begin{table}[H]
	\centering
	\caption{Accuracy study: Data used in the simulation}
	\label{tab:table5}
	\begin{tabular}{l c r | l}
          \hline
                                                                                                                                                              \\
          Material parameters      & $\epsilon$            & 2              & Initial configuration                                                           \\
                                   & $A$                   & 1              &                                                                                 \\
                                   & $\sigma$              & 1              & \multirow{4}{*}{\includegraphics[scale=0.125]{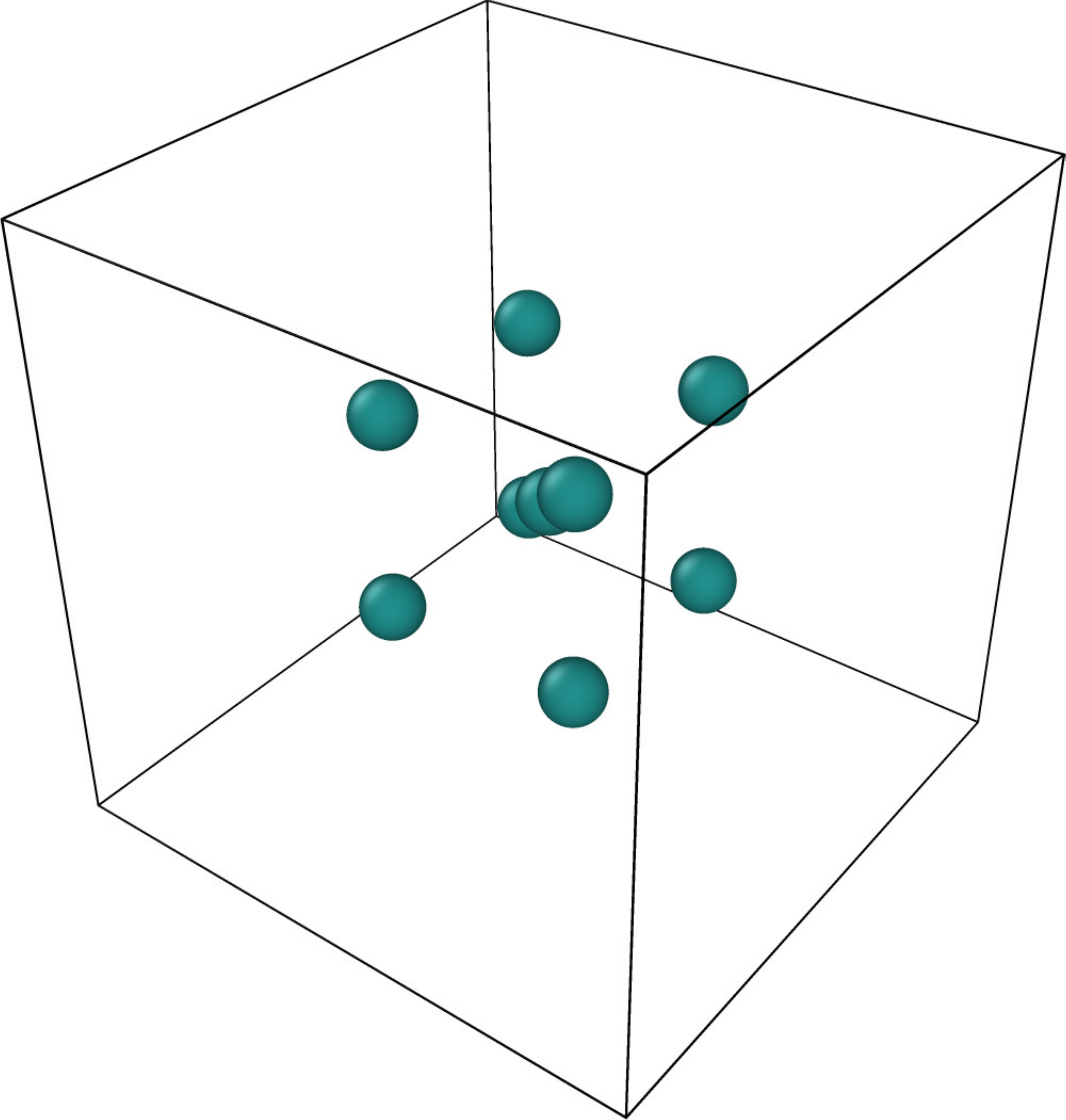}} \\
                                   & $\beta$               & 4              &                                                                                 \\
                                   & $Z_1$                 & 12             &                                                                                 \\
                                   & $r_c$                 & 3              &                                                                                 \\
          Mass                     & $m^{a}$               & 1              &                                                                                 \\
          Side length               & $L$                   & 6              &                                                                                 \\
          Newton tolerance         & \replaced[id=1r]{-}{$\epsilon$}            & 10$^{-6}$      &                                                                                 \\
          Simulation duration      & $T$                   & 0.5            &                                                                                 \\
          Reference time step size & $\Delta t_\text{ref}$ & 0.0001         &                                                                                 \\
          Time step size           & $\Delta t$            & 0.0005, 0.001, &                                                                                 \\
                                   &                       & 0.002, 0.0025, &                                                                                 \\
                                   &                       & 0.004, 0.005,  &                                                                                 \\
                                   &                       & 0.01, 0.02,    &                                                                                 \\
                                   &                       & 0.025          &                                                                                 \\
                                                                                                                                                              \\
          \hline
	\end{tabular}
\end{table}
Proceeding as in \Cref{sec:consistency-midpoint},
we perform an accuracy analysis of the EM integrator
using the midpoint rule as the reference solution. To study the convergence
of the numerical solutions, we employ ten time steps of decreasing size
and the error measures defined in \eqref{error-measures}.
\Cref{accuracy-study-position-eam,accuracy-study-momentum-eam} confirm again that
both the midpoint rule and the EM method are second order accurate approximations.
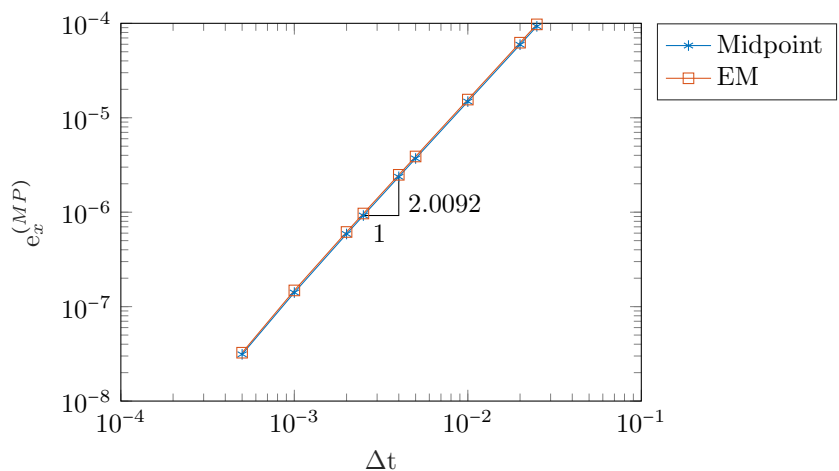
\begin{figure}[H]
	\centering
        \centering
	\setlength{\figH}{0.26\textheight}
	\setlength{\figW}{0.6\textwidth}
\definecolor{mycolor1}{rgb}{0.00000,0.44700,0.74100}%
\definecolor{mycolor2}{rgb}{0.85000,0.32500,0.09800}%
\begin{tikzpicture}

\begin{axis}[%
width=0.951\figW,
height=\figH,
at={(0\figW,0\figH)},
scale only axis,
xmode=log,
xmin=0.0001,
xmax=0.1,
xminorticks=true,
xlabel style={font=\color{white!15!black}},
xlabel={${\Delta}\text{t}$},
ymode=log,
ymin=1e-08,
ymax=0.0001,
yminorticks=true,
ylabel style={font=\color{white!15!black}},
ylabel={$\text{e}^{(MP)}_x$},
axis background/.style={fill=white},
legend style={legend pos = outer north east, legend cell align=left, align=left, draw=white!15!black}
]
\addplot [color=mycolor1, mark=asterisk, mark options={solid, mycolor1}]
  table[row sep=crcr]{%
0.0005	3.12624020624733e-08\\
0.001	1.41018307255814e-07\\
0.002	5.87644149018352e-07\\
0.0025	9.22057232207728e-07\\
0.004	2.37064483021028e-06\\
0.005	3.70677868318536e-06\\
0.01	1.48419894708889e-05\\
0.02	5.94008825082532e-05\\
0.025	9.28457786610436e-05\\
};
\addlegendentry{Midpoint}

\addplot [color=mycolor2, mark=square, mark options={solid, mycolor2}]
  table[row sep=crcr]{%
0.0005	3.25147061911456e-08\\
0.001	1.48455914724916e-07\\
0.002	6.1722390059182e-07\\
0.0025	9.68359655320297e-07\\
0.004	2.4882384162795e-06\\
0.005	3.89089076856158e-06\\
0.01	1.55760837755834e-05\\
0.02	6.23460216753594e-05\\
0.025	9.74568450476572e-05\\
};
\addlegendentry{EM}

\addplot [color=black, forget plot]
  table[row sep=crcr]{%
0.0025	9.22057232207728e-07\\
0.004	9.22057232207728e-07\\
0.004	2.37064483021028e-06\\
};
\node[right, align=left]
at (axis cs:0.004,0.00000125) {2.0092};
\node[right, align=left]
at (axis cs:0.0025,0.0000006) {1};
\end{axis}
\end{tikzpicture}%
	\caption{Accuracy study: Relative error in the position w.r.t midpoint rule}
	\label{accuracy-study-position-eam}
\end{figure}
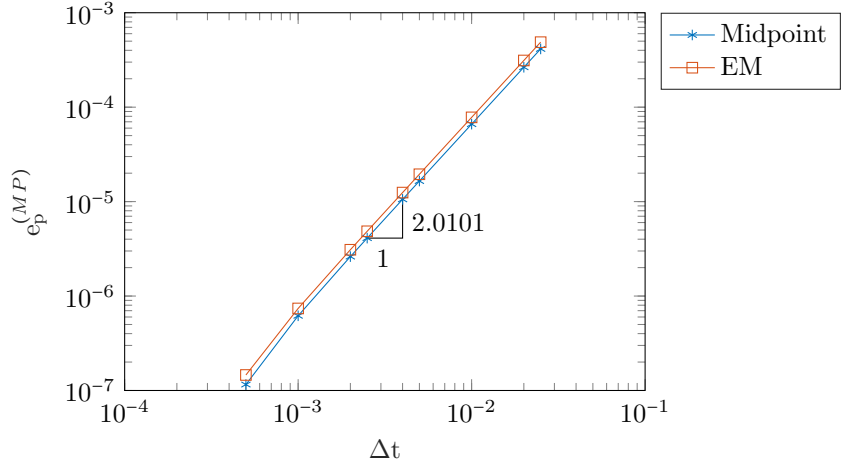
\begin{figure}[H]
	\centering
	\setlength{\figH}{0.26\textheight}
	\setlength{\figW}{0.6\textwidth}
\definecolor{mycolor1}{rgb}{0.00000,0.44700,0.74100}%
\definecolor{mycolor2}{rgb}{0.85000,0.32500,0.09800}%
\begin{tikzpicture}

\begin{axis}[%
width=0.951\figW,
height=\figH,
at={(0\figW,0\figH)},
scale only axis,
xmode=log,
xmin=0.0001,
xmax=0.1,
xminorticks=true,
xlabel style={font=\color{white!15!black}},
xlabel={${\Delta}\text{t}$},
ymode=log,
ymin=1e-07,
ymax=0.001,
yminorticks=true,
ylabel style={font=\color{white!15!black}},
ylabel={$\text{e}^{(MP)}_\text{p}\text{}$},
axis background/.style={fill=white},
legend style={legend pos = outer north east, legend cell align=left, align=left, draw=white!15!black}
]
\addplot [color=mycolor1, mark=asterisk, mark options={solid, mycolor1}]
  table[row sep=crcr]{%
0.0005	1.15582637128698e-07\\
0.001	6.17938900151803e-07\\
0.002	2.61546383681904e-06\\
0.0025	4.10660148978583e-06\\
0.004	1.05630229246447e-05\\
0.005	1.65167380798272e-05\\
0.01	6.61337988819561e-05\\
0.02	0.000264692726399661\\
0.025	0.000413742787792722\\
};
\addlegendentry{Midpoint}

\addplot [color=mycolor2, mark=square, mark options={solid, mycolor2}]
  table[row sep=crcr]{%
0.0005	1.45934503015496e-07\\
0.001	7.38974501109286e-07\\
0.002	3.09140636949256e-06\\
0.0025	4.85033173952639e-06\\
0.004	1.24611569485895e-05\\
0.005	1.94844319587346e-05\\
0.01	7.79916694787811e-05\\
0.02	0.000312163975970891\\
0.025	0.000487958146091766\\
};
\addlegendentry{EM}

\addplot [color=black, forget plot]
  table[row sep=crcr]{%
0.0025	4.10660148978583e-06\\
0.004	4.10660148978583e-06\\
0.004	1.05630229246447e-05\\
};
\node[right, align=left]
at (axis cs:0.004,0.000006) {2.0101};
\node[right, align=left]
at (axis cs:0.0025,0.0000025) {1};
\end{axis}
\end{tikzpicture}%
	\caption{Accuracy study: Relative error in the momentum w.r.t midpoint rule}
	\label{accuracy-study-momentum-eam}
\end{figure}

\subsection{Energy consistency study}
\begin{figure}[H]
        \centering
        \setlength{\figH}{0.26\textheight}
        \setlength{\figW}{0.6\textwidth}
        \input{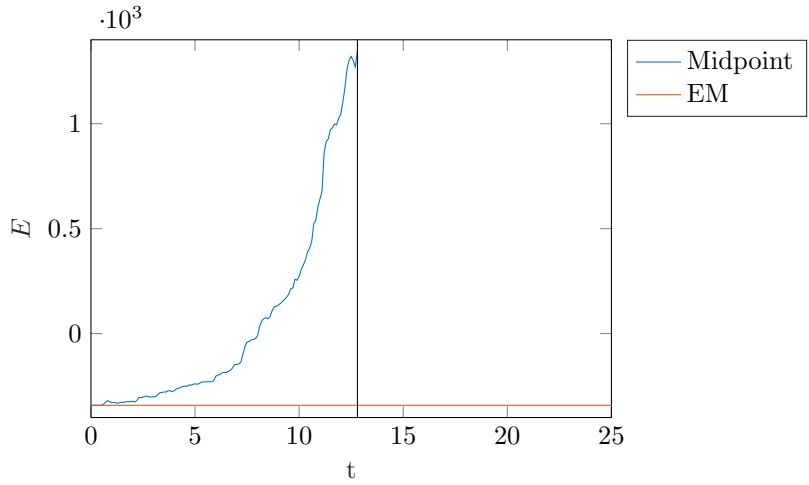}
	\caption{Energy consistency study: Total energy}
	\label{energy-consistency-study-eam-energy}
\end{figure}

\begin{figure}[H]
        \centering
        \setlength{\figH}{0.26\textheight}
        \setlength{\figW}{0.6\textwidth}
        \input{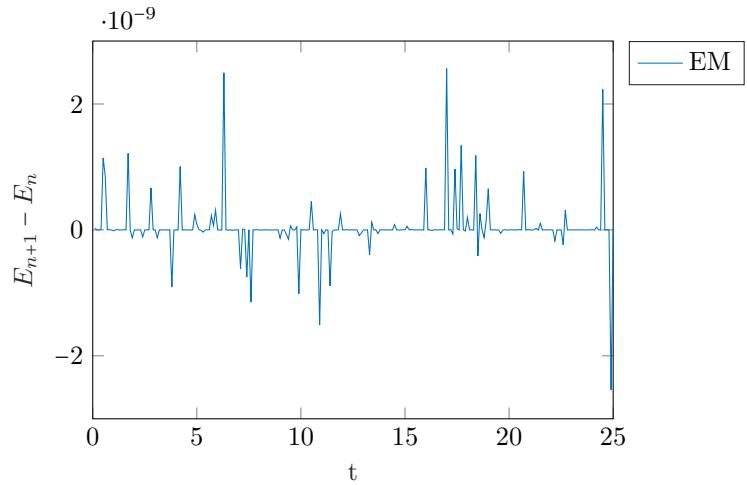}
	\caption{Energy consistency study: Total energy difference}
	\label{energy-consistency-study-eam-deltaEnergy}
\end{figure}

\begin{figure}[H]
        \centering
        \setlength{\figH}{0.26\textheight}
        \setlength{\figW}{0.6\textwidth}
        \input{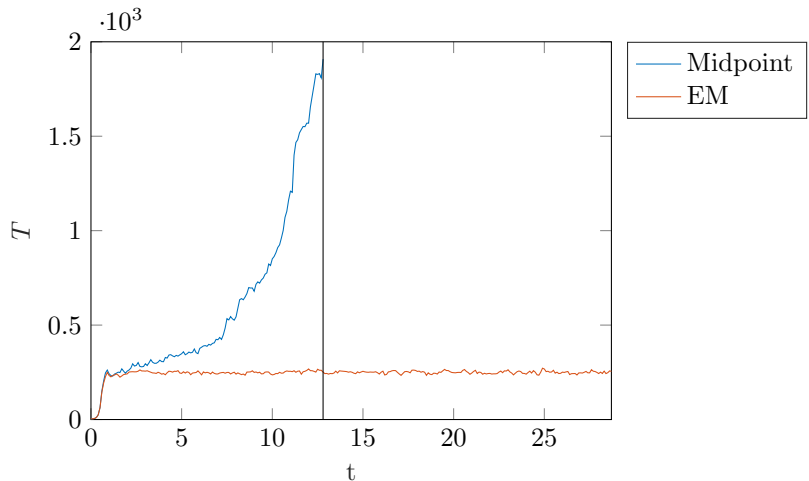}
        \caption{Energy consistency study: Kinetic energy}
	\label{accuracy-study-momentum-kinetic-energy}
\end{figure}

We consider in this last example
108 atoms inside a three-dimensional box $[-L/2,L/2]^3$,
where the atoms are arranged in a perfect face-centered cubic (FCC) lattice structure.
This FCC lattice is perturbed by the initial velocities of the atoms which are imparted
in such a way that the total initial kinetic energy is approximately $3.251$.
Further parameters of the simulation can be found
in \Cref{tab:table6}.
\begin{table}[H]
	\centering
	\caption{Energy consistency study: Data used in the simulation}
	\label{tab:table6}
	\begin{tabular}{l c r | l}
          \hline
                                                                                                                                 \\
          Material parameters & $\epsilon$ & 2         & Initial configuration                                                   \\
                              & $A$        & 1         &                                                                         \\
                              & $\sigma$   & 1         & \multirow{4}{*}{\includegraphics[scale=0.125]{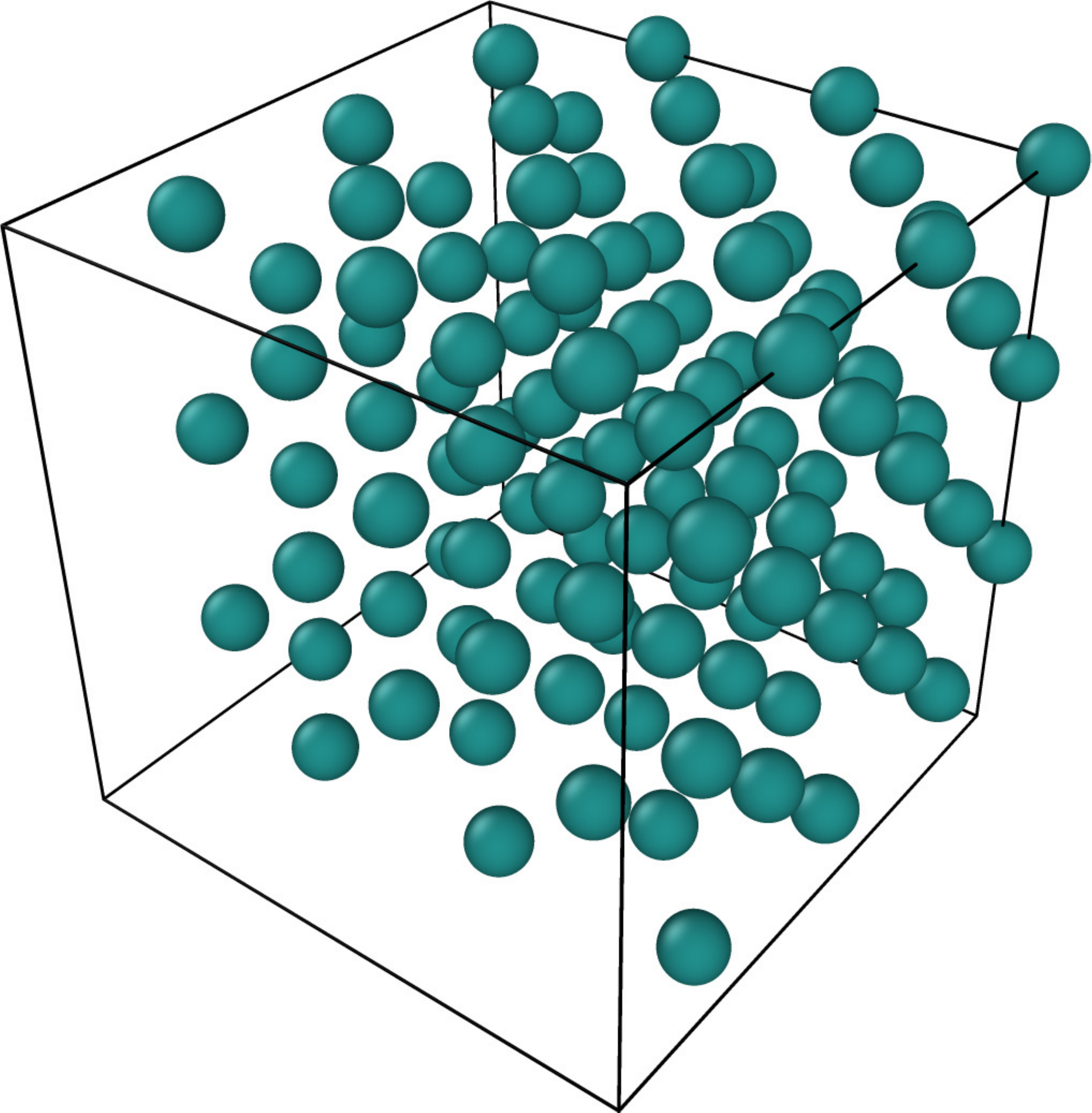}} \\
                              & $\beta$    & 4         &                                                                         \\
                              & $Z_1$      & 12        &                                                                         \\
                              & $r_c$      & 3         &                                                                         \\
          Mass                & $m^{a}$    & 1         &                                                                         \\
          Side length          & $L$        & 6         &                                                                         \\
          Newton tolerance    & \replaced[id=1r]{-}{$\epsilon$} & 10$^{-8}$ &                                                                         \\
          Simulation duration & $T$        & 40        &                                                                         \\
          Time step size      & $\Delta t$ & 0.1       &                                                                         \\
                                                                                                                                 \\
          \hline
	\end{tabular}
      \end{table}
      For the time step selected, the midpoint rule clearly violates the conservation of total energy
      (see \Cref{energy-consistency-study-eam-energy}). As in previous examples, this leads to an energy blow-up and finally to a termination of the simulation, indicated by the black line on the same figure.
In contrast, the EM method preserves the total energy up to round off errors, see \Cref{energy-consistency-study-eam-deltaEnergy}.
Finally, and as in the previous examples, \Cref{accuracy-study-momentum-kinetic-energy}
shows that the EM, but not the midpoint rule, is able to evolve the system of particles
up to thermodynamic equilibrium.

%%% Local Variables:
%%% TeX-command-extra-options: "-shell-escape"
%%% mode: latex
%%% TeX-master: "driver"
%%% End:

\section{Summary of main results}
\label{sec:conclusion}
Conserving schemes have been employed during more than four decades 
for the accurate and robust approximation of Hamiltonian problems
in mechanics. However, in the field of molecular dynamics, where time
integrators are routinely employed and are key to their
usefulness, the use of conserving schemes has barely been explored.

In this work we have analyzed energy and momentum conserving
schemes in the context of molecular dynamics. This second order,
implicit methods are an interesting alternative to other
commonly used integrators and we have proven than, in addition
to exhibiting exact energy conservation, are more robust
than the midpoint rule, the canonical second order implicit method.

The article has focused on the design of Energy-Momentum schemes for molecular dynamics in the view
of three issues that are characteristic and unique to these problems. First, the numerical treatment
of dynamics in periodic domains; second, the discretization of three-body potentials; last, the study
of interatomic functional potentials. Neither of these three topics had been previously studied in the
context of conserving schemes, to the authors' knowledge. However, the three of them are
key for their implementation and clearly have an impact on their performance.

Some of the most practical results of this work are new expressions for Energy-Momentum
approximations in fairly general problems in molecular dynamics. These approximations account for
the three key issues mentioned before, and can be shown to preserve linear momentum and energy,
exactly, while exhibiting a remarkable robustness.

%%% Local Variables:
%%% TeX-command-extra-options: "-shell-escape"
%%% mode: latex
%%% TeX-master: "driver"
%%% End:

\section{Acknowledgements}
Support for M.S. was provided by the Deutsche Forschungsgemeinschaft (DFG) under Grant BE
2285/13-1 and the Research Travel Grant of the Karlsruhe House of
Young Scientists (KYHS). This support is gratefully acknowledged.

\bibliographystyle{elsarticle-num}
\bibliography{bib}

\end{document}